\documentclass[showpacs,preprintnumbers,amsmath,amssymb]{revtex4}

\usepackage{epsfig}
\usepackage{graphicx}
\usepackage{dcolumn}
\usepackage{bm}

\begin{document}


\title{Finite-Energy Spectral-Weight Distributions of a 1D Correlated Metal}
\author{J. M. P. Carmelo}
\affiliation{Department of Physics, Massachusetts Institute of
Technology, Cambridge, Massachusetts 02139-4307, USA}
\affiliation{GCEP - Center of Physics, University of Minho, Campus
Gualtar, P-4710-057 Braga, Portugal}
\author{K. Penc}
\affiliation{Research Institute for Solid State Physics and
Optics, H-1525 Budapest, P.O.B. 49, Hungary}
\author{D. Bozi}
\affiliation{Department of Physics, Massachusetts Institute of
Technology, Cambridge, Massachusetts 02139-4307, USA}
\affiliation{GCEP and Department of Physics, University of Minho,
Campus Gualtar, P-4710-057 Braga, Portugal}
\date{24 May 2005}

\begin{abstract}
We derive general closed-form analytical expressions for the
finite-energy one- and two-electron spectral-weight distributions
of an one-dimensional correlated metal with on-site electronic
repulsion. Our results also provide general expressions for the
one- and two-atom spectral functions of a correlated quantum
system of cold fermionic atoms in a one-dimensional optical
lattice with on-site atomic repulsion. In the limit of zero spin
density our spectral-function expressions provide the correct
zero-spin density results. Our results reveal the dominant
non-perturbative microscopic many-particle mechanisms behind the
exotic spectral properties observed in quasi-one-dimensional
metals and correlated systems of cold fermionic atoms in
one-dimensional optical lattices.
\end{abstract}

\pacs{70}

\maketitle
\section{INTRODUCTION}
\label{SecI}

The main goal of this paper is to provide a detailed derivation of
the one-electron and two-electron spectral-weight distributions
recently used in Refs. \cite{spectral0,spectral1,super} in the 
unusual spectral properties of low-dimensional metals. Moreover,
our results will be used elsewhere in the study of the spectral
properties of the new quantum systems described by cold fermionic
atoms in optical lattices: following the experimental studies of
strongly correlated quantum systems of ultra cold bosonic atoms
\cite{Greiner}, new experiments involving cold fermionic atoms
(such as $^6Li$) in an optical lattice formed by interfering laser
fields are in progress \cite{NEW}.

We profit from symmetries and properties specific to the metallic
phase of the 1D Hubbard model \cite{Lieb} to derive closed-form
analytical expressions for the finite-energy one- and two-electron
spectral-weight distributions. The model also describes cold
fermionic atoms in optical lattices, provided that the electrons
are replaced by these atoms. The recently developed quantum
optical tools for manipulating atoms can be used to realize a 1D
lattice or any other lattice described by the Hubbard model
\cite{NEW,cold,Honerkamp}. Such correlated quantum systems have
applications in quantum computing \cite{cold}. Our study involves
the use of the holon and spinon description of Ref. \cite{I} and
of the general pseudofermion dynamical theory introduced in Ref.
\cite{V}: we are able to explicitly calculate the pseudfermion
spectral weights for the metallic phase. The pseudofermion
description was introduced in Ref. \cite{IIIb} and is such that
the pseudofermions which describe the elementary excitations are
freely propagating, and the information about the interactions is
encoded in their overall phase shifts \cite{S-P}. The method used
here is a generalization for all values of the on-site repulsion
$U$ of that used in Refs. \cite{Penc9596,Penc97} for
$U\rightarrow\infty$. For finite values of $U$ there were no
closed-form analytical expressions for the finite-energy
spectral-weight distributions, in contrast to simpler models
\cite{Lee}.

Our general closed-form analytical expressions for the one- and
two-electron spectral functions are controlled by the dominant
microscopic processes which are behind nearly the whole spectral
weight for all values of the momentum $k$, energy $\omega$, and
on-site repulsion $U$. In addition to other spectral features, we
explicitly derive the spectral-weight distribution expressions in
the vicinity of the the singular and edge branch lines used in the
study of the photoemission spectra and phase diagram of quasi-1D
metals in Refs. \cite{spectral0,spectral1,super,spectral}. The
preliminary results of those papers confirm that the
spectral-weight distributions derived here describe
quantitatively, for the whole energy band width, the unusual
spectral properties observed in such materials.

The paper is organized as follows: In Sec. \ref{SecII} we
introduce the model and provide basic information about the
pseudofermion description needed for our studies. The specific
classes of elementary microscopic processes which generate the one
and two-electron spectral weight of the metallic phase is the
subject of Sec. \ref{SecIII}. In Sec. \ref{SecIV} we calculate the
general expressions for the finite-energy spectral-weight
distributions of the metallic phase. The characterization of the
dominant processes behind such distributions and the derivation of
the closed-form analytical expressions generated by these
processes for the same distributions is the subject of Sec
\ref{SecV}. Finally, the concluding remarks are presented in Sec.
\ref{SecVI}.

\section{THE MODEL, THE WEIGHT DISTRIBUTIONS, ROTATED ELECTRONS, AND
THE HOLON, SPINON, AND PSEUDOFERMION DESCRIPTION} \label{SecII}

The 1D Hubbard model reads,

\begin{equation}
\hat{H}=-t\sum_{j,\,\sigma}[c_{j,\,\sigma}^{\dag} c_{j+1,\,\sigma}
+ h. c.]+U\sum_{j}\hat{n}_{j,\uparrow}\hat{n}_{j,\downarrow} \, ,
\label{H}
\end{equation}
where $c_{j,\,\sigma}^{\dagger}$ and $c_{j,\,\sigma}$ are
spin-projection $\sigma =\uparrow ,\downarrow$ electron operators
at site $j=1,2,...,N_a$ and
$\hat{n}_{j,\,\sigma}=c_{j,\,\sigma}^{\dagger}\,c_{j,\,\sigma}$.
The model (\ref{H}) describes $N_{\uparrow}$ spin-up electrons and
$N_{\downarrow}$ spin-down electrons in a chain of $N_a$ sites. We
use periodic boundary conditions and units such that the Planck
constant and electronic lattice constant are one. We denote the
electronic number by $N=N_{\uparrow}+N_{\downarrow}$. The number
of lattice sites $N_a$ is even and very large and in the units
used here the chain length reads $L=N_a$. We consider electronic
densities $n=n_{\uparrow }+n_{\downarrow}$ and spin densities
$m=n_{\uparrow}-n_{\downarrow}$ in the range $0<n< 1$ and $0<m<n$,
respectively, where $n_{\sigma}=N_{\sigma}/L=N_{\sigma}/N_a$. In
the thermodynamic limit the Fermi momenta read $k_F=\pi n/2$ and
$k_{F\sigma}=\pi n_{\sigma}$. We denote the electronic charge by
$-e$.

We consider the following general finite-$\omega$
$\cal{N}$-electron spectral-weight distribution,

\begin{equation}
B_{{\cal{N}}}^{l} (k,\,\omega) = \sum_{f}\, \vert\langle f\vert\,
{\hat{O}}_{{\cal{N}}}^{l} (k) \vert GS\rangle\vert^2\,\delta\Bigl(
\omega - l[E_f - E_{GS}]\Bigr) \, ; \hspace{0.5cm} l\omega > 0 \,
; \hspace{0.5cm} l = \pm 1 \, . \label{ABON}
\end{equation}
Here the general $\cal{N}$-electron operators
${\hat{O}}_{{\cal{N}}}^{+1} (k)\equiv {\hat{O}}_{{\cal{N}}}^{\dag}
(k)$ and ${\hat{O}}_{{\cal{N}}}^{-1} (k) \equiv
{\hat{O}}_{{\cal{N}}} (k)$ carry momentum $k$, the $f$ summation
runs over the excited energy eigenstates, the energy $E_f$
corresponds to these states, and $E_{GS}$ is the ground-state
energy. The local operator ${\hat{O}}_{{\cal{N}},\,j}^{+1}\equiv
{\hat{O}}_{{\cal{N}},\,j}^{\dag}$ or
${\hat{O}}_{{\cal{N}},\,j}^{-1}\equiv {\hat{O}}_{{\cal{N}},\,j}$
is related to the corresponding momentum-representation operator
${\hat{O}}_{{\cal{N}}}^{l} (k)$ of Eq. (\ref{ABON}) by a Fourier
transform. (Examples of such $\cal{N}$-electron operators are
given in Ref. \cite{V}.) For simplicity, we use in expression
(\ref{ABON}) a momentum extended scheme such that $k\in
(-\infty,\,+\infty)$, yet it is a simple exercise to obtain the
corresponding spectral function expressions for the first
Brillouin zone. The general spectral-weight distribution
(\ref{ABON}) obeys the following sum rule,

\begin{equation}
{lN_a\over 2\pi}\int_{-\infty}^{+\infty} dk\int_{0}^{l\infty}
d\omega\,B_{{\cal{N}}}^{l} (k,\,\omega) =
\sum_{j=1}^{N_a}\,\langle GS\vert\,
{\hat{O}}_{{\cal{N}},\,j}^{-l}\,
{\hat{O}}_{{\cal{N}},\,j}^{l}\vert GS\rangle \, ; \hspace{0.5cm} l
= \pm 1 \, . \label{sum-ABON}
\end{equation}
When ${\hat{O}}_{{\cal{N}},\,j}^{l}$ is hermitian one has that
${\hat{O}}_{{\cal{N}},\,j}^{l}={\hat{O}}_{{\cal{N}},\,j}^{-l}$. In
this case the convention is that $l=+1$ and thus the energy
$\omega$ is positive.

The concept of a rotated electron plays a key role in the
pseudofermion description. Concerning its relation to the holons,
spinons, and $c0$ pseudofermions whose occupancy configurations
describe the energy eigenstates of the model (\ref{H}), see Ref.
\cite{IIIb}. The holons have $\eta$-spin $1/2$, $\eta$-spin
projections $\pm 1/2$, charge $\pm 2e$, and spin zero. The spinons
have spin $1/2$, spin projections $\pm 1/2$, and no charge degrees
of freedom. Throughout this paper we denote such holons and
spinons according to their value of $\eta$-spin projections $\pm
1/2$ and spin projections $\pm 1/2$, respectively. Furthermore,
the $c0$ pseudfermion is a composite quantum object of a chargeon
of charge $-e$ and an antichargeon of charge $+e$. It has no
$\eta$-spin and no spin degrees of freedom. The $2\nu$-holon  (and
$2\nu$-spinon) composite $c\nu$ pseudofermions (and $s\nu$
pseudofermions) are $\eta$-spin and spin zero (and spin zero)
quantum objects (and with no charge degrees of freedom). These
composite quantum objects contain an equal number $\nu$ of $-1/2$
holons and $+1/2$ holons (and $-1/2$ spinons and $+1/2$ spinons).
The $\pm 1/2$ holons (and $\pm 1/2$ spinons) which are not
associated with $2\nu$-holon $c\nu$ pseudofermions (and
$2\nu$-spinon $s\nu$ pseudofermions) are the $\pm 1/2$ Yang holons
(and $\pm 1/2$ HL spinons). We denote the numbers of $\alpha\nu$
pseudofermions and $\alpha\nu$ pseudofermion holes by
$N_{\alpha\nu}$ and $N^h_{\alpha\nu}$, respectively, where $\alpha
=c,\,s$ and $\nu =0,1,2,...$ for $\alpha =c$ and $\nu =1,2,...$
for $\alpha =s$. (The value of $N^h_{\alpha\nu}$ is given in Eqs.
(B7) and (B8) of Ref. \cite{I}.) Moreover, $L_{\alpha,\,\pm 1/2}$
denotes the number of $\pm 1/2$ Yang holons ($\alpha =c$) or $\pm
1/2$ HL spinons ($\alpha =s$).

Often in this paper we use the notation $\alpha\nu\neq c0,\,s1$
branches, which refers to all $\alpha\nu$ branches except the $c0$
and $s1$ branches. Moreover, the summations (and products)
$\sum_{\alpha\nu}$, $\sum_{\alpha\nu =c0,\,s1}$, and
$\sum_{\alpha\nu\neq c0,\,s1}$ (and $\prod_{\alpha\nu}$,
$\prod_{\alpha\nu =c0,\,s1}$, and $\prod_{\alpha\nu\neq c0,\,s1}$)
run over all $\alpha\nu$ branches with finite $\alpha\nu$
pseudofermion occupancy in the corresponding state or subspace,
the $c0$ and $s1$ branches only, and all $\alpha\nu$ branches with
finite $\alpha\nu$ pseudofermion occupancy in the corresponding
state or subspace except the $c0$ and $s1$ branches, respectively.

The pseudofermion description refers to a Hilbert subspace called
the {\it pseudofermion subspace} (PS) in Ref. \cite{IIIb}, in
which the $\cal{N}$-electron excitations
${\hat{O}}_{{\cal{N}}}^{l} (k)\vert GS\rangle$ are contained. The
PS is spanned by the initial ground state and the excited energy
eigenstates originated from it by creation, annihilation, and
particle-hole processes involving the generation of a finite
number of active pseudofermion scattering centers, $-1/2$ Yang
holons, and $-1/2$ HL spinons plus a vanishing or small density of
low-energy and small-momentum $\alpha\nu =c0,\,s1$ pseudofermion
particle-hole processes. All these processes are defined in Refs.
\cite{V,IIIb}.

The $\alpha\nu$-pseudofermion discrete canonical-momentum values
have a functional character and read, ${\bar{q}}_j = q_j +
Q^{\Phi}_{\alpha\nu} (q_j)/L = [2\pi/ L] I^{\alpha\nu}_j +
Q^{\Phi}_{\alpha\nu} (q_j)/L$ where $j = 1, 2, ...,
N^*_{\alpha\nu}$ and
$N^*_{\alpha\nu}=N_{\alpha\nu}+N_{\alpha\nu}^h$. In this paper we
use the designation {\it functional} for all quantities whose
value depends on the energy eigenstates bare-momentum occupancy
configurations. According to the studies of Ref. \cite{S-P},
$Q^{\Phi}_{\alpha,\,\nu}(q_j)/2$ is a $\alpha\nu$ pseudofermion
scattering phase shift given by,

\begin{equation}
Q^{\Phi}_{\alpha\nu} (q_j)/2 = \pi \sum_{\alpha'\nu'}\,
\sum_{j'=1}^{N^*_{\alpha'\nu'}}\,\Phi_{\alpha\nu,\,\alpha'\nu'}(q_j,q_{j'})\,
\Delta N_{\alpha'\nu'}(q_{j'}) \, ; \hspace{0.5cm} j = 1, 2, ...,
N^*_{\alpha\nu} \, . \label{qcan1j}
\end{equation}
Here $\Delta N_{\alpha\nu}(q_{j})=\Delta {\cal{N}}_{\alpha\nu}
({\bar{q}}_j)$ is the bare-momentum distribution function
deviation $\Delta N_{\alpha\nu} (q_j) = N_{\alpha\nu} (q_j) -
N^0_{\alpha\nu} (q_j)$ corresponding to the excited energy
eigenstate. (The ground-state densely-packed bare-momentum
distribution function $N^0_{\alpha\nu} (q_j)$ is defined in Eqs.
(C.1)-(C.3) of Ref. \cite{I}.) This deviation is expressed in
terms of the {\it bare-momentum} $q_j=[2\pi I^{\alpha\nu}_j]/L$,
which is carried by the $\alpha\nu$ pseudoparticles, where
$I^{\alpha\nu}_j$ are the quantum numbers provided by the
Bethe-ansatz solution \cite{I}. Although the $\alpha\nu$
pseudoparticles carry bare-momentum $q_j$, one can also label the
corresponding $\alpha\nu$ pseudofermions by such a bare-momentum.
When we refer to the pseudofermion bare-momentum $q_j$, we mean
that $q_j$ is the bare-momentum value that corresponds to the
pseudofermion canonical momentum ${\bar{q}}_j = q_j +
Q^{\Phi}_{\alpha\nu} (q_j)/L$. In contrast to the $\alpha\nu$
pseudoparticles, the $\alpha\nu$ pseudofermions have no
residual-interaction energy terms. Instead, under the ground-state
- excited-energy-eigenstate transitions the $\alpha\nu$
pseudofermions and $\alpha\nu$ pseudofermion holes undergo
elementary scattering events with the $\alpha'\nu'$ pseudofermions
and $\alpha'\nu'$ pseudofermion holes created in these transitions
\cite{S-P}. This leads to the elementary two-pseudofermion phase
shifts $\pi\,\Phi_{\alpha\nu,\,\alpha'\nu'}(q_j,{q'}_j)$ on the
right-hand side of the overall scattering phase shift
(\ref{qcan1j}), defined by Eqs. (15) and (A1)-(A13) of Ref.
\cite{IIIb}. Moreover, the overall $\alpha\nu$ pseudofermion or
hole phase shift,

\begin{equation}
Q_{\alpha\nu}(q_j)/2 = Q_{\alpha\nu}^0/2 + Q^{\Phi}_{\alpha\nu}
(q_j)/2 \, , \label{Qcan1j}
\end{equation}
plays an important role in the pseudofermion theory \cite{S-P}.
Here $Q_{\alpha\nu} (q_j)/L$ gives the shift in the discrete
canonical-momentum value ${\bar{q}}_j $ that arises due to the
transition from the ground state to an excited energy eigenstate
and $Q_{\alpha\nu}^0/2$ can have the values $Q_{\alpha\nu}^0/2
=0,\,\pm\pi/2$ and is defined in Eq. (11) of Ref. \cite{IIIb}. In
this paper we use boundary conditions such that $Q_{\alpha\nu}^0/2
=0,\,-{\rm sgn}(k)\,\pi/2$, where $k$ is the excited-state
momentum relative to that of the initial ground state. Here we
assume that for the latter state $N/2$ and $N$ are odd and even
numbers, respectively. $Q_{\alpha\nu}^0/L$ gives the shift in the
discrete bare-momentum value $q_j $ that arises as a result of the
same transition. In reference \cite{S-P} it is confirmed that
$Q_{\alpha\nu} (q_j)/2$, Eq. (\ref{Qcan1j}), is a $\alpha\nu$
pseudofermion or hole overall phase shift associated with the
ground-state - excited-energy-eigenstate transition. Its
scattering part is the functional $Q^{\Phi}_{\alpha\nu} (q_j)/2$
given in Eq. (\ref{qcan1j}). (In contrast to the pseudofermions,
the $-1/2$ Yang holons and $-1/2$ HL spinons are scattering-less
objects \cite{S-P}.) The $\alpha\nu$ pseudofermion creation and
annihilation operators $f^{\dag}_{{\bar{q}}_j,\,\alpha\nu}$ and
$f_{{\bar{q}}_j,\,\alpha\nu}$, respectively, have exotic
anticommutation relations, provided in Ref. \cite{IIIb}. These
anticommutators involve the overall phase shifts (\ref{Qcan1j})
and play a key role in the spectral properties.

For the ground state the pseudofermion, $-1/2$ Yang holon, and
$-1/2$ HL spinon numbers are given by $N_{c0}=N$,
$N_{s1}=N_{\downarrow}$,
$N_{\alpha\nu}=L_{c,\,-1/2}=L_{s,\,-1/2}=0$ for $\alpha\nu\neq
c0,\, s1$. We call $N^0_{c0}$ and $N^0_{s1}$ the ground-state $c0$
and $s1$ pseudofermion numbers, respectively. The ground-state
$\alpha\nu =c0,\,s1$ bare-momentum distribution functions are such
that there is pseudofermion occupancy for $\vert q\vert\leq
q^0_{F\alpha\nu}$ and unoccupancy for $q^0_{F\alpha\nu}<\vert
q\vert\leq q^0_{\alpha\nu}$, where in the thermodynamic limit the
{\it Fermi-point} values are given by,

\begin{equation}
q^0_{Fc0} = 2k_F \, ; \hspace{0.5cm} q^0_{Fs1} = k_{F\downarrow}
\, . \label{q0Fcs}
\end{equation}
Moreover, for that state the limiting bare-momentum values of both
the $\alpha\nu =c0,\,s1$ and $\alpha\nu\neq c0,\,s1$ bands read,

\begin{equation}
q^0_{c0} = \pi \, ; \hspace{0.5cm} q^0_{s1} = k_{F\uparrow} \, ;
\hspace{0.5cm} q^0_{c\nu} = [\pi -2k_F] \, , \hspace{0.3cm} \nu
>0 \, ; \hspace{0.5cm} q^0_{s\nu} =
[k_{F\uparrow}-k_{F\downarrow}] \, , \hspace{0.3cm} \nu >1 \, .
\label{qcanGS}
\end{equation}
Within the thermodynamic limit, when in this paper we refer to
creation (or annihilation) of $\alpha\nu =c0,\,s1$ pseudofermions
with bare-momentum values at the {\it Fermi points}, we mean that
such a process leads to a densely packed bare-momentum
pseudofermion distribution for the final state, where the created
(or annihilated) pseudofermions occupy neighboring discrete
bare-momentum values $\pm q^0_{F\alpha\nu}+2j\pi/L$ such that
$j=\pm 1,\pm 2,...$ (or $j=0, \mp 1,\mp 2,...$). Similarly, when
we refer to creation of $\alpha\nu \neq c0,\,s1$ pseudofermions at
the limiting bare-momentum values, we mean that such a process
leads to a densely packed bare-momentum pseudofermion hole
distribution for the final state, where the created pseudofermions
occupy neighboring discrete bare-momentum values $\pm
q^0_{\alpha\nu}+2j\pi/L$ such that $j=0, \mp 1,\mp 2,...$. As
further discussed below, the $c\nu$ pseudofermions of limiting
bare-momentum value $q=\pm q_{c\nu}^0=\pm [\pi -2k_F]$ and $s\nu$
pseudofermions of limiting bare-momentum value $q=\pm
q_{s\nu}^0=\pm [k_{F\uparrow}-k_{F\downarrow}]$ can be described
in terms of $2\nu$ independent holons and $2\nu$ independent
spinons, respectively \cite{S-P}.

Our study of the metallic phase of the model (\ref{H}) involves
several numbers and number deviations. $N^{phNF}_{\alpha\nu}$ is
the number of finite-momentum and finite-energy
$\alpha\nu=c0,\,s1$ pseudofermion particle-hole processes. The
quantum number $\iota ={\rm sgn} (q) 1=\pm 1$ refers to the right
pseudfermion movers ($\iota =+1$) and left pseudfermion movers
($\iota =-1$) and $\Delta N^F_{\alpha\nu ,\,\iota}$ such that
$\Delta N^F_{\alpha\nu ,\,\pm 1}$ is the deviation in the number
of $\alpha\nu$ pseudofermions at the right $(+1)$ and left $(-1)$
{\it Fermi points}. In turn, the deviation in the number of
$\alpha\nu=c0,\,s1$ pseudofermions created or annihilated away
from these points is denoted by $\Delta N^{NF}_{\alpha\nu}$. The
actual number of $\alpha\nu$ pseudofermions created or annihilated
at the right $(+1)$ and left $(-1)$ {\it Fermi points} is denoted
by $\Delta N^{0,F}_{\alpha\nu ,\,\pm 1}$. It is such that $\Delta
N^F_{\alpha\nu,\,\iota}= \Delta
N^{0,F}_{\alpha\nu,\,\iota}+\iota\,Q^ 0_{\alpha\nu}/2\pi$, where
$Q_{\alpha\nu}^0/2$ is the scattering-less phase shift on the
right-hand side of Eq. (\ref{Qcan1j}). Furthermore,
$N_{\alpha\nu,\,\iota}^{F}$ refers to the $\alpha\nu\neq c0,\, s1$
branches and is the number of $\alpha\nu$ pseudofermions of
limiting bare momentum $q=\iota\,q_{\alpha\nu}^0$ such that $\iota
=\pm 1$. The number of $\alpha\nu$ pseudofermions created away
from the limiting bare-momentum values is called
$N_{\alpha\nu}^{NF}$.

\section{SPECTRAL-WEIGHT ELEMENTARY PROCESSES FOR DENSITIES $0<n<1$ AND
$0<m<n$}\label{SecIII}

In this section we study the pseudofermion occupancy
configurations of the PS energy eigenstates. As in reference
\cite{IIIb}, we classify the pseudofermion elementary processes
that generate the PS from the ground state into three types:
\vspace{0.25cm}

(A) Finite-energy and finite-momentum elementary $c0$ and $s1$
pseudofermion processes away from the corresponding {\it Fermi
points} involving creation or annihilation of a finite number of
pseudofermions plus creation of $\alpha\nu\neq c0,\,s1$
pseudofermions with bare-momentum values different from the
limiting bare-momentum values $\pm q_{\alpha\nu}^0$;
\vspace{0.25cm}

(B) Zero-energy and finite-momentum processes that change the
number of $c0$ and $s1$ pseudofermions at the $\iota={\rm sgn}
(q)\,1=+1$ right and $\iota={\rm sgn} (q)\,1=-1$ left $c0$ and
$s1$ {\it Fermi points} - these processes transform the
ground-state densely packed bare-momentum occupancy configuration
into an excited-state densely packed bare-momentum occupancy
configuration. Furthermore, creation of a finite number of
independent $-1/2$ holons and independent $-1/2$ spinons,
including $-1/2$ Yang holons, $-1/2$ HL spinons, and $-1/2$ holons
and $-1/2$ spinons associated with $c\nu$ pseudofermions of
limiting bare momentum $q=\pm q_{c\nu}^0=\pm [\pi -2k_F]$ and
$s\nu$ pseudofermions of limiting bare momentum $q=\pm
q_{s\nu}^0=\pm [k_{F\uparrow}-k_{F\downarrow}]$,
respectively;\vspace{0.25cm}

(C) Low-energy and small-momentum elementary $c0$ and $s1$
pseudofermion particle-hole processes in the vicinity of the
$\iota={\rm sgn} (q)\,1=+1$ right and $\iota={\rm sgn} (q)\,1=-1$
left $c0$ and $s1$ {\it Fermi points}, relative to the
excited-state $\alpha\nu= c0,\,s1$ pseudofermion densely packed
bare-momentum occupancy configurations generated by the above
elementary processes (B).\vspace{0.25cm}

It is found in reference \cite{S-P} that the invariance under the
electron - rotated-electron unitary transformation of the
$\alpha\nu$ pseudofermions created at limiting bare momentum
$q=\pm q_{\alpha\nu}^0$ and belonging to $\alpha\nu\neq c0,\,s1$
branches implies that each of such $c\nu$ pseudofermions (and
$s\nu$ pseudofermions) separates into $2\nu$ independent holons
(and $2\nu$ independent spinons) and a $c\nu$ (and $s\nu$) FP
scattering center. By independent holons and spinons it is meant
those which remain invariant under the electron - rotated-electron
unitary transformation. (The Yang holons and HL spinons are also
independent holons and spinons, respectively.) The above
designation FP stands for {\it Fermi points}. Indeed, it is found
in the same reference that the $c0$ and $s1$ pseudofermion and
pseudofermion hole scatterers feel the created $c\nu$ (and $s\nu$)
FP scattering centers as being $c0$ (and $c0$ and $s1$)
pseudofermion scattering centers at the {\it Fermi points}.
Therefore, creation of $\alpha\nu$ pseudofermions of limiting bare
momentum $q=\pm q_{\alpha\nu}^0$ and belonging to $\alpha\nu\neq
c0,\,s1$ branches is included above in the elementary processes
(B).

In addition to the PS, the following subspaces play an important
role in our studies:\vspace{0.25cm}

A {\it CPHS ensemble subspace} is a subspace of a canonical
ensemble subspace with fixed $N_{\uparrow}$ and $N_{\downarrow}$
electronic numbers, which is spanned by all energy eigenstates
with fixed values for the $-1/2$ Yang holon number $L_{c,\,-1/2}$,
$-1/2$ HL spinon number $L_{s,\,-1/2}$, $c0$ pseudofermion number
$N_{c0}$, and for the sets of $\alpha\nu\neq c0$ pseudofermion
numbers $\{N_{c\nu}\}$ and $\{N_{s\nu}\}$ corresponding to the
$\nu=1,2,...$ branches. Here CPHS stands for $c0$ pseudofermion,
holon, and spinon;\vspace{0.25cm}

A {\it J-CPHS ensemble subspace} is a subspace of a CPHS ensemble
subspace spanned by the excited energy eigenstates with the same
values for the numbers $N^{phNF}_{c0}$, $N^{phNF}_{s1}$, $\Delta
N^F_{c0 ,\,+1}$, $\Delta N^F_{c0 ,\,-1}$, $\Delta N^F_{s1 ,\,+1}$,
$\Delta N^F_{s1 ,\,-1}$, and sets of numbers $\{N^F_{\alpha\nu
,\,+1}\}$ and $\{N^F_{\alpha\nu ,\,-1}\}$ for the $\alpha\nu\neq
c0,\,s1$ branches with finite pseudofermion occupancy in the CPHS
ensemble subspace;\vspace{0.25cm}

A {\it J-CPHS subspace} is a J-CPHS ensemble subspace whose
numbers $N^{phNF}_{c0}$ and $N^{phNF}_{s1}$ have finite
values;\vspace{0.25cm}

A {\it reduced J-CPHS subspace} is a subspace of a J-CPHS subspace
spanned by all the energy eigenstates of the latter subspace which
are generated from the ground state by elementary processes (A)
and (B) only.\vspace{0.25cm}

The CPHS ensemble subspace dimension is given by,

\begin{equation}
D_{CPHS} = {N_a\choose N_{c0}}{N_0
-N_{s1}-2\sum_{\nu''>1}N_{s\nu''}\choose N_{s1}}\, \prod_{\nu>0}
{N^*_{c\nu}\choose N_{c\nu}}\,\prod_{\nu'>1} {N^*_{s\nu'}\choose
N_{s\nu'}}  \, . \label{D-CPHS}
\end{equation}
It is a product of the number of occupancy configurations of the
$c0$ pseudofermions, $s1$ pseudofermions, and $\alpha\nu\neq
c0,\,s1$ pseudofermions belonging to branches with CPHS ensemble
subspace finite occupancy. Once the ground-state numbers
$N^0_{c0}=N^0$ and $N^0_{s1}=N^0_{\downarrow}$ are known and the
corresponding excited-state values of these numbers are given by
$N_{c0}=N^0_{c0} + \Delta N_{c0}$ and $N_{s1}=N^0_{s1} + \Delta
N_{s1}$, one can define a CPHS ensemble subspace by the set of
deviation number values $\Delta N_{c0}$ and $\Delta N_{s1}$ and
set of number values $L_{c,\,-1/2}$, $L_{s,\,-1/2}$, and
$\{N_{\alpha\nu}\}$ for the $\alpha\nu\neq c0,\,s1$ branches.

Since the PS excited energy eigenstates involve the creation of
none or of a finite number of $\alpha\nu$ pseudofermions belonging
to the $\alpha\nu\neq c0,\,s1$ branches such that $\Delta
N_{\alpha\nu} = N_{\alpha\nu}$, all the ${N^*_{\alpha\nu}\choose
N_{\alpha\nu}}$ CPHS ensemble subspace occupancy configurations of
each of the latter $\alpha\nu$ bands correspond to PS excited
energy eigenstates. The same applies to the CPHS ensemble subspace
$c0$ (and $s1$) pseudofermion occupancy configurations provided
that $n=1$ (and $m=0$) for the initial ground state. Indeed, for
such a density there are no $c0$ (and $s1$) pseudofermion holes
for that state, and again all excited-state CPHS ensemble subspace
configurations can be reached by a finite number of $c0$ (and
$s1$) pseudofermion processes. However, for densities $0<n<1$ and
$0<m<n$ the ground state has both "particles" and "holes" in the
$c0$ and $s1$ pseudofermion bands, whose numbers are such that
$N_{\alpha\nu}\rightarrow\infty$ and
$N^h_{\alpha\nu}\rightarrow\infty$ as $N_a\rightarrow\infty$.
Thus, reaching from the ground state many of the excited-state
CPHS ensemble subspace $c0$ and $s1$ pseudofermion occupancy
configurations involves in the thermodynamic limit an infinite
number of finite-momentum and finite-energy pseudofermion
processes. It follows that for the metallic phase at finite spin
density only the J-CPHS subspaces of each CPHS ensemble subspace
are contained in the PS. Indeed, J-CPHS ensemble subspaces with
$N^{phNF}_{c0}$ and/or $N^{phNF}_{s1}$ infinite, do not belong to
the PS and have no finite overlap with the one- and two-electron
excitations.

\subsection{GENERAL DISTRIBUTION-FUNCTION DEVIATIONS
FOR DENSITIES $0<n<1$ AND $0<m<n$}

A J-CPHS subspace is spanned by the excited energy eigenstates
contained in the corresponding reduced J-CPHS subspace and by all
excited energy eigenstates generated from the former states by the
elementary processes (C). Here we study the pseudofermion
occupancy configurations of the excited energy eigenstates that
span a J-CPHS subspace. There is a particular type of reduced
J-CPHS subspace which we call {\it point-subspace}. A
point-subspace is spanned by a single excited energy eigenstate
whose deviation numbers and numbers are such that,

\begin{equation}
\Delta N_{\alpha\nu} = \Delta N^F_{\alpha\nu} = \sum_{\iota =\pm
1} \Delta N_{\alpha\nu,\,\iota}^{F} \, , \hspace{0.25cm} \alpha\nu
= c0,\,s1 \, ; \hspace{0.50cm} N_{\alpha\nu} = N_{\alpha\nu}^{F} =
\sum_{\iota =\pm 1} N_{\alpha\nu,\,\iota}^{F} \, , \hspace{0.25cm}
\alpha\nu\neq c0,\,s1 \, . \label{om0-k0}
\end{equation}
The number deviation $\Delta N^{F}_{\alpha\nu}$ obeys the relation
$\Delta N_{\alpha\nu} = \Delta N^{NF}_{\alpha\nu} + \Delta
N^{F}_{\alpha\nu}$, where $\Delta N^{NF}_{\alpha\nu}$ and $\Delta
N^{F}_{\alpha\nu}$ refer to the $c0$ and $s1$ pseudofermion number
deviations generated by the elementary processes (A) and (B),
respectively. For the $\alpha\nu = c0,\,s1$ branches, the
deviation number $\Delta N^{F}_{\alpha\nu,\,\iota}$ is related to
the current number deviation, $\Delta J^F_{\alpha\nu} = {1\over
2}\sum_{\iota =\pm 1}(\iota)\,\Delta N^F_{\alpha\nu ,\,\iota}$
such that $\Delta N^F_{\alpha\nu ,\,\iota} = \Delta
N^F_{\alpha\nu}/2 + \iota\, \Delta J^F_{\alpha\nu}$. Furthermore,
for the $\alpha\nu\neq c0,\,s1$ branches the number
$N^{F}_{\alpha\nu,\,\iota}$ is associated with the current number
$J^F_{\alpha\nu} = {1\over 2}\sum_{\iota =\pm
1}(\iota)\,N^F_{\alpha\nu ,\,\iota}$ such that $N^F_{\alpha\nu
,\,\iota} = N^F_{\alpha\nu}/2 + \iota\, J^F_{\alpha\nu}$.

The excited energy eigenstates that span a reduced J-CPHS subspace
have well-defined $c0$ and $s1$ pseudofermion right ($\iota =+1$)
and left ($\iota =-1$) bare-momentum and canonical-momentum {\it
Fermi points},

\begin{equation}
q_{F\alpha\nu ,\,\iota} = \iota\,q^0_{F\alpha\nu} + \Delta
q_{F\alpha\nu ,\,\iota} \, ; \hspace{1cm} \Delta q_{F\alpha\nu
,\,\iota} = \iota\,{2\pi\over L} \Delta N^F_{\alpha\nu ,\,\iota}
\, ; \hspace{0.5cm} \alpha\nu = c0,\,s1 \, ; \hspace{0.5cm} \iota
= \pm 1 \, , \label{qiFan}
\end{equation}
and

\begin{equation}
{\bar{q}}_{F\alpha\nu ,\,\iota} = \iota\,q^0_{F\alpha\nu} + \Delta
{\bar{q}}_{F\alpha\nu ,\,\iota} \, ; \hspace{1cm} \Delta
{\bar{q}}_{F\alpha\nu ,\,\iota} = \iota\,{2\pi\over L} \Delta
N^F_{\alpha\nu ,\,\iota} + {Q^{\Phi}_{\alpha\nu}
(\iota\,q^0_{F\alpha\nu})\over L} \, ; \hspace{0.5cm} \alpha\nu =
c0,\,s1 \, ; \hspace{0.3cm} \iota = \pm 1 \, , \label{barqanF}
\end{equation}
respectively, where $q^0_{F\alpha\nu}$ is the initial ground-state
{\it Fermi-point} value given in Eq. (\ref{q0Fcs}). Here
$Q^{\Phi}_{\alpha\nu}(q)/2$ is the $\alpha\nu$ pseudofermion
overall scattering phase shift (\ref{qcan1j}). We call
$2\Delta_{\alpha\nu}^{\iota}$ the square of the $\alpha\nu
,\,\iota$ {\it Fermi-point} value shift in units of $2\pi/L$,

\begin{equation}
2\Delta_{\alpha\nu}^{\iota}\equiv \left({\Delta
{\bar{q}}_{F\alpha\nu ,\,\iota}\over [2\pi/L]}\right)^2 =
\left(\iota\,\Delta N_{\alpha\nu,\,\iota}^{0,F}+ {Q_{\alpha\nu}
(\iota\,q^0_{F\alpha\nu})\over 2\pi}\right)^2 =
\left(\iota\,\Delta N_{\alpha\nu,\,\iota}^F+ {Q^{\Phi}_{\alpha\nu}
(\iota\,q^0_{F\alpha\nu})\over 2\pi}\right)^2 \, ; \hspace{0.25cm}
\alpha\nu = c0,\,s1 \, ; \hspace{0.15cm} \iota = \pm 1 \, .
\label{Delta}
\end{equation}
This quantity plays a key-role in the spectral properties, as
discussed in Secs. IV and V. It has the form given in Eq.
(\ref{Delta}) for $\cal{N}$-electron spectral functions leading to
deviations in the electronic numbers, $\Delta N$, such that
$\vert\Delta N\vert =0,1,2$. Our analysis is limited to such
spectral functions. Our studies of these functions involve the use
of specific expressions for the pseudofermion bare-momentum
distribution function deviations. The $c0$ and $s1$ pseudofermion
bare-momentum distribution-function deviations of the excited
energy eigenstates that span a J-CPHS subspace have the following
general form,

\begin{equation}
\Delta N_{\alpha\nu} (q) = \Delta N^{NF}_{\alpha\nu} (q) + \Delta
N^{F}_{\alpha\nu} (q) + \Delta N^{phF}_{\alpha\nu} (q) \, ;
\hspace{0.5cm} \alpha\nu = c0,\,s1 \, . \label{DN-gen}
\end{equation}

The bare-momentum distribution-function deviation $\Delta
N^{NF}_{\alpha\nu} (q)$ corresponds to the above elementary
processes (A) and is given by,

\begin{equation}
\Delta N^{NF}_{\alpha\nu} (q) = {\rm sgn}(\Delta
N^{NF}_{\alpha\nu})\,{2\pi\over L}\,\sum_{i=1}^{\vert \Delta
N^{NF}_{\alpha\nu}\vert}\delta (q-q_i) + {2\pi\over
L}\,\sum_{j=1}^{N^{phNF}_{\alpha\nu}}\Bigl[\delta (q-q_j)-\delta
(q-{q'}_j)\Bigl] \, ; \hspace{0.5cm} \alpha\nu = c0,\,s1 \, ,
\label{DN-NF}
\end{equation}
where ${\rm sgn}(\Delta N^{NF}_{\alpha\nu}) = {\rm sgn}(\Delta
N_{\alpha\nu})$ and $q_i$ with $i=1,...,\vert \Delta
N^{NF}_{\alpha\nu}\vert$ gives the set of bare-momentum values
associated with creation ($\Delta N^{NF}_{\alpha\nu}>0$) or
annihilation ($\Delta N^{NF}_{\alpha\nu}<0$) of $\alpha\nu$
pseudofermions away from the Fermi points. Moreover, $q_j$ and
${q'}_j$ with $j=1,...,N^{phNF}_{\alpha\nu}$ denote the
bare-momentum values associated with the particles and holes,
respectively, of the finite-energy and finite-momentum $\alpha\nu$
pseudofermion particle-hole processes.

The bare-momentum distribution function deviation $\Delta
N^{F}_{\alpha\nu} (q)$ results from the above elementary processes
(B) and reads,

\begin{equation}
\Delta N^{F}_{\alpha\nu} (q) = {2\pi\over L}\,\delta (\vert q\vert
-q^0_{F\alpha\nu})\,\Delta N^F_{\alpha\nu ,\,\iota}\,\delta_{\iota
1,\,{\rm sgn}(q) 1} \, ; \hspace{0.5cm} \alpha\nu = c0,\,s1 \, .
\label{DN-le-DF}
\end{equation}

Finally, the bare-momentum distribution function deviation $\Delta
N^{phF}_{\alpha\nu} (q)$ is associated with the elementary
processes (C) and is given by,

\begin{equation}
\Delta N^{phF}_{\alpha\nu} (q) = {2\pi\over L}\,\sum_{\iota =\pm
1}\sum_{j_{\iota}=1}^{N^{phF}_{\alpha\nu,\,\iota}}\Bigl[\delta
(q-q_{j_{\iota}})-\delta (q-{q'}_{j_{\iota}})\Bigl] \, ;
\hspace{0.5cm} \alpha\nu = c0,\,s1 \, , \hspace{0.25cm} \iota
={\rm sgn} (q)\,1 \, . \label{DN-le}
\end{equation}
In the latter equation $q_{j_{\iota}}$ and ${q'}_{j_{\iota}}$ with
$j_{\iota}=1,...,N^{phF}_{\alpha\nu,\,\iota}$ denote the
bare-momentum values associated with the "particles" and "holes",
respectively, in the vicinity of the {\it Fermi point} of Eq.
(\ref{qiFan}) corresponding to the $\alpha\nu ,\,\iota$ sub-branch
index value $\iota ={\rm sgn} (q)\,1=\pm 1$. Such bare-momentum
values are ordered as $\iota\,q_{1_{\iota}}< \iota\, q_{2_{\iota}}
< ... < \iota\, q_{{N^{phF}_{\alpha\nu,\,\iota}}_{\iota}}$ and
$\iota\, {q'}_{1_{\iota}}< \iota\, {q'}_{2_{\iota}} < ... <
\iota\, {q'}_{{N^{phF}_{\alpha\nu,\,\iota}}_{\iota}}$ for
$\alpha\nu = c0,\,s1$ and $\iota =\pm 1$. Once the elementary
processes (C) generate the J-CPHS subspaces from the corresponding
reduced subspaces, the set of "particle" and "hole" bare-momentum
values $\{q_{j_{\iota}}\}$ and $\{{q'}_{j_{\iota}}\}$ are such
that $q=q_{j_{\iota}}$ and $q={q'}_{j_{\iota}}$ was unoccupied and
occupied in the corresponding initial excited energy eigenstate
belonging to the latter subspace, respectively. The number
$N^{phF}_{\alpha\nu,\,\iota}$ is such that
$N^{phF}_{\alpha\nu,\,\iota}\leq m_{\alpha\nu,\,\iota}$, where
$m_{\alpha\nu,\,\iota}$ is the number of $\alpha\nu = c0,\,s1$
pseudofermion elementary particle-hole processes of momentum
$\iota [2\pi/L]$ involved in the deviation $\Delta
N^{phF}_{\alpha\nu} (q)$ of Eq. (\ref{DN-le}). Different such
deviations may correspond to the same number
$m_{\alpha\nu,\,\iota}$.

The $\alpha\nu\neq c0,\,s1$ branches have no finite pseudofermion
occupancy for the ground state. The $\alpha\nu\neq c0,\,s1$
pseudofermion occupancy configurations of the excited energy
eigenstates which span a J-CPHS subspace can be defined by the
values of the sets of numbers $\{N_{\alpha\nu,\,\iota}^{F}\}$
associated with the independent holons or spinons and the
$\alpha\nu$ FP scattering centers and the following bare-momentum
distribution-function deviations,

\begin{equation}
\Delta N_{\alpha\nu}^{NF} (q) = {2\pi\over
L}\,\sum_{i=1}^{N_{\alpha\nu}^{NF}}\delta (q-{q'}_i) \, ;
\hspace{0.5cm} \alpha\nu\neq c0,\,s1 \, . \label{DN-NF-an}
\end{equation}
Here ${q'}_i$, with $i=1,...,N_{\alpha\nu}^{NF}$, gives the set of
bare-momentum values of the pseudofermions created by the
elementary processes (A) away from the limiting bare-momentum
values $\pm q^0_{\alpha\nu}$.

\subsection{THE ENERGY AND MOMENTUM SPECTRA FOR DENSITIES $0<n<1$ AND
$0<m<n$}

Our studies of Secs. IV and V require an analysis of the energy
and momentum spectra beyond that of Ref. \cite{V}. The
pseudofermion dispersions $\epsilon_{\alpha\nu} (q)$ determine the
form of the energy excitation spectrum of the weight distributions
studied in these sections. These energy bands are plotted in Figs.
6-9 of Ref. \cite{II} as a function of $q$ for $m\rightarrow 0$
and several values of $U/t$ and $n$. By suitable manipulation of
Eqs. (C15)-(C21) of Ref. \cite{I}, we find the following
dependence on the pseudofermion bare-momentum $q$,

\begin{equation}
\epsilon_{c0} (q) = \mu - \mu_0 H -{U\over 2} -2t\cos k^{0}(q) +
2t\int_{-Q}^{+Q}dk\,\widetilde{\Phi }_{c0,\,c0}
\left(k,k^{0}(q)\right)\,\sin k \, , \label{epc}
\end{equation}

\begin{equation}
\epsilon_{c\nu} (q) = 2\mu\nu + \epsilon^0_{c\nu} (q) = (2\mu
-U)\,\nu + 4t\,{\rm Re}\,\Bigl\{ \sqrt{1 -
\Bigl(\Lambda^{0}_{c\nu}(q) - i\nu {U\over 4t}\Bigr)^2}\Bigr\} +
2t\int_{-Q}^{+Q}dk\,\widetilde{\Phi }_{c0,\,c\nu}
\left(k,\Lambda^{0}_{c\nu}(q)\right)\,\sin k \, , \label{epcn}
\end{equation}
and

\begin{equation}
\epsilon_{s\nu} (q) = 2\nu\mu_0 H + \epsilon^0_{s\nu} (q) =
2\nu\mu_0 H+ 2t\int_{-Q}^{+Q}dk\, \widetilde{\Phi }_{c0,\,s\nu}
\left(k,\Lambda^{0}_{s\nu}(q)\right)\,\sin k \, . \label{epsn}
\end{equation}
Here the two-pseudofermion phase shifts $\widetilde{\Phi }$ are
given in Eqs. (\ref{tilPcc}) and (\ref{tilPanan}) of Appendix A,
$Q$ is the parameter defined by Eq. (\ref{Beq}) of that Appendix,
and $\mu =\mu (n,m,U/t)$ and $H =H (n,m,U/t)$ can be expressed in
terms of the same phase shifts as,

\begin{equation}
\mu = {U\over 2} +2t\cos Q - t\int_{-Q}^{+Q}dk\,[2\widetilde{\Phi
}_{c0,\,c0} \left(k,Q\right)+\widetilde{\Phi }_{c0,\,s1}
\left(k,B\right)]\,\sin k \, ; \hspace{0.5cm} \mu_0 H =
-t\int_{-Q}^{+Q}dk\, \widetilde{\Phi }_{c0,\,s1}
\left(k,B\right)\,\sin k \, . \label{mu}
\end{equation}
The ground-state rapidity functions $k^{0}(q)$,
$\Lambda^{0}_{c\nu}(q)$, and $\Lambda^{0}_{s\nu}(q)$ appearing on
the right-hand side of Eqs. (\ref{epc})-(\ref{epsn}) are defined
in terms of their inverse functions in Eqs.
(\ref{kcGS})-(\ref{GcnGS}) of Appendix A. (Analytical expressions
for the rapidity functions $k^{0}(q)$, $\Lambda^{0}_{c\nu}(q)$,
and $\Lambda^{0}_{s\nu}(q)$ as $m\rightarrow 0$ and both
$U/t\rightarrow 0$ and $U/t>>1$ are provided in Ref.
\cite{spectral}.) The zero-energy, $2\mu\nu$-energy, and
$2\nu\mu_0 H$-energy levels of the pseudofermion energy bands
(\ref{epc})-(\ref{epsn}) relative to the ground-state energy are
such that,

\begin{eqnarray}
\epsilon_{c0} (\pm 2k_F) = \epsilon_{s1} (\pm k_{F\downarrow}) & =
& 0 \, ; \hspace{0.50cm} \epsilon_{c\nu} (\pm [\pi -2k_F]) =
2\nu\mu \, , \hspace{0.25cm} \nu > 0 \, ; \hspace{0.50cm}
\epsilon_{s\nu} (\pm [k_{F\uparrow} -k_{F\downarrow}]) = 2\nu\mu_0
H \, , \hspace{0.25cm} \nu > 1 \, ; \nonumber \\
\epsilon^0_{c\nu} (\pm [\pi -2k_F]) & = & 0 \, , \hspace{0.25cm}
\nu > 0 \, ; \hspace{0.50cm} \epsilon^0_{s\nu} (\pm [k_{F\uparrow}
-k_{F\downarrow}]) = 0 \, , \hspace{0.25cm} \nu > 1 \, .
\label{levels}
\end{eqnarray}

Creation onto the ground state of one $c\nu$ pseudofermion or
$s\nu$ pseudofermion belonging to $\nu >0$ and $\nu >1$ branches,
respectively, is a finite-energy process whose minimal energy
value is given by $\epsilon_{c\nu} (\pm [\pi -2k_F]) =2\nu\mu$ or
$\epsilon_{s\nu} (0) = 2\nu\mu_0 H+\epsilon^0_{s\nu} (0)$,
respectively. Creation onto the ground state of one independent
$-1/2$ holon or independent $-1/2$ spinon requires energy $2\mu$
or $2\mu_0 H$, respectively \cite{I}. In the limit $n\rightarrow
1$ (and $m\rightarrow 0$) both the energy and bare-momentum band
widths of the dispersions $\epsilon_{c\nu} (q)$ for $\nu>0$ (and
$\epsilon_{s\nu} (q)$ for $\nu>1$) vanish. Moreover, while the
energy band width $4t$ and bare-momentum band width $2\pi$ of the
dispersion $\epsilon_{c0} (q)$ remain the same for all values of
$U/t$, the energy band width of the dispersions of all other
pseudofermion branches vanishes as $U/t\rightarrow\infty$. This
effect follows from the localized character reached by the
$\alpha\nu$ pseudofermions of $\nu=1,2,...$ branches as
$U/t\rightarrow\infty$ \cite{I,II}. In that limit the $c\nu\neq
c0$ pseudofermions and $s\nu$ pseudofermions separate into $2\nu$
independent holons and spinons, respectively.

The group velocity $v_{\alpha\nu}(q)$ and the {\it light} velocity
$v_{\alpha\nu}$ are given by,

\begin{equation}
v_{\alpha\nu}(q) = {\partial\epsilon_{\alpha\nu}(q)\over{\partial
q}} \, , \hspace{0.25cm} {\rm all}\hspace{0.15cm}{\rm branches} \,
; \hspace{0.5cm} v_{\alpha\nu}\equiv
v_{\alpha\nu}(q^0_{F\alpha\nu}) \, , \hspace{0.25cm} \alpha\nu =
c0,\,s1 \, , \label{v0}
\end{equation}
and play an important role in the spectral-weight distribution
expressions obtained in Sec. V.

The $\alpha\nu$-branch finite-energy spectrum associated with the
elementary processes (A) corresponds to the reduced J-CPHS
subspaces and reads,

\begin{eqnarray}
\Delta E_{\alpha\nu} (\{q_i\},\{q_j\},\{{q'}_j\}) & = & {\rm
sgn}(\Delta N^{NF}_{\alpha\nu})\,\sum_{i=1}^{\vert \Delta
N^{NF}_{\alpha\nu}\vert} \epsilon_{\alpha\nu} (q_i) +
\sum_{j=1}^{N^{phNF}_{\alpha\nu}}\Bigl[\,\epsilon_{\alpha\nu}
(q_j)-\epsilon_{\alpha\nu} ({q'}_j)\Bigr] \, ; \hspace{0.5cm}
\alpha\nu =
c0,\,s1 \, ; \nonumber \\
\Delta E_{\alpha\nu} (\{{q'}_i\}) & = &
\sum_{i=1}^{N^{NF}_{\alpha\nu}} \epsilon_{\alpha\nu} ({q'}_i) \, ;
\hspace{0.5cm} \alpha\nu \neq c0,\,s1 \, , \label{DE-NF}
\end{eqnarray}
where the pseudofermion energy bands are given in Eqs.
(\ref{epc})-(\ref{epsn}).

The finite-momentum spectrum has contributions both from the
elementary processes (A) and (B) and also corresponds to the
reduced J-CPHS subspaces. It is given by,

\begin{eqnarray}
\Delta P_{\alpha\nu} (\{q_i\},\{q_j\},\{{q'}_j\}) & = & \Delta
P^{F}_{\alpha\nu} + \Delta P^{NF}_{\alpha\nu}
(\{q_i\},\{q_j\},\{{q'}_j\}) \, ; \hspace{0.5cm} \alpha\nu =
c0,\,s1 \, ; \nonumber \\
\Delta P_{c\nu} (\{{q'}_i\}) & = & k_{c\nu} + \Delta P^{NF}_{c\nu}
(\{{q'}_i\}) \, ; \hspace{0.5cm} c\nu \neq c0 \, ; \hspace{0.5cm}
\Delta P_{s\nu} (\{{q'}_i\}) = \sum_{i=1}^{N^{NF}_{s\nu}} {q'}_i
\, ; \hspace{0.5cm} s\nu\neq s1 \, . \label{DP-an}
\end{eqnarray}
Here $\Delta P^{F}_{\alpha\nu}$ is given in Eq. (67) of Ref.
\cite{V},

\begin{equation}
\Delta P^{NF}_{\alpha\nu} (\{q_i\},\{q_j\},\{{q'}_j\}) = {\rm
sgn}(\Delta N^{NF}_{\alpha\nu})\,\sum_{i=1}^{\vert \Delta
N^{NF}_{\alpha\nu}\vert} q_i+
\sum_{j=1}^{N^{phNF}_{\alpha\nu}}\Bigl[q_j -{q'}_j\Bigr] \, ;
\hspace{0.5cm} \alpha\nu = c0,\,s1 \, , \label{DP-NF}
\end{equation}

\begin{equation}
\Delta P^{NF}_{c\nu} (\{{q'}_i\}) = -\sum_{i=1}^{N^{NF}_{c\nu}}
{q'}_i \, ; \hspace{0.5cm} k_{c\nu} = (1+\nu )\pi\,N^{NF}_{c\nu}
\, ; \hspace{0.5cm} c\nu\neq c0 \, , \label{DP-NI}
\end{equation}
and $q^0_{F\alpha\nu}$ is the $c0$ and $s1$ pseudofermion
ground-state {\it Fermi-point} value given in Eq. (\ref{q0Fcs}).
The general finite-energy and finite-momentum spectrum generated
by the elementary processes (A) and (B) reads,

\begin{equation}
\Delta E = \Delta E\,(\{q_i\},\{{q'}_i\},\{q_j\},\{{q'}_j\}) =
\sum_{\alpha\nu=c0,\,s1} \Delta E_{\alpha\nu}
(\{q_i\},\{q_j\},\{{q'}_j\}) + \sum_{\alpha\nu\neq c0,\,s1} \Delta
E_{\alpha\nu} (\{{q'}_i\}) + \sum_{\alpha=c,\,s} E_{\alpha} \, ,
\label{DE-T}
\end{equation}
and

\begin{eqnarray}
\Delta P & = & \Delta P\,(\{q_i\},\{{q'}_i\},\{q_j\},\{{q'}_j\}) =
P_0 + \sum_{\alpha\nu=c0,\,s1}\Delta P^{NF}_{\alpha\nu}
(\{q_i\},\{q_j\},\{{q'}_j\}) \nonumber \\
& + & \sum_{c\nu\neq c0}\Delta P^{NF}_{c\nu} (\{{q'}_i\}) +
\sum_{s\nu\neq s1}\Delta P_{s\nu} (\{{q'}_i\}) \, , \label{DP-T}
\end{eqnarray}
where

\begin{equation}
P_0 = \Bigl[\,k_0^F + \sum_{c\nu\neq c0}k_{c\nu} +
\sum_{\alpha=c,\,s} P_{\alpha}\Bigr] \, ; \hspace{0.5cm} k_0^F
=\sum_{\alpha\nu =c0,\,s1}\Delta P^{F}_{\alpha\nu} \, ,
\label{P0-k0F}
\end{equation}
and the energy $E_{\alpha}$ and momentum $P_{\alpha}$ correspond
to the independent holons $(\alpha =c)$ and spinons $(\alpha =c)$
and are given in Eqs. (66) and (67) of Ref. \cite{V},
respectively. Both the spectra (\ref{DE-T}) and (\ref{DP-T}) are
associated with reduced J-CPHS subspaces.

Generation of the J-CPHS subspaces from the corresponding reduced
subspaces involves the elementary processes (C). Given the linear
$\alpha\nu =c0,\,s1$ pseudofermion energy dispersion near the {\it
Fermi points}, these processes lead to small momentum and energy
values such that,

\begin{equation}
k' = \sum_{\alpha\nu=c0,\,s1}\sum_{\iota =\pm 1}\iota\,{2\pi\over
L} \,m_{\alpha\nu,\,\iota} \, ; \hspace{0.5cm} \omega' =
\sum_{\alpha\nu=c0,\,s1}\sum_{\iota =\pm 1}{2\pi\over
L}\,v_{\alpha\nu} \,m_{\alpha\nu,\,\iota} \, .
\label{DE-DP-phF-summ}
\end{equation}
Here $m_{\alpha\nu,\,\iota}$ is the number of elementary
$\alpha\nu$ pseudofermion particle-hole processes of momentum
$\iota [2\pi/L]$ defined above and $v_{\alpha\nu}$ is the
$\alpha\nu =c0,\,s1$ pseudofermion {\it light} velocity given in
Eq. (\ref{v0}). Thus, the elementary processes (C) generate a set
of excited energy eigenstates with energy and momentum given by
those of the initial reduced-subspace state, Eqs. (\ref{DE-T}) and
(\ref{DP-T}), respectively, plus the small energy and momentum
given in Eq. (\ref{DE-DP-phF-summ}). According to the results of
Ref. \cite{IIIb}, the elementary processes (C) have a
non-interacting character in terms of $\alpha\nu =c0,\,s1$
pseudoparticles. In turn, for the remaining low-energy excitations
such pseudoparticles have residual interactions, in contrast to
the corresponding $\alpha\nu =c0,\,s1$ pseudofermions. This
pseudoparticle non-interacting character implies that the energy
spectrum of Eq. (\ref{DE-DP-phF-summ}) remains linear in
$m_{\alpha\nu,\,\iota}$ for small finite values of
$m_{\alpha\nu,\,\iota}/N_a$ as $N_a\rightarrow\infty$.

The energy $\omega_0$ for transitions from the ground state to a
point-subspace plays an important role in the spectral-weight
distributions. Such an energy is that of the excited energy
eigenstate whose deviation numbers are given in Eq. (\ref{om0-k0})
and reads,

\begin{equation}
\omega_0 = \sum_{\alpha=c,\,s} E_{\alpha} = 2\mu\, M_{c,\,-1/2} +
2\mu_0 H\, [M_{s,\,-1/2}-N_{s1}] \, , \label{omega0}
\end{equation}
where $N_{s1} = N^0_{s1} +\Delta N_{s1}$. We note that each CPHS
ensemble subspace contains at least one point-subspace and that
all point-subspaces of a CPHS ensemble subspace have the same
energy, given in Eq. (\ref{omega0}). Furthermore, the smallest
excitation energy for transitions from the ground state to a CPHS
ensemble subspace reads,

\begin{equation}
E_{CPHS}^0 = \omega_0 + \sum_{s\nu\neq s1}\epsilon^0_{s\nu}
(0)\,N_{s\nu} \, , \label{E0}
\end{equation}
where $\omega_0$ is the energy (\ref{omega0}) of the
point-subspace(s) contained in the CPHS ensemble subspace and
$\epsilon^0_{s\nu} (0)<0$ for $m>0$ and $\epsilon^0_{s\nu}
(0)\rightarrow 0$ as $m\rightarrow 0$. Thus, for $m\rightarrow 0$,
one has that $E_{CPHS}^0 = \omega_0 = 2\mu\, M_{c,\,-1/2}$. In
this case the energy value $E_{CPHS}^0=\omega_0 =0$ corresponds to
a $M_{c,\,-1/2}=0$ CPHS ensemble subspace, $E_{CPHS}^0=\omega_0
=2\mu$ to the $M_{c,\,-1/2}=1$ first-Hubbard band CPHS ensemble
subspace, $E_{CPHS}^0=\omega_0 =4\mu$ to the $M_{c,\,-1/2}=2$
second-Hubbard band CPHS ensemble subspace, and so on. We recall
that the number of $-1/2$ holons, $M_{c,\,-1/2}$, equals that of
rotated-electron doubly occupied sites \cite{I}.

The excited energy eigenstate that spans a point-subspace has
energy $\omega =l\omega_0$ and momentum $k=lk_0$ relative to the
initial ground state, where $k_0$ is given by,

\begin{equation}
k_0 = k_0^F + \sum_{\alpha=c,\,s} P_{\alpha} =
\pi[L_{c,\,-1/2}+\sum_{\nu =1}^{\infty}\nu N^F_{c\nu}]+
4k_F\Bigl[\,\Delta J^F_{c0} + \sum_{\nu =1}^{\infty}\,J^F_{c\nu} +
\sum_{\nu =2}^{\infty}\,J^F_{s\nu}\Bigr] +
2k_{F\downarrow}\Bigl[\,\Delta J^F_{s1} - 2\sum_{\nu
=2}^{\infty}\,J^F_{s\nu}\Bigr] \, . \label{k0-J}
\end{equation}
Here $k_0^F$ is provided in Eq. (\ref{P0-k0F}). A property of the
one- and two-electron spectral-weight distributions is that for
small values of the energy deviation $(\omega -l\omega_0)$ and
$m\rightarrow 0$ there is spectral weight only for values of
momentum $k$ such that $(k-lk_0)$ is also small. While the
momentum spectrum $\pi\,[L_{c,\,-1/2}+\sum_{\nu =1}^{\infty}\nu
N^F_{c\nu}]$ is additive in the momentum $\pi$ of each independent
$-1/2$ holon, the momenta $4k_F\Bigl[\,\Delta J^F_{c0} + \sum_{\nu
=1}^{\infty}\,J^F_{c\nu} + \sum_{\nu
=2}^{\infty}\,J^F_{s\nu}\Bigr]$ and $2k_{F\downarrow}[\,\Delta
J^F_{s1} - 2\sum_{\nu =2}^{\infty}\,J^F_{s\nu}\Bigr]$ refer to the
$c0$ and $s1$ pseudofermion {\it Fermi points} current number
deviations. Note that although the current numbers $J^F_{c\nu}$
and $J^I_{s\nu}$ arise from the $c\nu\neq c0$ and $s\nu\neq s1$ FP
scattering centers, they lead to momentum contributions associated
with the $c0$ {\it Fermi points} and $c0$ and $s1$ {\it Fermi
points}, respectively. These current numbers correspond to phase
shifts whose scattering centers are felt by the $c0$ and $s1$
pseudofermions has being at the $\alpha\nu=c0,\,s1$ {\it Fermi
points} \cite{S-P}. Point-subspaces contained in the same CPHS
ensemble subspace have the same energy but different momentum
values. For a given CPHS ensemble subspace, we call {\it first
point-subspaces} and {\it second point-subspaces} the subspaces
with the momentum (\ref{k0-J}) given by $k_0=\pm\vert k_0\vert$
which correspond to the minimum value of $\vert k_0\vert$ and the
second smallest value, respectively.

\subsection{THE ELEMENTARY PROCESSES AND THE ACTIVE SCATTERING CENTERS}

Active scattering centers are those which contribute to the
scattering phase shift (\ref{qcan1j}). We find below that only the
elementary processes (A) and (B) generate active scattering
centers. For the excited energy eigenstates described by the
bare-momentum distribution function deviations given in Eqs.
(\ref{DN-gen})-(\ref{DN-NF-an}) the scattering phase shift
(\ref{qcan1j}) can be written as,

\begin{equation}
Q^{\Phi}_{\alpha\nu} (q)/2 = Q^{\Phi (NF)}_{\alpha\nu} (q)/2 +
Q^{\Phi (F)}_{\alpha\nu} (q)/2 \, . \label{qcan1j-J}
\end{equation}
Here the phase shifts $Q^{\Phi (NF)}_{\alpha\nu} (q)/2$ and
$Q^{\Phi (F)}_{\alpha\nu} (q)/2$ result from active scattering
centers generated by the elementary processes (A) and (B),
respectively. The spectral properties are mainly controlled by the
scattering phase shifts $Q^{\Phi}_{\alpha\nu} (\iota
q^0_{F\alpha\nu})/2$ of the $\alpha\nu = c0,\,s1$ pseudofermion
and hole scatterers of bare momentum values $q\approx \pm
q^0_{F\alpha\nu}$. The scattering phase shift $Q^{\Phi
(NF)}_{\alpha\nu} (\iota q^0_{F\alpha\nu})/2$, where $\alpha\nu =
c0,\,s1$ and $\iota = \pm 1$, can be written as,

\begin{eqnarray}
& & {Q^{\Phi (NF)}_{\alpha\nu} (\iota q^0_{F\alpha\nu},
\{q_i\},\{{q'}_i\},\{q_j\},\{{q'}_j\})\over 2} = \pi
\Bigl\{\sum_{\alpha'\nu'=c0,\,s1}\,\Bigl({\rm sgn}(\Delta
N^{NF}_{\alpha'\nu'})\sum_{i=1}^{\vert \Delta
N^{NF}_{\alpha'\nu'}\vert}\Phi_{\alpha\nu,\,\alpha'\nu'}(\iota
q^0_{F\alpha\nu},q_i) \nonumber \\
& + &  \sum_{j=1}^{N^{phNF}_{\alpha'\nu'}}\Bigl[
\Phi_{\alpha\nu,\,\alpha'\nu'}(\iota q^0_{F\alpha\nu},q_j)-
\Phi_{\alpha\nu,\,\alpha'\nu'}(\iota
q^0_{F\alpha\nu},{q'}_j)\Bigr]\Bigr) +\sum_{\alpha'\nu'\neq
c0,\,s1}\,\sum_{i=1}^{N_{\alpha'\nu'}^{NF}}\Phi_{\alpha\nu,\,\alpha'\nu'}(\iota
q^0_{F\alpha\nu},{q'}_i)\Bigr\} \, . \label{qcan1j-J-NF}
\end{eqnarray}
In turn, the phase shift $Q^{\Phi (F)}_{\alpha\nu} (\iota
q^0_{F\alpha\nu})/2 $ was derived in Ref. \cite{S-P} and reads,

\begin{eqnarray}
{Q^{\Phi (F)}_{\alpha\nu} (\iota q^0_{F\alpha\nu})\over 2} & = &
\pi\Bigl(\iota\, \xi^0_{\alpha\nu\,c0}\,{\Delta N^F_{c0}\over 2} +
\iota\, \xi^0_{\alpha\nu\,s1}\,{\Delta N^F_{s1}\over 2} - \iota\,
{\Delta N^F_{\alpha\nu}\over 2} - \Delta J^F_{\alpha\nu} +
\xi^1_{\alpha\nu\,c0}\Bigl[\,\Delta J^F_{c0} + \sum_{\nu
=1}^{\infty}\,J^F_{c\nu} + \sum_{\nu
=2}^{\infty}\,J^F_{s\nu}\Bigr] \nonumber \\
& + & \xi^1_{\alpha\nu\,s1}\Bigl[\,\Delta J^F_{s1} - 2\sum_{\nu
=2}^{\infty}\,J^F_{s\nu}\Bigr]\Bigr) \, . \label{qcan1j-J-F}
\end{eqnarray}
Here the {\it Fermi-point} parameters
$\xi^j_{\alpha\nu\,\alpha'\nu'}$ are symmetrical ($j=0$) and
antisymmetrical ($j=1$) combinations in the $\alpha'\nu' =
c0,\,s1$ pseudofermion scattering centers right and left {\it
Fermi-point} values of the elementary $\alpha\nu = c0,\,s1$
two-pseudofermion phase shifts in units of $\pi$ given by,

\begin{equation}
\xi^j_{\alpha\nu\,\alpha'\nu'}=
\delta_{\alpha,\,\alpha'}\,\delta_{\nu,\,\nu'} + \sum_{\iota=\pm
1}(\iota^j)\,\Phi_{\alpha\nu,\,\alpha'\nu'}
(q^0_{F\alpha\nu},\iota\,q^0_{F\alpha'\nu'}) \, ; \hspace{0.5cm}
j=0,\,1 \, ; \hspace{0.5cm} \alpha\nu = c0,\, s1 \, ;
\hspace{0.5cm} \alpha'\nu' = c0,\, s1 \, . \label{xi}
\end{equation}
The parameters (\ref{xi}) appear in the spectral-function
exponents derived in Secs. IV and V. In the limit $m\rightarrow
0$, they are given by $\xi^0_{c0\,c0}=1/\xi_0$,
$\xi^0_{c0\,s1}=0$, $\xi^0_{s1\,c0}=-1/\sqrt{2}$,
$\xi^0_{s1\,s1}=\sqrt{2}$, $\xi^1_{c0\,c0}=\xi_0$,
$\xi^1_{c0\,s1}=\xi_0/2$, $\xi^1_{s1\,c0}=0$, and
$\xi^1_{s1\,s1}=1/\sqrt{2}$. Here $\xi_0$ is the parameter defined
in Eq. (74) of Ref. \cite{92} and in the text above that equation.
It is such that $\xi_0\rightarrow\sqrt{2}$ and $\xi_0\rightarrow
1$ as $U/t\rightarrow 0$ and $U/t\rightarrow\infty$, respectively.
As further discussed in Ref. \cite{S-P}, the form of the general
phase-shift expression (\ref{qcan1j-J-F}) confirms that although
the current numbers $J^F_{c\nu}$ and $J^F_{s\nu}$ are associated
with the $c\nu\neq c0$ and $s\nu\neq s1$ FP scattering centers,
the $c0$ and $s1$ pseudofermions feel such centers as being at the
$c0$ {\it Fermi points} and $c0$ and $s1$ {\it Fermi points},
respectively.

The following functional plays a major role in the spectral
properties,

\begin{equation}
\zeta_0 = \zeta_0 (\{q_i\},\{{q'}_i\},\{q_j\},\{{q'}_j\}) =
2\Delta_{c0}+2\Delta_{s1} \, ; \hspace{0.5cm} 2\Delta_{\alpha\nu}
= 2\Delta_{\alpha\nu}^{+1}+2\Delta_{\alpha\nu}^{-1} \, ;
\hspace{0.25cm} \alpha\nu = c0,\, s1 \, . \label{zeta0}
\end{equation}
Moreover, by use of of the general expressions (\ref{qcan1j-J-NF})
and (\ref{qcan1j-J-F}) one finds that the four functionals
$2\Delta_{\alpha\nu}^{\iota}$, Eq. (\ref{Delta}), can be written
as follows,

\begin{eqnarray}
2\Delta_{\alpha\nu}^{\iota} & = & 2\Delta_{\alpha\nu}^{\iota}
(\{q_i\},\{{q'}_i\},\{q_j\},\{{q'}_j\}) = \Bigl({Q^{\Phi
(NF)}_{\alpha\nu} (\iota q^0_{F\alpha\nu})\over 2\pi} + \iota\,
\xi^0_{\alpha\nu\,c0}\,{\Delta N^F_{c0}\over 2} +
\iota\, \xi^0_{\alpha\nu\,s1}\,{\Delta N^F_{s1}\over 2}\nonumber \\
& + & \xi^1_{\alpha\nu\,c0}\Bigl[\,\Delta J^F_{c0} + \sum_{\nu
=1}^{\infty}\,J^F_{c\nu} + \sum_{\nu
=2}^{\infty}\,J^F_{s\nu}\Bigr]+
\xi^1_{\alpha\nu\,s1}\Bigl[\,\Delta J^F_{s1} - 2\sum_{\nu
=2}^{\infty}\,J^F_{s\nu}\Bigr]\Bigr)^2 \, ; \hspace{0.5cm}
\alpha\nu = c0,\,s1 \, , \hspace{0.25cm} \iota = \pm 1 \, ,
\label{Delta0}
\end{eqnarray}
where the phase shift $Q^{\Phi (NF)}_{\alpha\nu} (\iota
q^0_{F\alpha\nu})$ is that of Eq. (\ref{qcan1j-J-NF}).

Analysis of the above general phase-shift expressions confirms
that the $c0$ and $s1$ elementary processes (C) associated with
the bare-momentum distribution function deviations $\Delta
N^{phF}_{\alpha\nu} (q)$ of Eq. (\ref{DN-le}) do not generate
active scattering centers. Let us consider a "hole" and "particle"
created at bare-momentum values $q'$ and $q' + [2\pi/L] N_f$,
respectively, where $N_f=\iota 1,\,\iota 2,...$ is a finite
integer number and $\iota =\pm 1$. According to Eq.
(\ref{qcan1j}), such a elementary process leads to the scattering
canonical-momentum shift $Q^{\Phi}_{\alpha\nu} (q)/L= [2\pi/L][
\Phi_{\alpha\nu,\,\alpha'\nu'}(q,q' + [2\pi/L]
N_f)-\Phi_{\alpha\nu,\,\alpha'\nu'}(q,q')]$ such that
$Q^{\Phi}_{\alpha\nu} (q)/L\approx [2\pi/ L]^2\,N_f\,[\partial
\Phi_{\alpha\nu,\,\alpha'\nu'}(q,q')/\partial q']$, for all
$\alpha\nu$ pseudofermions and holes whose branches have finite
pseudofermion occupancy in the excited energy eigenstate. Here the
second expression corresponds to the leading-order contribution in
$1/L$. Since for the pseudofermion description discrete
bare-momentum and canonical-momentum contributions of order
$[1/L]^j$ such that $j>1$ have no physical significance
\cite{IIIb}, we thus conclude that $Q^{\Phi}_{\alpha\nu} (q)/L=0$.

Once all the $N^{phF}_{\alpha\nu,\,\iota}$ "particle" and "hole"
pairs of the bare-momentum distribution function deviation $\Delta
N^{phF}_{\alpha\nu} (q)$ of Eq. (\ref{DN-le}) are such that
$q_{j_{\iota}}-{q'}_{j_{\iota}}=[2\pi/L] N_f$ where $N_f=\iota
1,\,\iota 2,...$ is a finite integer number, such a deviation does
not contribute to the scattering phase shifts. Therefore, up to
first order in $1/L$, the scattering phase-shift contributions
from the deviations involving the summations of the
$N^{phF}_{\alpha\nu,\,\iota}$ "particle" $\delta$-functions
$[2\pi/L]\,\delta (q-q_{j_{\iota}})$ of the deviation
(\ref{DN-le}) are exactly canceled by those involving the
$N^{phF}_{\alpha\nu,\,\iota}$ "hole" $\delta$-functions
$-[2\pi/L]\,\delta (q-{q'}_{j_{\iota}})$. That the $\alpha\nu =
c0,\,s1$ pseudofermion "particle" and "hole" scattering centers
generated by the elementary processes (C) are not active
scattering centers implies the validity of the following property:
The overall scattering phase shift (\ref{qcan1j}), overall phase
shift (\ref{Qcan1j}), and functional (\ref{Delta0}) have the same
value for the whole set of J-CPHS subspace energy eigenstates
generated by the elementary processes (C) from a given energy
eigenstate of the corresponding reduced J-CPHS subspace. This
property plays an important role in our study.

\section{FINITE-ENERGY WEIGHT DISTRIBUTIONS FOR DENSITIES $0<n<1$ AND $0<m<n$} \label{SecIV}

Our starting point for the derivation of closed-form analytical
expressions for the one- and two-electron weight distributions
obtained in the ensuing section is the general
${\cal{N}}$-electron spectral function given in Eq. (36) of Ref.
\cite{V},

\begin{equation}
B_{{\cal{N}}}^{l} (k,\,\omega) =
\sum_{i=0}^{\infty}c^l_i\sum_{\{\Delta
N_{\alpha\nu}\},\,\{L_{\alpha,\,-1/2}\}}\Bigl[\sum_{\{N^{phNF}_{\alpha\nu}\},\,\{\Delta
N^F_{\alpha\nu ,\,\iota}\},\,\{N^F_{\alpha\nu
,\,\iota}\}}\,B^{l,i} (k,\,\omega)\Bigr] \, ; \hspace{0.5cm} c^l_0
= 1 \, , \hspace{0.5cm} l=\pm 1 \, , \label{ABONjl-J-CPHS}
\end{equation}
where for the present case of densities $0<n<1$ and $0<m<n$ the
summation over the numbers $N^{phNF}_{\alpha\nu}$ is limited to
finite values and thus refers to reduced J-CPHS subspaces only. On
the right-hand side of Eq. (\ref{ABONjl-J-CPHS}), $c^l_i$ is the
constant of the operator expressions given in Eqs. (32)-(34) of
Ref. \cite{V} such that $c^l_i\rightarrow 0$ as
$U/t\rightarrow\infty$ for $i>0$ and the function $B^{l,i}
(k,\,\omega)$ is defined in Eq. (68) of that reference. Let us
rewrite the $\alpha\nu$ pseudofermion spectral functions
introduced in Eq. (63) of Ref. \cite{V} as follows,

\begin{eqnarray}
B^{l,NF,i}_{\alpha\nu} (k,\omega) & = & N_a \Bigl\{\Bigl(\Theta
(\Delta N_{\alpha\nu}^{NF})\,\prod_{i=1}^{\vert \Delta
N_{\alpha\nu}^{NF}\vert}\Bigl[{1\over
N_a}\Bigl(\sum_{q_i=-q^0_{\alpha\nu}}^{-q^0_{F\alpha\nu}} +
\sum_{q_i=q^0_{F\alpha\nu}}^{q^0_{\alpha\nu}}\Bigr)\Bigr] + \Theta
(-\Delta N_{\alpha\nu}^{NF})\,\prod_{i=1}^{\vert \Delta
N_{\alpha\nu}^{NF}\vert}\Bigl[{1\over
N_a}\sum_{q_i=-q^0_{F\alpha\nu}}^{q^0_{F\alpha\nu}}
\Bigr]\Bigr)\nonumber \\
& \times & \prod_{j=1}^{N^{phNF}_{\alpha\nu}} \Bigl[{1\over
N_a}\Bigl(\sum_{q_j=-q^0_{\alpha\nu}}^{-q^0_{F\alpha\nu}}
+\sum_{q_j=q^0_{F\alpha\nu}}^{q^0_{\alpha\nu}}\Bigr) \,{1\over
N_a}\sum_{{q'}_j=-q^0_{F\alpha\nu}}^{q^0_{F\alpha\nu}}\Bigr]
\Bigl\}\,\delta (\omega -l\Delta E_{\alpha\nu})\,\delta_{k,\,
l\Delta P_{\alpha\nu}} \, ; \hspace{0.5cm} \alpha\nu = c0,\,s1 \, ;
\nonumber \\
B^{l,NF,i}_{\alpha\nu} (k,\omega) & = &
N_a\Bigr[\prod_{i=1}^{N_{\alpha\nu}^{NF}}{1\over
N_a}\sum_{{q'}_i=-q^0_{\alpha\nu}}^{q^0_{\alpha\nu}} \Bigr]\,
\delta (\omega - l\,\Delta E_{\alpha\nu})\,\delta_{k,\,l\Delta
P_{\alpha\nu}}\, ; \hspace{0.35cm} \alpha\nu\neq c0,\, s1 \, ;
\hspace{0.25cm} l = \pm 1 \, ; \hspace{0.25cm} i=0,1,2,... \, ,
\label{2B}
\end{eqnarray}
where the energy and momentum spectra are given in Eqs.
(\ref{DE-NF}) and (\ref{DP-an}), respectively. On the right-hand
side of Eq. (\ref{2B}) and in all expressions given below the
$\Theta $ function is such that $\Theta (x)=0$ for $x<0$ and
$\Theta (x)=1$ for $x\geq 0$. Moreover, here we have explicitly
written the summation $\sum_{J-CPHS-\alpha\nu-(A)}$ of expression
(63) of Ref. \cite{V}, which runs over the reduced J-CPHS subspace
$\alpha\nu$ pseudofermion occupancy configurations generated by
the elementary processes (A). The number of such occupancy
configurations is given by $\prod_{\alpha\nu}D_{\alpha\nu}$, where
for the $\alpha\nu=c0,\,s1$ branches the dimension $D_{\alpha\nu}$
reads,

\begin{equation}
D_{\alpha\nu} = \Bigl\{\Theta \Bigl({\rm sgn}(\Delta
N^{NF}_{\alpha\nu})\Bigr)\,{N^h_{\alpha\nu}\choose \Delta
N^{NF}_{\alpha\nu}} + \Theta \Bigl(-{\rm sgn}(\Delta
N^{NF}_{\alpha\nu})\Bigr)\,{N_{\alpha\nu}\choose -\Delta
N^{NF}_{\alpha\nu}}\Bigr\}\,{N^h_{\alpha\nu}\choose
N^{phNF}_{\alpha\nu}}\,{N_{\alpha\nu}\choose N^{phNF}_{\alpha\nu}}
\, , \label{Dan-0}
\end{equation}
and is given in Eq. (64) of Ref. \cite{V} for the $\alpha\nu\neq
c0,\,s1$ branches. We recall that Eq. (\ref{Dan-0}) refers to
densities $0<n<1$ and $0<m<n$ and thus the numbers $N_{c0}$,
$N^h_{c0}$, $N_{s1}$, and $N^h_{s1}$ are such that
$N_{c0}\rightarrow\infty$, $N^h_{c0}\rightarrow\infty$,
$N_{s1}\rightarrow\infty$, and $N^h_{s1}\rightarrow\infty$ as
$N_a\rightarrow\infty$, whereas for the PS excited energy
eigenstates the numbers $\Delta N^{NF}_{\alpha\nu}$, $\Delta
N^{F}_{\alpha\nu}$, and $N^{phNF}_{\alpha\nu}$ have finite values.
Thus, in expression (\ref{Dan-0}) we have not considered small
finite corrections to the numbers $N_{c0}$, $N^h_{c0}$, $N_{s1}$,
and $N^h_{s1}$, which lead to higher order corrections that vanish
in the thermodynamic limit. Moreover, in expression (\ref{2B}) we
could treat the bare-momentum summations as independent for each
created or annihilated pseudofermion. It follows that for
$N_a\rightarrow\infty$, exclusion of the occupancy configurations
where two or several created or annihilated pseudofermions {\it
meet} at the same discrete bare-momentum value leads to vanishing
corrections to expression (\ref{2B}).

From use of the spectral function expression (\ref{2B}) in the second expression of Eq.
(68) of Ref. \cite{V}, we arrive to the following expression for the function $B^{l,i}
(k,\,\omega)$ appearing on the right-hand side of Eq. (\ref{ABONjl-J-CPHS}),

\begin{eqnarray}
& & B^{l,i} (k,\,\omega) = \Bigr\{\prod_{\alpha\nu
=c0,\,s1}\Bigl[\Bigl(\Theta (\Delta
N_{\alpha\nu}^{NF})\,\prod_{i=1}^{\vert \Delta
N_{\alpha\nu}^{NF}\vert}\Bigl[{1\over
N_a}\Bigl(\sum_{q_i=-q^0_{\alpha\nu}}^{-q^0_{F\alpha\nu}} +
\sum_{q_i=q^0_{F\alpha\nu}}^{q^0_{\alpha\nu}}\Bigr)\Bigr]+\Theta
(-\Delta N_{\alpha\nu}^{NF})\,\prod_{i=1}^{\vert \Delta
N_{\alpha\nu}^{NF}\vert}\Bigl[{1\over
N_a}\sum_{q_i=-q^0_{F\alpha\nu}}^{q^0_{F\alpha\nu}}
\Bigr]\Bigr)\nonumber \\
& \times & \prod_{j=1}^{N^{phNF}_{\alpha\nu}} \Bigl[{1\over
N_a}\Bigl(\sum_{q_j=-q^0_{\alpha\nu}}^{-q^0_{F\alpha\nu}}
+\sum_{q_j=q^0_{F\alpha\nu}}^{q^0_{\alpha\nu}}\Bigr) \,{1\over
N_a}\sum_{{q'}_j=-q^0_{F\alpha\nu}}^{q^0_{F\alpha\nu}}\Bigr]
\Bigl]\Bigl\}\Bigr[\prod_{\alpha\nu \neq
c0,\,s1}\prod_{i=1}^{N_{\alpha\nu}^{NF}}{1\over
N_a}\sum_{{q'}_i=-q^0_{\alpha\nu}}^{q^0_{\alpha\nu}} \Bigr]
{\Theta\Bigl(\Omega -l[\omega-l\Delta E]\Bigr)\over
C_c\,C_s}\nonumber \\
& \times &\Theta\Bigl(l[\omega-l\Delta
E]\Bigr)\,\Theta\left({l[\omega-l\Delta E]\over \vert k -l\Delta
P\vert} -v_{s1}\right){\breve{B}}^{l,i} \left(\omega-l\Delta
E,\,{\omega-l\Delta E\over k -l\Delta P}\right) \, ;
\hspace{0.5cm} i=0,1,2,... \, , \hspace{0.25cm} l= \pm 1 \, .
\label{B-PAR-J-CPHS-sum-0}
\end{eqnarray}
Here the general energy functional $\Delta E=\Delta E
(\{q_i\},\{{q'}_i\},\{q_j\},\{{q'}_j\})$ and momentum functional
$\Delta P = \Delta P (\{q_i\},\{{q'}_i\},\{q_j\},\{{q'}_j\})$ are
given in Eqs. (\ref{DE-T}) and (\ref{DP-T}), respectively, and in
the thermodynamic limit the value of the small positive energy
$\Omega$, which corresponds to the energy range of the elementary
processes (C), is controlled by the functional $\zeta_0$, Eq.
(\ref{zeta0}), of the corresponding initial reduced-subspace
state. (The energy $\Omega$ vanishes when $\zeta_0\rightarrow 0$
and the weight distribution becomes $\delta$-function like, as
further discussed in Sec. V.) In numerical studies of the
spectral-function expressions derived in this paper the small
energy $\Omega$ is treated as a parameter whose value for fixed
finite values of $n$, $m$, and $U/t$ is determined by imposing the
exact sum rule (\ref{sum-ABON}) in the end of the calculations.
The bare-momentum summations of expression
(\ref{B-PAR-J-CPHS-sum-0}) run over all
$\prod_{\alpha\nu}D_{\alpha\nu}$ occupancy configurations
generated by the elementary processes (A). For the whole
$(k,\,\omega)$-plane and except for $k$ and $\omega$ values such
that $\omega\approx \iota\,v_{\alpha\nu}(k-lk_0)+l\omega_0$, where
$\alpha\nu = c0,\,s1$, $\iota =\pm 1$, $\omega_0$ is provided in
Eq. (\ref{omega0}), and $k_0$ in Eq. (\ref{k0-J}), the function
${\breve{B}}^{l,i} (\Delta\omega,v)$ on right-hand side of Eq.
(\ref{B-PAR-J-CPHS-sum-0}) is the convolution of the $c0$ and $s1$
spectral functions given in Eq. (62) of Ref. \cite{V}. These
functions read,

\begin{eqnarray}
{\breve{B}}^{l,i} (\Delta\omega,v) & = & {1\over N_a}
\int_{-\infty}^{+\infty}d\omega'\sum_{k'} B^{l,i}_{Q_{s1}}
\Bigl(\Delta\omega/v -
k',\Delta\omega-\omega'\Bigr)\,B^{l,i}_{Q_{c0}}
\Bigl(k',\omega'\Bigr)\nonumber \\
& \approx & {{\rm sgn} (v)\over
2\pi}\int_{0}^{\Delta\omega}d\omega'\int_{-{\rm sgn}
(v)\Delta\omega/v_{c0}}^{+{\rm sgn} (v)\Delta\omega/v_{c0}}dk'
B^{l,i}_{Q_{s1}} \Bigl(\Delta\omega/v -
k',\Delta\omega-\omega'\Bigr)\,B^{l,i}_{Q_{c0}}
\Bigl(k',\omega'\Bigr) \, ; \nonumber \\
B^{l,i}_{Q_{\alpha\nu}} (k',\omega') & = &
\sum_{J-CPHS-\alpha\nu-(C)}\vert\langle 0\vert
F_{J-GS,\,\alpha\nu}F_{p-h,\,\alpha\nu}
F_{-GS,\,\alpha\nu}^{\dag}\vert 0\rangle\vert^2\delta
\Bigl(\omega'-l{2\pi\over L}\,v_{\alpha\nu}
m_{\alpha\nu}\Bigr)\delta_{k',\, l\sum_{\iota}\iota\,{2\pi\over
L}m_{\alpha\nu,\,\iota}} \, . \label{B-l-i-breve}
\end{eqnarray}
In this equation $i=0,1,2,...$, $l= \pm 1$,
$\Delta\omega=(\omega-l\Delta E)$, $\Delta k=(k -l\Delta P)$, the
summation $\sum_{J-CPHS-\alpha\nu-(C)}$ runs over the J-CPHS
ensemble subspace $\alpha\nu =c0,\,s1$ pseudofermion occupancy
configurations generated by the elementary processes (C), the
matrix-element generators are given in Appendix B, and
$m_{\alpha\nu}=\sum_{\iota =\pm 1}m_{\alpha\nu,\,\iota}$. The
identity of the first and second expressions of Eq.
(\ref{B-l-i-breve}) follows from the structure of the spectrum of
Eq. (\ref{DE-DP-phF-summ}), which is generated by the elementary
processes (C). However, we note that the validity of the
second expression and of the spectral-function expressions given
below is limited to electronic densities
such that $v_{c0}>v_{s1}$. We emphasize that this
inequality holds for all electronic densities $n$ of the
metallic phase except for a small domain in the
vicinity of $n=1$. The velocity $v$ appearing in the argument of the
function (\ref{B-l-i-breve}) plays an important role in our study
and is given by,

\begin{equation}
v = {\Delta\omega\over \Delta k} = {(\omega-l\Delta E)\over
(k-l\Delta P)} \, ; \hspace{1cm} {\rm sgn} (v)\,1 = {\rm sgn}
(\Delta k)\,l \, ; \hspace{1cm} \vert v\vert
>v_{s1}\, . \label{V-ok}
\end{equation}
The inequality $\vert v\vert >v_{s1}$ and the theta-functions
appearing on the right-hand side of Eq. (\ref{B-PAR-J-CPHS-sum-0})
follow again from the structure of the spectrum of Eq.
(\ref{DE-DP-phF-summ}).

Each excited energy eigenstate generated from the ground state by
the elementary processes (A) and (B) corresponds to one point,
$(l\Delta P,\,l\Delta E)$, of the finite-weight $(k,\,\omega
)$-plane region associated with such a subspace. The set of all
such points generated by the bare-momentum summations on the
right-hand side of Eq. (\ref{B-PAR-J-CPHS-sum-0}) span a
well-defined finite-weight $(k,\,\omega)$-plane point-like,
line-like, or surface-like domain. From the reduced J-CPHS
subspace excited energy eigenstates corresponding to points
$(l\Delta P,\,l\Delta E)$ in the vicinity of the
$(k,\,\omega)$-plane point that the spectral-function expression
(\ref{B-PAR-J-CPHS-sum-0}) refers to, the elementary processes (C)
generate excited energy eigenstates whose momentum and energy
relative to the initial ground state is $k$ and $\omega$,
respectively. Thus, the momentum $l\Delta P$ and energy $l\Delta
E$ of the reduced J-CPHS subspace excited energy eigenstates
relative to the initial ground state are such that
$(\omega-l\Delta E)$ and $(k -l\Delta P)$ are small and $\vert
v\vert=\vert(\omega-l\Delta E)/(k -l\Delta P)\vert$ obeys the
inequality $\vert v\vert>v_{s1}$.

The $\alpha\nu =c0,\,s1$ pseudofermion spectral function $B^{l,i}_{Q_{\alpha\nu}}
(k',\omega')$ of Eq. (\ref{B-l-i-breve}) can be expressed as the convolution (with an
extra pre-factor of $N_a/2$) of the two following $\alpha\nu,\,\iota$ spectral functions
for the $\iota =+1$ and $\iota =-1$ sub-branches, respectively,

\begin{equation}
B^{l,\iota,i}_{Q_{\alpha\nu}} (k',\omega') =
\sum_{J-CPHS-\alpha\nu,\,\iota-(C)}\vert\langle 0\vert
F_{J-GS,\,\alpha\nu,\,\iota}\,F_{p-h,\,\alpha\nu,\,\iota}\,
F_{-GS,\,\alpha\nu,\,\iota}^{\dag}\vert 0\rangle\vert^2 \delta
\Bigl(\omega' -l{2\pi\over L}\,v_{\alpha\nu}
\,m_{\alpha\nu,\,\iota}\Bigr) \delta_{k',\,{\iota\,\omega'\over
v_{\alpha\nu}}} \, , \label{Blikom-iota}
\end{equation}
where $\alpha\nu = c0,\, s1$ and $l = \pm 1$. Both the function
$B^{l,i}_{Q_{\alpha\nu}} (k',\omega')$ of Eq. (\ref{B-l-i-breve})
and the function (\ref{Blikom-iota}) refer to the subspace spanned
by the tower of states generated by the elementary processes (C)
from each initial reduced J-CPHS subspace excited state. The
summation $\sum_{J-CPHS-\alpha\nu,\,\iota-(C)}$ on the right-hand
side of Eq. (\ref{Blikom-iota}) runs over all $\alpha\nu=c0,\,s1$
pseudofermion occupancy configurations generated by the
small-momentum and low-energy elementary processes (C) for the
$\alpha\nu$ pseudofermions belonging to the $\iota ={\it sgn}
(q)\,1=\pm 1$ sub-branch.

The contributions of the elementary processes (A) to the
matrix-element overlaps of the above spectral functions are easy
to evaluate, and lead to the matrix-element expressions given in
Eq. (59) of Ref. \cite{V}. After such contributions are accounted
for, the problem is reduced to the overlap of the $c0$ (and $s1$)
pseudofermion probability amplitudes provided in Eq. (69) of that
reference. That amplitude is associated with the matrix element
$\langle 0\vert F_{J-GS,\,\alpha\nu}\,F_{p-h,\,\alpha\nu}\,
F_{-GS,\,\alpha\nu}^{\dag}\vert 0\rangle$ given in Eq. (60) of
Ref. \cite{V} and the corresponding $\alpha\nu= c0,\,s1$
pseudofermion spectral function $B^{l,i}_{Q_{\alpha\nu}}
(k',\omega')$ of Eq. (\ref{B-l-i-breve}). The generators appearing
in the matrix elements given in Eqs. (59) and (60) of Ref.
\cite{V} are expressed in terms of the sets of canonical-momentum
and bare-momentum values of the distribution function deviations
(\ref{DN-gen})-(\ref{DN-NF-an}) in Eqs.
(\ref{J-NF})-(\ref{F-non-c0-s1}) of Appendix B. In contrast to the
matrix elements of Eq. (59) of that reference, the evaluation of
the probability amplitude $\vert\langle 0\vert
F_{J-GS,\,\alpha\nu}\,F_{p-h,\,\alpha\nu}\,
F_{-GS,\,\alpha\nu}^{\dag}\vert 0\rangle\vert^2$  is a complex
problem, which we address in the ensuing section. The states
$F_{-GS,\,\alpha\nu}^{\dag}\vert 0\rangle$ and
$F^{\dag}_{p-h,\,\alpha\nu}\,F^{\dag}_{J-GS,\,\alpha\nu}\vert
0\rangle$ involved in that probability amplitude describe
$N^0_{\alpha\nu}+\Delta N^F_{\alpha\nu}$ pseudofermions, whose
discrete canonical-momentum values are those of the ground-state
and excited-state J-CPHS subspace, respectively. Also the state
$F^{\dag}_{J-GS,\,\alpha\nu}\vert 0\rangle$ describes
$N^0_{\alpha\nu}+\Delta N^F_{\alpha\nu}$ pseudofermions, whose
discrete canonical-momentum values are those of the excited-state
reduced J-CPHS subspace. Both the states
$F_{-GS,\,\alpha\nu}^{\dag}\vert 0\rangle$ and
$F^{\dag}_{J-GS,\,\alpha\nu}\vert 0\rangle$ refer to densely
packed canonical-momentum occupancy configurations. The
probability amplitude $A^{(0,0)}_{\alpha\nu}\equiv \vert\langle
0\vert F_{J-GS,\,\alpha\nu}\, F_{-GS,\,\alpha\nu}^{\dag}\vert
0\rangle\vert^2$ associated with such canonical-momentum densely
packed configurations gives the $\alpha\nu$ pseudofermion
spectral-function lowest-peak weight. By use of the general
$\alpha\nu=c0,\,s1$ pseudofermion determinants given in Eqs. (71)
and (72) of Ref. \cite{V}, we find the following expression for
such a weight,

\begin{eqnarray}
A^{(0,0)}_{\alpha\nu} & = & \Big({1\over
N_{\alpha\nu}^*}\Bigr)^{2[N^{0}_{\alpha\nu}+\Delta
N_{\alpha\nu}^F]}\, \prod_{j=1}^{N_{\alpha\nu}^*}\,
\sin^2\Bigl({N_{\alpha\nu}^{-0}(q_j)[Q_{\alpha\nu}(q_j)
-\pi]+\pi\over 2}\Big) \, \prod_{j=1}^{N_{\alpha\nu}^*-1}\,
\Big[\sin\Bigl({\pi j\over N_{\alpha\nu}^*}\Bigr)\Bigr]^{2[N_{\alpha\nu}^* -j]} \nonumber \\
& \times &
\prod_{i=1}^{N_{\alpha\nu}^*}\prod_{j=1}^{N_{\alpha\nu}^*}\,\theta
(j-i)\nonumber \\
& \times & \sin^2\Bigl({N_{\alpha\nu}^{-0}({q}_j)
N_{\alpha\nu}^{-0}({q}_i)[Q_{\alpha\nu}({q}_j) -
Q_{\alpha\nu}({q}_i) + 2\pi (j-i)-\pi N_{\alpha\nu}^*] + \pi
N_{\alpha\nu}^*\over
2N_{\alpha\nu}^*}\Bigr) \nonumber \\
& \times &
\prod_{i=1}^{N_{\alpha\nu}^*}\prod_{j=1}^{N_{\alpha\nu}^*}\,{1\over
\sin^2\Bigl(N_{\alpha\nu}^{-0}({q}_i)
N_{\alpha\nu}^{-0}({q}_j)[{\pi (j-i)\over N_{\alpha\nu}^*} +
{Q_{\alpha\nu}({q}_j)\over 2N_{\alpha\nu}^*} -{\pi\over 2}] +
{\pi\over 2}\Bigr)} \, ; \hspace{0.5cm} \alpha\nu = c0,\,s1 \, ,
\label{A00}
\end{eqnarray}
where $N_{\alpha\nu}^{-0}(q)= N_{\alpha\nu}^{0}(q) + \Delta
N_{\alpha\nu}^{F}(q)$, $N_{\alpha\nu}^{0} (q_j)$ is the
ground-state bare-momentum distribution function provided in Eqs.
(C.1)-(C.3) of Ref. \cite{I}, and the deviation $\Delta
N_{\alpha\nu}^{F}(q)$ is given in Eq. (\ref{DN-le-DF}). Moreover,
we find that in the thermodynamic limit the lowest-peak weight
(\ref{A00}) has the following approximate behavior,

\begin{equation}
A^{(0,0)}_{\alpha\nu}\approx \prod_{\iota =\pm 1}
A^{(0,0)}_{\alpha\nu,\,\iota}\Bigl[1+{\cal{O}}\Bigl({1\over
N_a}\Bigr)\Bigr]\, ; \hspace{0.50cm} A^{(0,0)}_{\alpha\nu,\,\iota}
= {f_{\alpha\nu,\,\iota} \over \Bigl(N_a\,
S^0_{\alpha\nu}\Bigr)^{-1/2+ 2\Delta_{\alpha\nu}^{\iota}}} \, ;
\hspace{0.25cm} \alpha\nu = c0,\,s1 \, ; \hspace{0.15cm} \iota =
\pm 1 \, . \label{A002}
\end{equation}
Here $2\Delta_{\alpha\nu}^{\iota}$ is the functional given in Eq.
(\ref{Delta0}) and $f_{\alpha\nu,\,\iota}$ reads,

\begin{equation}
f_{\alpha\nu,\,\iota} = \sqrt{f\Bigl(Q_{\alpha\nu}(\iota
q^0_{F\alpha\nu})+{\rm sgn} (k)\pi\Bigr)} \, ; \hspace{0.50cm}
f_{\alpha\nu} = \prod_{\iota =\pm 1}f_{\alpha\nu,\,\iota} \, ;
\hspace{0.25cm} \alpha\nu = c0,\,s1 \, , \label{fan}
\end{equation}
where $k$ is the excited-state momentum relative to the initial
ground state, $f (Q)=f(-Q)$ is the function defined in Ref.
\cite{Penc97}, which appears on the right-hand side of Eq. (24) of
that reference, and $f_{\alpha\nu}$ appears in spectral-function
expressions introduced below. Moreover, $S^0_{\alpha\nu}$ is a
$n$, $m$, and $U/t$ dependent constant such that
$S^0_{c0}\,S^0_{s1}\rightarrow 1$ both for $U/t\rightarrow 0$ and
for $U/t\rightarrow \infty$ and $m\rightarrow 0$. (From  Ref.
\cite{Penc97} we learn that $S^0_{c0}\rightarrow\sin (\pi n)$ for
$U/t\rightarrow \infty$ and $m\rightarrow 0$, and thus
$S^0_{s1}\rightarrow 1/\sin (\pi n)$ in such a limit.)

The momentum and energy dependence of the spectral-function
expression (\ref{B-PAR-J-CPHS-sum-0}) is controlled by the
convolution function, Eq. (\ref{B-l-i-breve}), of the $c0$ and
$s1$ pseudofermion spectral functions $B^{l,i}_{Q_{\alpha\nu}}
(k',\omega')$ given in the same equation. The main mechanism of
such a control is the exotic matrix-element overlap associated
with the $c0$ and $s1$ pseudofermion elementary processes (C),
which generate $c0$ and $s1$ pseudofermion "particles" and "holes"
in the vicinity of the {\it Fermi points}. Although such
"particles" and "holes" are not active scattering centers, they
are active scatterers whose overall phase shifts originate the
orthogonality catastrophe leading to the unusual overlaps of the
matrix elements $\langle 0\vert
F_{J-GS,\,\alpha\nu}\,F_{p-h,\,\alpha\nu}\,
F_{-GS,\,\alpha\nu}^{\dag}\vert 0\rangle$. Interestingly, only the
overall phase shifts of the scatterers which are not active
scattering centers and are generated by the elementary processes
(C) contribute to the relative weights associated with such an
orthogonality catastrophe. On the other hand, the value of the
corresponding overall phase shifts is solely determined by the
occupancy configurations of the initial reduced J-CPHS space
states generated by the elementary processes (A) and (B) from the
ground state. Indeed, the latter processes generate the active
scattering centers that determine the value of the overall phase
shifts of the "particles" and "holes" created by the elementary
processes (C). Therefore, the unusual spectral properties result
from the interplay of the elementary processes (A) and (B), which
generate the active scattering centers, with the elementary
processes (C), which generate the active scatterers, as far as the
overall phase shift associated with the relative weights and the
corresponding orthogonality catastrophe that controls these
properties are concerned. All the $\alpha\nu =c0,\,s1$
pseudofermions with bare momentum away from the {\it Fermi points}
are also scatterers, but their overall phase shifts only
contribute to the lowest-peak weight probability amplitude
$A^{(0,0)}_{\alpha\nu}$ given in Eq. (\ref{A00}).

Next, let us calculate the convolution function and spectral
functions given in Eqs. (\ref{B-l-i-breve}) and
(\ref{Blikom-iota}). The expressions of these functions fully
define the general spectral functions (\ref{ABONjl-J-CPHS}) and
(\ref{B-PAR-J-CPHS-sum-0}). The canonical-momentum occupancy
configuration of the state $F_{J-GS,\,\alpha\nu}^{\dag}\vert
0\rangle$ associated with the lowest-peak weight probability
amplitude $A^{(0,0)}_{\alpha\nu}$ is densely packed and thus such
that $(m_{\alpha\nu,\,+1}, m_{\alpha\nu,\,+1})=(0,0)$. Following
the form of the momentum and energy spectra provided in Eq.
(\ref{DE-DP-phF-summ}), the summation on the right-hand side of
the $B^{l,i}_{Q_{\alpha\nu}} (k',\omega')$ expression of Eq.
(\ref{B-l-i-breve}) (and Eq. (\ref{Blikom-iota})) over all
occupancy configurations generated by the elementary $\alpha\nu
=c0,\,s1$ pseudofermion particle-hole processes (C) with the same
energy and momentum, is equivalent to consider such a summation
for all processes with the same value of
$m_{\alpha\nu,\,\iota}=[lL/4\pi v_{\alpha\nu}](\omega' +\iota
v_{\alpha\nu}\,k')$ (and $m_{\alpha\nu,\,\iota}=[lL/2\pi
v_{\alpha\nu}]\omega'=[l\iota L/2\pi]\,k'$). By performing that
summation we reach the relative weight
$A^{(0,0)}_{\alpha\nu}/\vert\langle 0\vert
F_{J-GS,\,\alpha\nu}\,F_{p-h,\,\alpha\nu}\,
F_{-GS,\,\alpha\nu}^{\dag}\vert 0\rangle\vert^2$ corresponding to
excited energy eigenstates with $N^{phF}_{\alpha\nu,\,\iota}$
$\alpha\nu$ pseudofermion particle-hole processes. In Appendix B
we consider states generated by $N^{ph}_{\alpha\nu}$ $\alpha\nu
=c0,\,s1$ pseudofermion particle-hole processes whose "particle"
and "hole" bare-momenta have arbitrary values. (The number
$N^{ph}_{\alpha\nu}$ is not related to the number
$N^{phNF}_{\alpha\nu}$ of $\alpha\nu =c0,\,s1$ pseudofermion
finite-momentum and finite-energy particle-hole processes
generated by the elementary processes (A). Instead, it is such
that for $N_a\rightarrow\infty$ its values only contribute to the
$\alpha\nu$ pseudofermion relative weights when the sum rule
$N^{ph}_{\alpha\nu}=\sum_{\iota =\pm
1}N^{phF}_{\alpha\nu,\,\iota}$ is obeyed.) For
$N_a\rightarrow\infty$ only the relative weights associated with
the small-momentum and low-energy $\alpha\nu = c0,\,s1$
pseudofermion elementary particle-hole processes (C) in the
vicinity of the {\it Fermi points} $\pm q^0_{F\alpha\nu}$ lead to
finite contributions to the $\alpha\nu = c0,\,s1$ pseudofermion
spectral function $B^{l,i}_{Q_{\alpha\nu}} (k',\omega')$ of Eq.
(\ref{B-l-i-breve}). Each of the excited energy eigenstates
described by the numbers $(m_{\alpha\nu,\,+1},
m_{\alpha\nu,\,-1})$ can be multiply degenerate. Therefore, we
need to sum up the relative weights
$a^{[N^{ph}_{\alpha\nu}]}_{\alpha\nu}$, Eq. (\ref{aNN}) of
Appendix B, for all the $\alpha\nu$ pseudofermion particle-hole
configurations with the same values of $(m_{\alpha\nu,\,+1},
m_{\alpha\nu,\,-1})$. Fortunately, for these processes, the
general relative weight expressions given in Eqs.
(\ref{aNN})-(\ref{a20}) of Appendix B simplify. This follows in
part from the symmetry found in Sec. III that all the excited
energy eigenstates generated by the elementary processes (C) from
a given initial reduced-subspace state have the same value for the
four functionals $2\Delta_{\alpha\nu}^{\iota}$ defined in Eq.
(\ref{Delta0}). We find the following general expression for the
relative weights in the tower of excited energy eigenstates,

\begin{equation}
a_{\alpha\nu}(m_{\alpha\nu,\,+1},
m_{\alpha\nu,\,-1})=\Bigl[\prod_{\iota =\pm 1}
a_{\alpha\nu,\,\iota}(m_{\alpha\nu,\,\iota})\Bigr]
\Bigl[1+{\cal{O}}\Bigl({\ln N_a\over N_a}\Bigr)\Bigr] \, ;
\hspace{0.5cm} \alpha\nu = c0,\,s1 \, . \label{aNNDP}
\end{equation}
Here $a_{\alpha\nu,\,\iota}(m_{\alpha\nu,\,\iota})$ is the relative
weight associated of the spectral function (\ref{Blikom-iota}). It
reads,

\begin{equation}
a_{\alpha\nu,\,\iota}(m_{\alpha\nu,\,\iota}) = \Theta
(m_{\alpha\nu,\,\iota})\,\prod_{j=1}^{m_{\alpha\nu,\,\iota}}
{(2\Delta_{\alpha\nu}^{\iota} + j -1)\over j} = \Theta
(m_{\alpha\nu,\,\iota})\,\frac{\Gamma (m_{\alpha\nu,\,\iota} +
2\Delta_{\alpha\nu}^{\iota})}{\Gamma (m_{\alpha\nu,\,\iota}+1)\,
\Gamma (2\Delta_{\alpha\nu}^{\iota})} \, ; \hspace{0.5cm}
\alpha\nu = c0,\,s1 \, ; \hspace{0.25cm} \iota =\pm 1 \, ,
\label{aNDP}
\end{equation}
where $\Gamma (x)$ is the usual gamma function. (While, by
construction, the integer $m_{\alpha\nu,\,\iota}$ is such that
$m_{\alpha\nu,\,\iota}\geq 0$, for later use it is convenient to
include the theta-function in the relative weight general
expression (\ref{aNDP}).) It follows from Eq. (\ref{aNDP}) that,

\begin{equation}
a_{\alpha\nu,\,\iota}(1) = 2\Delta_{\alpha\nu}^{\iota} \, ;
\hspace{0.5cm} \alpha\nu = c0,\,s1 \, ; \hspace{0.25cm} \iota =\pm
1 \, , \label{a10DP-iota}
\end{equation}
what reveals that the functional (\ref{Delta}) is the relative
weight associated with the $m_{\alpha\nu,\,\iota}=1$ peak of the
$\alpha\nu,\,\iota$ pseudofermion spectral function
(\ref{Blikom-iota}). Moreover,

\begin{equation}
a_{\alpha\nu} (1,\,0)= 2\Delta_{\alpha\nu}^{+1} \, ; \hspace{1cm}
a_{\alpha\nu} (0,\,1) = 2\Delta_{\alpha\nu}^{-1} \, ;
\hspace{0.5cm} \alpha\nu = c0,\,s1 \, . \label{a10DP}
\end{equation}
The function (\ref{aNDP}) has the following asymptotic behavior,

\begin{equation}
a_{\alpha\nu,\,\iota} (m_{\alpha\nu,\,\iota}) \approx\Theta
(m_{\alpha\nu,\,\iota})\,\frac{1}{\Gamma
(2\Delta_{\alpha\nu}^{\iota})}
\Bigl(m_{\alpha\nu,\,\iota}\Bigr)^{2\Delta_{\alpha\nu}^{\iota}-1}
\, ; \hspace{0.50cm} 2\Delta_{\alpha\nu}^{\iota}\neq 0 \, ;
\hspace{0.25cm} \alpha\nu = c0,\,s1 \, ; \hspace{0.25cm} \iota =
\pm 1 \, . \label{f}
\end{equation}

Use of the above results in the pseudofermion spectral functions
given in Eqs. (\ref{B-l-i-breve}) and (\ref{Blikom-iota}) leads to
the following expressions for these functions,

\begin{eqnarray}
B^{l,i}_{Q_{\alpha\nu}} (k',\omega') & = &
\sum_{m_{\alpha\nu,\,+1},\,m_{\alpha\nu,\,-1}}\,A^{(0,0)}_{\alpha\nu}\,a_{\alpha\nu}(m_{\alpha\nu,\,+1},\,
m_{\alpha\nu,\,-1})\,\delta \Bigl(\omega' -l{2\pi\over
L}\,v_{\alpha\nu}\sum_{\iota
=\pm1}m_{\alpha\nu,\,\iota}\Bigr)\,\delta_{k',\,l{2\pi\over
L}\,\sum_{\iota
=\pm1}\iota\,m_{\alpha\nu,\,\iota}} \nonumber \\
& = & {L\over 4\pi
v_{\alpha\nu}}\,A^{(0,0)}_{\alpha\nu}\,\prod_{\iota =\pm
1}\,a_{\alpha\nu,\,\iota}\Bigl({l[\omega'
+\iota\,v_{\alpha\nu}\,k']\over 4\pi v_{\alpha\nu}/L}\Bigr) \, ;
\hspace{0.5cm} \alpha\nu = c0,\,s1 \, , \label{B-J-i-sum-GG}
\end{eqnarray}
and

\begin{eqnarray}
B^{l,\iota,i}_{Q_{\alpha\nu}} (k',\omega') & = &
\sum_{m_{\alpha\nu,\,\iota}}A^{(0,0)}_{\alpha\nu,\,\iota}\,
a_{\alpha\nu,\,\iota}(m_{\alpha\nu,\,\iota}) \,\delta
\Bigl(\omega' -l{2\pi\over
L}\,v_{\alpha\nu}\,m_{\alpha\nu,\,\iota}\Bigr)\,
\delta_{k',\,{\iota\,\omega'\over v_{\alpha\nu}}}\nonumber \\
& = & {L\over 2\pi v_{\alpha\nu}}\,
A^{(0,0)}_{\alpha\nu,\,\iota}\,\,a_{\alpha\nu,\,\iota}
\Bigl({l\,\omega'\over 2\pi
v_{\alpha\nu}/L}\Bigr)\,\delta_{k',\,{\iota\,\omega'\over
v_{\alpha\nu}}} \, ; \hspace{0.5cm} \alpha\nu = c0,\, s1 \, ;
\hspace{0.25cm} \iota = \pm 1 \, , \label{Bkom-sum-G}
\end{eqnarray}
respectively. In order to distinguish the momentum and energy
values of these pseudofermion spectral functions from those of the
spectral function (\ref{ABONjl-J-CPHS}), $k$ and $\omega$, in Eqs.
(\ref{B-l-i-breve}), (\ref{Blikom-iota}), (\ref{B-J-i-sum-GG}),
and (\ref{Bkom-sum-G}) we denote the former momentum and energy
values by $k'$ and $\omega'$, respectively. It follows from Eq.
(\ref{f}) that for $2\Delta_{\alpha\nu}^{\iota}\neq 0$ the
spectral-function expressions (\ref{B-J-i-sum-GG}) and
(\ref{Bkom-sum-G}) have the following asymptotic behavior,

\begin{eqnarray}
B^{l,i}_{Q_{\alpha\nu}} (k',\omega') & \approx &
{f_{\alpha\nu}\over
4\pi\,v_{\alpha\nu}\,S^0_{\alpha\nu}}\,\prod_{\iota =\pm
1}\,{1\over \Gamma (2\Delta_{\alpha\nu}^{\iota})}\,\Theta
(l[\omega' +\iota\,v_{\alpha\nu}\,k'])\Bigl({l[\omega'
+\iota\,v_{\alpha\nu}\,k']\over 4\pi
\,v_{\alpha\nu}\,S^0_{\alpha\nu}}\Bigr)^{2\Delta_{\alpha\nu}^{\iota}-1}
\, ; \hspace{0.5cm} \alpha\nu = c0,\, s1 \, ; \nonumber \\
B^{l,\iota,i}_{Q_{\alpha\nu}} (k',\omega') & \approx &
{\sqrt{N_a\over S^0_{\alpha\nu}}}{f_{\alpha\nu,\,\iota}\over
2\pi\, v_{\alpha\nu}\,\Gamma
(2\Delta_{\alpha\nu}^{\iota})}\,\Theta (l\omega')
\Bigr({l\omega'\over 2\pi\,
v_{\alpha\nu}\,S^0_{\alpha\nu}}\Bigl)^{2\Delta_{\alpha\nu}^{\iota}-1}\,
\delta_{k',\,{\iota\,\omega'\over v_{\alpha\nu}}} \, ;
\hspace{0.5cm} \alpha\nu = c0,\, s1 \, ; \hspace{0.15cm} \iota
=\pm 1 \, , \label{Bkom-sum}
\end{eqnarray}
for small finite values of $l[\omega' +\iota\,v_{\alpha\nu}\,k']$
and $l\omega'$, respectively.

Let us consider the convolution function (\ref{B-l-i-breve}) for
vanishing energy values of the order of $1/L$. For instance, the
six smallest discrete energy values read $l\Delta\omega =0$,
$l\Delta\omega =[2\pi/L]v_{s1}$, $l\Delta\omega =[2\pi/L]v_{c0}$,
$l\Delta\omega =[4\pi/L]v_{s1}$, $l\Delta\omega
=[2\pi/L][v_{c0}+v_{s1}]$, and $l\Delta\omega =[4\pi/L]v_{c0}$
such that $l\Delta\omega\leq [4\pi/L]\,v_{c0}$. For vanishing
energy values we define that convolution function as
${\breve{B}}^{l,i} (\Delta\omega,v) = {\bar{B}}^{l,i}
(\Delta\omega,\Delta k)$, where by use of the first expression of
Eq. (\ref{B-l-i-breve}) we find,

\begin{eqnarray}
{\bar{B}}^{l,i} (\Delta\omega,\Delta k) & = & 2\pi\left({1\over
N_a}\right)^{\zeta_0}D_0\,\delta (\Delta\omega)\,\delta (\Delta k)
\, ; \hspace{0.5cm} D_0 = \prod_{\alpha\nu =
c0,\,s1}{S^0_{\alpha\nu}\,f_{\alpha\nu}\over
(S^0_{\alpha\nu})^{2\Delta_{\alpha\nu}}} \, ; \hspace{0.5cm}
l\Delta\omega = 0 \, ; \hspace{0.25cm} l\Delta
k= 0 \, ; \nonumber \\
{\bar{B}}^{l,i} (\Delta\omega,\Delta k) & = & 2\pi\left({1\over
N_a}\right)^{\zeta_0}D_0\,a_0\,\delta (l\Delta\omega-\omega
(C))\,\delta (l\Delta k-k (C)) \, ; \hspace{0.5cm} l\Delta\omega =
\omega (C) \, ; \hspace{0.25cm} v = {\omega (C)\over k (C)} \, .
\label{B-l-i-breve-int-0}
\end{eqnarray}
Here $\zeta_0$ and $2\Delta_{\alpha\nu}$ are the functionals given in Eq. (\ref{zeta0})
and for the above five energy values such that $l\Delta\omega>0$ the relative weight
$a_0$ reads: $a_0 =2\Delta_{\alpha\nu}^{\iota}$ for $\omega (C) =[2\pi/L]v_{\alpha\nu}$,
$k (C) =\iota\,[2\pi/L]$, and $v=\iota\,v_{\alpha\nu}$;
$a_0=2\Delta_{\alpha\nu}^{\iota}\,2\Delta_{\alpha\nu}^{-\iota}$ for $\omega (C)
=[4\pi/L]v_{\alpha\nu}$, $k (C) =0$, and $1/v=0$;
$a_0=2\Delta_{\alpha\nu}^{\iota}\,2\Delta_{{\bar{\alpha}\bar{\nu}}}^{\iota}$ for $\omega
(C) =[2\pi/L][v_{c0}+v_{s1}]$, $k (C) =\iota\,[4\pi/L]$, and $v=\iota\,[v_{c0}+v_{s1}]/2$
where we introduced the index ${\bar{\alpha}\bar{\nu}}$ such that ${\bar{c}\bar{0}}=s1$
and ${\bar{s}\bar{1}}=c0$; $a_0
=2\Delta_{\alpha\nu}^{\iota}\,2\Delta_{{\bar{\alpha}\bar{\nu}}}^{-\iota}$ for $\omega (C)
=[2\pi/L][v_{c0}+v_{s1}]$, $k (C) =0$, and $1/v=0$; and $a_0
=2\Delta_{\alpha\nu}^{\iota}[2\Delta_{\alpha\nu}^{\iota}+1]/2$ for $\omega (C)
=[4\pi/L]v_{\alpha\nu}$, $k (C) =\iota\,[4\pi/L]$, and $v=\iota\,v_{\alpha\nu}$. All
these expressions refer to $\alpha\nu = c0,\, s1$ and $\iota =\pm 1$. Similar $a_0$
expressions can be derived for larger energy values of order $1/L$.

In turn, for small finite values of $l\Delta\omega$ we use the
$\alpha\nu$ pseudofermion spectral function expression
(\ref{B-J-i-sum-GG}) and the variables $x=\omega'/\Delta\omega$
and $y={\rm sgn} (v)k'\,[v_{c0}/\Delta\omega]$ where $\Delta\omega
=(\omega -l\Delta E)$ in the second expression of Eq.
(\ref{B-l-i-breve}) and find,

\begin{eqnarray}
{\breve{B}}^{l,i} (\Delta\omega,v) & \approx &
{A^{(0,0)}_{c0}A^{(0,0)}_{s1}\over 2\pi\,v_{c0}}
\Bigl({l\Delta\omega\over 4\pi\sqrt{v_{c0}\,v_{s}}/L} \Bigr)^2
\int_{0}^1dx\int_{-1}^{+1}dy\prod_{\iota =\pm 1}
a_{c0,\,\iota}\left({l\Delta\omega\over
4\pi\sqrt{v_{c0}\,v_{s1}}/L}\Bigl[\sqrt{{v_{s1}\over
v_{c0}}}\,\Bigl(x+\iota\,y\Bigr)\Bigr]\right)\nonumber \\
& \times & a_{s1,\,\iota}\left({l\Delta\omega\over
4\pi\sqrt{v_{c0}\,v_{s1}}/L}\Bigl[\sqrt{{v_{c0}\over
v_{s1}}}\,\Bigl(1 -x+\iota\,\Bigl[{v_{s1}\over\vert
v\vert}-{v_{s1}\over v_{c0}}\,y\Bigr]\Bigr) \Bigr]\right) \, ;
\hspace{0.50cm} i=0,1,2,... \, , \hspace{0.25cm} l= \pm 1 \, .
\label{B-l-i-breve-int}
\end{eqnarray}
The expressions given in Eqs. (\ref{B-l-i-breve-int-0}) and
(\ref{B-l-i-breve-int}) combined with Eqs. (\ref{ABONjl-J-CPHS})
and (\ref{B-PAR-J-CPHS-sum-0}) provide the desired general
spectral function expression, except for $k$ and $\omega$ values
such that $\omega\approx \iota\,v_{\alpha\nu}(k-lk_0)+l\omega_0$,
where $\alpha\nu = c0,\,s1$ and $\iota =\pm 1$. In the ensuing
section we use the function (\ref{B-PAR-J-CPHS-sum-0}) to derive
closed-form analytical expressions for the finite-energy one- and
two-electron spectral-weight distributions.

For small finite values of $l\Delta\omega =l(\omega -l\Delta E)$,
use of the relative weight asymptotic expression (\ref{f}) in Eq.
(\ref{B-l-i-breve-int}) for the whole $x$ and $y$ integration
domains leads to the following asymptotic behavior for the
convolution function (\ref{B-l-i-breve-int}),

\begin{equation}
{\breve{B}}^{l,i} (\Delta\omega,v) \approx {F_0 (1/v)\over
4\pi\sqrt{v_{c0}\,v_{s1}}}\,\Theta
\Bigl(l\Delta\omega\Bigr)\,\Bigl({l\Delta\omega\over
4\pi\sqrt{v_{c0}\,v_{s1}}} \Bigr)^{-2+\zeta_0}  \, ;
\hspace{0.5cm} i=0,1,2,... \, , \hspace{0.25cm} l= \pm 1 \, .
\label{B-breve-asym}
\end{equation}
Here $\zeta_0$ is the functional given in Eq. (\ref{zeta0}) and
the function $F_0 (z)$ reads,

\begin{eqnarray}
F_0 (z) & = & 2D_0\,\sqrt{{v_{s1}\over
v_{c0}}}\int_0^1dx\int_{-1}^{+1}dy\,\prod_{\iota =\pm
1}{\Theta\Bigl(1 -x+{\rm sgn} (z)\,\iota\,\Bigl[v_{s1}\vert
z\vert-{v_{s1}\over v_{c0}}\,y\Bigr]\Bigr)\over \Gamma
(2\Delta_{s1}^{\iota})}\,{\Theta \Bigl(x+{\rm sgn}
(z)\,\iota\,y\Bigr)
\over \Gamma (2\Delta_{c0}^{\iota})}\nonumber \\
& \times & \left(\sqrt{{v_{c0}\over v_{s1}}}\,\Bigl[1 -x+{\rm sgn}
(z)\,\iota\,\Bigl[v_{s1}\vert z\vert-{v_{s1}\over
v_{c0}}\,y\Bigr]\Bigr]\right)^{2\Delta_{s1}^{\iota}-1}\,\left(\sqrt{{v_{s1}\over
v_{c0}}}\,\Bigl[x+{\rm sgn}
(z)\,\iota\,y\Bigr]\right)^{2\Delta_{c0}^{\iota}-1} \, ,
\label{C-v}
\end{eqnarray}
where $2\Delta_{s1}$ is the functional given in Eq. (\ref{zeta0})
and $D_0$ is defined in Eq. (\ref{B-l-i-breve-int-0}). Note that
both the values of the functional $\zeta_0 = \zeta_0
(\{q_i\},\{{q'}_i\},\{q_j\},\{{q'}_j\})$ defined in Eq.
(\ref{zeta0}) and of the function $F_0 (z)=C_0
(z,\,\{q_i\},\{{q'}_i\},\{q_j\},\{{q'}_j\})$ of Eq. (\ref{C-v})
depend on the set of bare-momentum values
$\{q_i\},\{{q'}_i\},\{q_j\},\{{q'}_j\}$ which define the
active-scattering-centers occupancy configurations of the initial
reduced-subspace excited energy eigenstates generated by the
elementary processes (A). Such a dependence follows from the
functional character of the quantity $2\Delta_{\alpha\nu}^{\iota}
= 2\Delta_{\alpha\nu}^{\iota}
(\{q_i\},\{{q'}_i\},\{q_j\},\{{q'}_j\})$ defined by Eq.
(\ref{Delta0}). The dependence of the value of the latter
functional on the set of bare-momentum values
$\{q_i\},\{{q'}_i\},\{q_j\},\{{q'}_j\}$ occurs through the
scattering phase shift $Q^{\Phi (NF)}_{\alpha\nu} (\iota
q^0_{F\alpha\nu},\{q_i\},\{{q'}_i\},\{q_j\},\{{q'}_j\})/2$
appearing in Eq. (\ref{Delta0}), which is defined in Eq.
(\ref{qcan1j-J-NF}). Thus, we should keep in mind that the values
of the functionals $2\Delta_{\alpha\nu}^{\iota}$ and $\zeta_0$
appearing in the spectral-function expressions derived below are
different for each pseudofermion bare-momentum occupancy
configuration defining a specific excited energy eigenstate
generated from the ground state by the elementary processes (A)
and (B).

\section{WEIGHT DISTRIBUTION EXPRESSIONS GENERATED BY THE DOMINANT PROCESSES}
\label{SecV}

In this section we derive analytical closed-form expressions for
the one- and two-electron weight distributions associated with the
function (\ref{ABONjl-J-CPHS}) for $(k,\,\omega)$-plane regions
whose weight is generated by the dominant processes. These
processes are characterized below in terms of rotated-electron
processes. Our study includes the derivation of the weight
distribution expressions valid for $k$ and $\omega$ values such
that $\omega\approx \iota\,v_{\alpha\nu}(k-lk_0)+l\omega_0$, where
$\alpha\nu = c0,\,s1$ and $\iota =\pm 1$. Some of the analytical
expressions found below refer to the vicinity of the weight
distribution singularities and edges. Importantly, the spectral
features observed in real experiments correspond to such
singularities \cite{blinde,spectral1,super}.

\subsection{THE DOMINANT PROCESSES AND THE FUNCTIONAL $\beth_0 (k,\,\omega)$}

The spectral weight distribution associated with the terms of the
general finite-energy spectral function (\ref{ABONjl-J-CPHS}) of
index $i>0$ decreases very rapidly for increasing values of the
number $i$ of extra rotated-electron pairs. Indeed, for the
excited energy eigenstates associated with the spectral-function
contributions of increasing $i$ value the functional $\zeta_0$ of
Eq. (\ref{zeta0}) also has increasingly larger values. Typically,
the contributions of orders $i$ larger than $i=1$ are beyond
numerical measurability, once the spectral weight associated with
the $i=0$ term of the spectral function (\ref{ABONjl-J-CPHS})
corresponds in general to over 99\% of the whole spectral weight.
(This is confirmed for the one-electron spectral function in Ref.
\cite{1-spec}.) Thus, for practical applications it is in general
enough to consider the elementary processes associated with the
$i=0$ term only.

The smallness of the weight associated with the $i>0$ terms of the
spectral function (\ref{ABONjl-J-CPHS}) results from the form of
the function (\ref{B-PAR-J-CPHS-sum-0}). Indeed, for increasing
values of the index $i$ the expression of the operator
${\tilde{\Theta}}_{k,\,\alpha\nu}^{l,i}$ associated with the
operator ${\tilde{\Theta}}_{k,\,\alpha\nu}^{l,NF,i}$, Eq. (58) of
Ref. \cite{V}, has in terms of rotated-electron operators an
increasing number of rotated-electron particle-hole pair operators
\cite{V}. Thus, for increasing values of $i$ application of such
an operator onto the ground state produces elementary processes
(A) and (B) which generate an increasing number of active
scattering centers. Since the value of the functional $\zeta_0$
given in Eq. (\ref{zeta0}) is in general an increasing function of
the number of generated active scattering centers, it follows that
the $i>0$ contributions to the spectral function are very small.
Below it is confirmed that most singularities of the latter
function occur at $(k,\,\omega)$-plane isolated points and lines,
are of power-law form, and are controlled by processes which
generate excited states whose $\zeta_0$ values are smaller than
two and one, respectively. Thus, a quite good approximation
corresponds to replacing the general spectral-function expression
(\ref{ABONjl-J-CPHS}) by its $i=0$ term. The
excited-energy-eigenstate deviations generated by the dominant
processes associated with the $i=0$ operator
${\tilde{\Theta}}_{{{\cal{N}}_0},\,j}^{l}$ on the right-hand side
of Eq. (32) of Ref. \cite{V} obey the two selection rules given in
Eq. (41) of the same reference. The numbers
${\bar{\cal{N}}}_{-1,\,l_s}^{l}={\bar{\cal{N}}}_{-1,\,l_s}^{l,0}$
in that selection rule are those of Eq. (30) of Ref. \cite{V}
specific to the corresponding operator
${\hat{\Theta}}_{{\cal{N}},\,j}^{l}$ defined in Eqs. (27) and (28)
of the same reference. Furthermore, all excited energy eigenstates
whose deviations do not obey that selection rule are generated by
the $i>0$ operators on the right-hand side of the latter equation.
It follows that a good approximation for the spectral function
consists in using expression (\ref{ABONjl-J-CPHS}) for $i=0$ and
the summation on the right-hand side of Eq. (\ref{ABONjl-J-CPHS})
over the J-CPHS subspaces whose number deviations obey the sum
rules (18), (19), (39), and (43) of Ref. \cite{V} and selection
rules given in Eqs. (21) and (41) of the same reference.

Moreover, for the one- and two-electron spectral functions nearly
the whole weight is concentrated in the $(k,\,\omega)$-plane
regions associated with creation by the elementary processes (A)
of none, one, and two active scattering centers away from the
$\alpha\nu =c0,\,s1$ {\it Fermi points} and $\alpha\nu\neq
c0,\,s1$ limiting bare-momentum values. Concerning singular weight
features, we find below that for finite values of the on-site
repulsion $U$ only exceptionally and for some spectral functions
and isolated $(k,\,\omega)$-plane points are these features of
$\delta$-function type. In general, such singularities are instead
of power-law shape. Power-law spectral-function behavior occurs
for $(k,\,\omega)$-plane regions in the vicinity of points or
lines corresponding to different reduced J-CPHS subspaces than
that associated with such regions. In order to describe such an
effect it is convenient to associate each $(k,\,\omega)$-plane
point with the minimum value of the functional $\zeta_0$, Eq.
(\ref{zeta0}), of the excited energy eigenstates generated by the
elementary processes (A) and (B) that correspond to such a point.
We thus introduce the functional,

\begin{equation}
\beth_0 (k,\,\omega)\equiv {\rm min}\,\zeta_0 \, . \label{Omega0}
\end{equation}
In general, for the same regions of the $(k,\,\omega)$-plane the
spectral function (\ref{ABONjl-J-CPHS}) has contributions from
different functions $B^{l,i} (k,\,\omega)$, associated with
different J-CPHS subspaces of the same CPHS ensemble subspace. The
value of the functional (\ref{Omega0}) corresponds to the smallest
value of $\zeta_0$ of all states associated with the
$(k,\,\omega)$-plane point into consideration. For two-dimensional
$(k,\,\omega)$-plane regions associated with a given reduced
J-CPHS subspace, the value of the functional $\zeta_0$, Eq.
(\ref{zeta0}), is in general a continuous and smooth function of
$k$ and $\omega$. Therefore, in that case the value of the
functional $\beth_0 (k,\,\omega)$ is also in general a continuous
and smooth function of $k$ and $\omega$. Importantly, the spectral
weight power-law features correspond to the points or lines where
$\beth_0 (k,\,\omega)$ shows discontinuities. (The inverse is not
always true.) Typically, such discontinuities occur in the border
lines of the $(k,\,\omega)$-plane domains associated with the
reduced J-CPHS subspaces. Such subspaces can correspond to a
$(k,\,\omega)$-plane isolated point, line, or two-dimensional
domain. In general, the corresponding value of the the functional
$\zeta_0$, Eq. (\ref{zeta0}), is smallest for the point-subspaces
whose deviation numbers and numbers are given in Eq.
(\ref{om0-k0}). It is also in general smaller for the line-like
reduced J-CPHS subspaces than for the subspaces corresponding to a
$(k,\,\omega)$-plane two-dimensional domain. Therefore, when the
reduced J-CPHS subspace $(k,\,\omega)$-plane domains correspond to
isolated points or lines, the discontinuities of the value of the
functional (\ref{Omega0}) occur at these points or lines. It also
shows discontinuities at the $(k,\,\omega)$-plane border lines
defining the limits of the reduced J-CPHS subspace two-dimensional
domains.

The reason why the spectral function has power-law behavior in the
vicinity of the $(k,\,\omega)$-plane isolated points and lines,
where the value of the functional (\ref{Omega0}) has
discontinuities, is the small energy gap between these regions and
the point or line corresponding to the discontinuities. The value
of the functional (\ref{zeta0}) is in general different at the
point or line than in its vicinity. Moreover, in general that
value is smaller for the former line or point than in its
proximity. Thus, the small gap gives the minimal excitation energy
of the tower states associated with the $(k,\,\omega)$-plane point
where one is calculating the spectral function and which are
generated by the elementary processes (C) relative to that of the
initial states corresponding to the point or line where the value
of the functional (\ref{Omega0}) has a discontinuity. We recall
that all states generated by the elementary processes (C) have the
same value for the functional $\zeta_0$, Eq. (\ref{zeta0}), as the
initial state, which corresponds to the point or line where the
value of the functional (\ref{Omega0}) has a discontinuity. The
occurrence of the small gap implies that the corresponding
convolution function (\ref{B-l-i-breve-int}) has the power-law
asymptotic form (\ref{B-breve-asym}). Note that such a power-law
behavior is controlled by the exponent $-2+\zeta_0$, which is
negative provided that $\zeta_0<2$. Furthermore, once the value of
the functional (\ref{zeta0}) is smaller for these excited states
than for states generated by elementary processes (C) from initial
states corresponding to $(k,\,\omega)$-plane points located closer
to the point where we are calculating the spectral function, the
spectral function itself has a power-law shape, as confirmed
below.

In turn, if in the vicinity of a $(k,\,\omega)$-plane point the
value of the functional (\ref{Omega0}) has no discontinuities, the
spectral function expression (\ref{B-PAR-J-CPHS-sum-0}) at that
point involves integrations over gapless contributions from
transitions generated by the elementary processes (C) from initial
states corresponding to a small two-dimensional
$(k,\,\omega)$-plane domain just below ($l=+1$) or above ($l=-1$)
the point. Once the value of the functional $\beth_0 (k,\,\omega)$
is in this case a continuous and smooth function of $k$ and
$\omega$, its value remains the same in the infinitesimal vicinity
of the point. In this case the spectral function has no power-law
behavior, as confirmed below. The lack of such a behavior follows
from the gapless character of the excitations contributing to the
spectral function at that point. Below we consider the
spectral-weight distribution behavior generated by creation of
none, one, and two active scattering centers away from the
$\alpha\nu =c0,\,s1$ {\it Fermi points} and $\alpha\nu\neq
c0,\,s1$ limiting bare-momentum values. For one- and two-electron
spectral functions the weight generated by processes involving the
creation of more than two active pseudofermion or hole scattering
centers away from the $\alpha\nu =c0,\,s1$ {\it Fermi points} and
$\alpha\nu\neq c0,\,s1$ limiting bare-momentum values is very
small and can in general be neglected.

Our results reveal that the functional (\ref{Omega0}) plays a
major role in the control of the spectral weight distribution.
Increasing the number of created active pseudofermion and
pseudofermion hole scattering centers increases in general the
value of that functional. Moreover, the summations in expression
(\ref{B-PAR-J-CPHS-sum-0}) over the states generated by the
elementary processes (A) further increase the value of the
power-law exponent $-2+\zeta_0$ appearing in the convolution
function (\ref{B-breve-asym}). Indeed, these summations add
integer numbers to the exponent $-2+\zeta_0$: creation of each
active scattering center adds $1$ to that exponent, as confirmed
below. A central point is that most spectral weight singularities
are of power-law type and correspond to the discontinuities of the
value of the functional (\ref{Omega0}) in the
$(k,\,\omega)$-plane. Also the weight associated with creation of
two or more active scattering centers away from the $\alpha\nu
=c0,\,s1$ {\it Fermi points} and $\alpha\nu\neq c0,\,s1$ limiting
bare-momentum values is controlled by that functional.
Furthermore, the same functional controls the exceptional
occurrence of $\delta$-function singularities: such singularities
corresponds to the $(k,\,\omega)$-plane points where its value
vanishes.

Each spectral-weight distribution (\ref{B-PAR-J-CPHS-sum-0})
corresponds to a J-CPHS subspace. We start by considering the
spectral-weight distributions generated by creation of two active
pseudofermions and/or pseudofermion holes scattering centers away
from the $\alpha\nu =c0,\,s1$ {\it Fermi points} and
$\alpha\nu\neq c0,\,s1$ limiting bare-momentum values. Thereafter,
we consider spectral-weight distributions
(\ref{B-PAR-J-CPHS-sum-0}) associated with contributions from
processes involving creation of none and one active scattering
center away from the $\alpha\nu =c0,\,s1$ {\it Fermi points} and
$\alpha\nu\neq c0,\,s1$ limiting bare-momentum values. We find
below that the latter processes lead to important point- and
line-like singular spectral-weight features.

\subsection{TWO-DIMENSIONAL $(k,\,\omega)$-PLANE REGIONS WHERE THE SPECTRAL
WEIGHT IS CONTROLLED BY TWO ACTIVE-SCATTERING-CENTER CREATION}

The parametric equations that define the two-dimensional
$(k,\,\omega)$-plane domains generated by creation of two active
scattering centers away from the $\alpha\nu =c0,\,s1$ {\it Fermi
points} and $\alpha\nu\neq c0,\,s1$ limiting bare-momentum values
are of the general form,

\begin{equation}
k = lk_{\alpha\nu,\,\alpha'\nu'}(q,\,q') \, ; \hspace{1.0cm}
\omega = l\omega_{\alpha\nu,\,\alpha'\nu'}(q,\,q') \, .
\label{general}
\end{equation}
For instance,

\begin{eqnarray}
k & = & l[k_0 + c_1\,q+ c_1'\,q'] \, ; \hspace{0.5cm} \omega =
l\omega_{\alpha\nu,\,\alpha'\nu'}(q,\,q')= l[\omega_0
+c_1\,\epsilon_{\alpha\nu} (q)+c_1'\,\epsilon_{\alpha'\nu'} (q')]
\, , \hspace{0.25cm} \alpha\nu = c0,\,s1 \, , \hspace{0.25cm}
\alpha'\nu' = c0,\,s1 \, ; \nonumber \\
k & = & l[k_0 + c_1\,q+ k_{c\nu} - q'] \, ; \hspace{0.5cm} \omega
= l\omega_{\alpha\nu,\,\alpha'\nu'}(q,\,q')= l[\omega_0
+c_1\,\epsilon_{\alpha\nu} (q)+\epsilon^0_{c\nu'} (q')] \, ,
\hspace{0.25cm} \alpha\nu = c0,\,s1 \, , \hspace{0.25cm}
\nu'>0 \, ; \nonumber \\
k & = & l[k_0 + c_1\,q+q'] \, ; \hspace{0.5cm} \omega =
l\omega_{\alpha\nu,\,\alpha'\nu'}(q,\,q')= l[\omega_0
+c_1\,\epsilon_{\alpha\nu} (q)+\epsilon^0_{s\nu'} (q')] \, ,
\hspace{0.25cm} \alpha\nu = c0,\,s1 \, , \hspace{0.25cm} \nu'>1 \,
. \label{*plane}
\end{eqnarray}
Here the index $l=\pm 1$ is that of the corresponding spectral
function $B_{{\cal{N}}}^{l} (k,\,\omega)$ of Eq.
(\ref{ABONjl-J-CPHS}), the momentum $k_{c\nu}$, energy $\omega_0$,
and momentum $k_0$ are given in Eqs. (\ref{DP-NI}),
(\ref{omega0}), and (\ref{k0-J}), respectively, and the constant
$c_1$ [and $c_1'$] is such that $c_1=+1$ (and $c_1=-1$) for
creation of a $\alpha\nu$ pseudofermion (and a $\alpha\nu$
pseudofermion hole). For simplicity, in Eq. (\ref{*plane}) we
considered that the two created active scattering centers are both
$\alpha\nu=c0,\,s1$ pseudofermions and/or holes, a
$\alpha\nu=c0,\,s1$ pseudofermion or hole and a $c\nu\neq c0$
pseudofermion, and a $\alpha\nu=c0,\,s1$ pseudofermion or hole and
a $s\nu\neq s1$ pseudofermion. Generalization to other
active-scattering-center choices is straightforward.

The summation over the excited energy eigenstates generated by the
elementary processes (A) which contribute to the spectral weight
at the $(k,\,\omega)$-plane point defined by Eq. (\ref{*plane}) is
performed in Appendix B. The momentum and energy of these states
relative to the ground state is close to $k$ and $\omega$,
respectively. These contributions occur through generation from
these states, by the elementary processes (C), of new excited
energy eigenstates whose momentum and energy relative to the
initial ground state are precisely $k$ and $\omega$, respectively.
Since the former excited energy eigenstates generated by the
elementary processes (A) have very close momentum and energy,
except for vanishing contributions of order $1/N_a$ the
corresponding value of the functional (\ref{Omega0}) is the same.
For finite values of that functional it is found in Appendix B
that the spectral function has following general form,

\begin{equation}
B_{{\cal{N}}}^{l} (k,\,\omega) \approx {1\over
\pi\,C_c\,C_s}\left[\int_{-1/v_{s1}}^{+1/v_{s1}} dz\,F_0
(z)\right]\,\Bigl({\Omega\over 4\pi\sqrt{v_{c0}\,v_{s1}}}
\Bigr)^{\zeta_0 (q,\,q')}\,{\sqrt{v_{c0}\,v_{s1}}\over \zeta_0
(q,\,q')\,\vert v_{\alpha\nu} (q)-v_{\alpha'\nu'} (q')\vert}\, \,
; \hspace{0.5cm} l= \pm 1 \, . \label{B-2D-inte}
\end{equation}
While this expression and all other spectral-weight distribution
expressions obtained in this paper correspond to densities $0<n<1$
and $0<m<n$, taking the limit $m\rightarrow 0$ in such expressions
leads to the correct $m=0$ results \cite{spectral}. The expression
(\ref{B-2D-inte}) is not valid in the vicinity of isolated
$(k,\,\omega)$-plane points whose weight is associated with
point-subspaces and in the proximity of the branch lines studied
below. In the vicinity of these points and lines one must consider
the contributions from processes involving the creation of none
and one active scattering center away from the $\alpha\nu
=c0,\,s1$ {\it Fermi points} and $\alpha\nu\neq c0,\,s1$ limiting
bare-momentum values, respectively. On the right-hand side of Eq.
(\ref{B-2D-inte}), $\zeta_0=\zeta_0 (q,\,q')$ is the value of the
general functional (\ref{zeta0}) specific to the two
active-scattering-center excitation under consideration, $F_0 (z)$
is the function (\ref{C-v}), and the small but finite energy
$\Omega$ corresponds to the range of the elementary processes (C).
We recall that $\Omega$ is treated as a parameter whose value for
fixed finite values of $n$, $m$, and $U/t$ is determined by
imposing the exact sum rule (\ref{sum-ABON}) to the full
spectral-function expression in the end of the calculations.
$\Omega$ also appears in the spectral-function contributions from
none and one active scattering center creation away from the
$\alpha\nu =c0,\,s1$ {\it Fermi points} and $\alpha\nu\neq
c0,\,s1$ limiting bare-momentum values, as confirmed below. The
general spectral-function expression (\ref{B-2D-inte}) is valid
for finite values of the functional (\ref{Omega0}). For regions of
the $(k,\,\omega)$-plane where the value of that functional
vanishes the energy $\Omega$ is such that $\Omega\rightarrow 0$
and the spectral-function expression has a different form, as
discussed below.

The $k$ and $\omega$ dependence of the weight-distribution
expression (\ref{B-2D-inte}) occurs mainly through the $q$ and
$q'$ dependence of the value of the functional $\zeta_0 (q,\,q')$
and through the absolute value of $v_{\alpha\nu}
(q)-v_{\alpha'\nu'} (q')$. The created active-scattering-center
bare momentum values $q$ and $q'$ are related to the
weight-distribution value of $k$ and $\omega$ by Eq.
(\ref{*plane}). We emphasize that there are two types of
$(k,\,\omega)$-plane points: (i) those associated with one pair of
$(q,\,q')$ values and (ii) those corresponding to two different
pairs of such bare-momentum values. In the case (ii) the spectral
function at the $(k,\,\omega)$-plane point is given by the sum of
two terms of the general form given in Eq. (\ref{B-2D-inte}). The
shape of the two-dimensional sub-domains of type (i) and (ii)
depends on the specific form of the parametric equations
(\ref{*plane}). While the goal of this paper is the derivation of
general analytical expressions for the one- and two-electron
spectral-weight distributions of the metallic phase, examples of
such sub-domains will be given elsewhere for specific one- and
two-electron spectral functions.

The two-active-scattering-center $(k,\,\omega)$-plane
two-dimensional domain is bounded by well defined lines. Also the
two sub-domains (i) and (ii) are bounded by well defined lines. An
important general property is that the part of the limiting line
of the two-dimensional domain which is also a limiting line for
the sub-domain (ii) [this excludes the limiting line between the
sub-domains (i) and (ii)] is such that $v_{\alpha\nu}
(q)=v_{\alpha'\nu'} (q')$. We call such a line {\it border line}.
The shape $\omega =\omega_{BL} (k)$ of a border line is defined by
the following parametric equations,

\begin{equation}
\omega_{BL} (k) = l[\,\omega_0 +c_1\,\epsilon_{\alpha\nu}
(q)+c_1'\,\epsilon_{\alpha'\nu'} (q')]\,\delta_{v_{\alpha\nu}
(q),\,v_{\alpha'\nu'} (q')} \, ; \hspace{0.5cm} k =
lk_{\alpha\nu,\,\alpha'\nu'}(q,\,q')\,\delta_{v_{\alpha\nu}
(q),\,v_{\alpha'\nu'} (q')} \, . \label{b-lines}
\end{equation}
According to expression (\ref{B-2D-inte}), the weight distribution
has singular behavior at that line. Note that such an expression
defines the $k$ and $\omega$ dependence of the weight
distribution for the regions just below ($l=+1$) or above
($l=-1$) the border line. We recall that the border line
corresponds to a discontinuity in the values of the functional
(\ref{Omega0}). Thus, the weight distribution has power-law
behavior in the region just above ($l=+1$) or below
($l=-1$) the border line. The $k$ and $\omega$ dependence
of the weight distribution in that region is not given
by Eq. (\ref{B-2D-inte}). The study of that dependence proceeds
much as for the branch lines considered below and will
be carried out for specific weight distributions elsewhere.
The same applies to the region just above ($l=+1$) or below 
($l=-1$) the remaining limiting lines of the two-dimensional
two-active-scattering-center $(k,\,\omega)$-plane domain.

We emphasize that depending on the specific weight distribution,
the dominant processes can correspond to different choices of
active-scattering-center pairs. Thus, it can occur that the same
$(k,\,\omega)$-plane region involves contributions from several
expressions of general form given in Eq. (\ref{B-2D-inte}). In
that case the weight distribution at a $(k,\,\omega)$-plane point
is given by the sum of several terms of the general form provided
in that equation. Moreover, one must also add the spectral
features generated by creation of none and one active scattering
center away from the $\alpha\nu =c0,\,s1$ {\it Fermi points} and
$\alpha\nu\neq c0,\,s1$ limiting bare-momentum values, which we
derive below.

\subsection{THE POINT-LIKE AND LINE-LIKE SPECTRAL-WEIGHT DISTRIBUTION FEATURES}

The smallest reduced subspaces are point-subspaces whose deviation
numbers and numbers are given in Eq. (\ref{om0-k0}). Analysis of
the general spectral-function expression (\ref{ABONjl-J-CPHS}) for
$i=0$ reveals that the most divergent singular power-law spectral
features appear in the vicinity of the $(k,\,\omega)$-plane
points, $(lk_0,\,l\omega_0)$, corresponding to such subspaces.
Such a weight results from processes that do not involve creation
away from the $\alpha\nu =c0,\,s1$ {\it Fermi points} and
$\alpha\nu\neq c0,\,s1$ limiting bare-momentum values of active
scattering centers. For $(k,\,\omega)$-plane regions in the
vicinity of the corresponding $(lk_0,\,l\omega_0)$ points the
dominant contributions to the general expression
(\ref{B-PAR-J-CPHS-sum-0}) are generated by elementary processes
(C) whose initial state corresponds to the point-subspace. Here
$\omega_0$ and $k_0$ are given in Eqs. (\ref{omega0}) and
(\ref{k0-J}), respectively. Such contributions lead to
$B_{{\cal{N}}}^{l} (k,\,\omega)\approx\Theta (\Omega
-l[\omega-l\omega_0])\,{\breve{B}}^{l,0}
(\omega-l\omega_0,\,[\omega-l\omega_0/ k -lk_0])$ and thus the
weight distribution has the following power-law behavior,

\begin{equation}
B_{{\cal{N}}}^{l} (k,\,\omega)\approx {F_0 (1/v)\over
4\pi\,\sqrt{v_{c0}\,v_{s1}}\,C_c\,C_s}\,\Theta \Bigl(\Omega
-l[\omega-l\omega_0]\Bigr)\,\Theta
\Bigl(l[\omega-l\omega_0]\Bigr)\Bigl({l[\omega-l\omega_0]\over
4\pi\sqrt{v_{c0}\,v_{s1}}} \Bigr)^{-2+\zeta_0}  \, ,
\label{B-PAR-V-L}
\end{equation}
where $l= \pm 1$ and the velocity $v =[\omega-l\omega_0]/[k
-lk_0]$ is always such that $\vert v\vert>v_{s1}$ and $v\neq \pm
v_{c0}$. Depending on the specific point-subspace and weight
distribution, the domain of $z=1/v$ values corresponding to the
small finite spectral-weight $(k,\,\omega)$-plane region where the
expression (\ref{B-PAR-V-L}) is valid is in general bounded by two
of the four values $-1/v_{c0}$, $-1/v_{s1}$, $1/v_{s1}$, and
$1/v_{c0}$. This expression refers to small finite values of
$[\omega-l\omega_0]$, the functional $\zeta_0$ and the function
$F_0 (z)$ are given in Eqs. (\ref{zeta0}) and (\ref{C-v}),
respectively, and the value of the corresponding functionals
$2\Delta_{c0}^{\pm 1}$ and $2\Delta_{s1}^{\pm 1}$ whose general
expression is given in Eq. (\ref{Delta0}) is that of the initial
point-subspace. It is not valid for $k$ and $\omega$ values such
that $\omega\approx \iota\,v_{\alpha\nu}(k-lk_0)+l\omega_0$, where
$\alpha\nu = c0,\,s1$, $\iota =\pm 1$. (The expression for such
$k$ and $\omega$ values is given below.) When the exponent
$-2+\zeta_0$ is negative, the weight distribution
(\ref{B-PAR-V-L}) has a singularity at the $(lk_0,\,l\omega_0)$
point. In general, the value of that exponent increases for
increasing value of $\vert k_0\vert $. Thus, it is larger for the
spectral function expression in the vicinity of the
$(lk_0,\,l\omega_0)$ points corresponding to the second
point-subspaces than in the vicinity of the points associated with
the first point-subspaces.

For the small $(k,\,\omega)$-plane region in the vicinity of the
point $(lk_0,\,l\omega_0)$, where the expression (\ref{B-PAR-V-L})
is valid, the contributions from functions of form
(\ref{B-PAR-J-CPHS-sum-0}) associated with J-CPHS subspaces other
than the point-subspace are very small. Therefore, in that region
the spectral function (\ref{ABONjl-J-CPHS}) has for $i=0$
approximately the form given in Eq. (\ref{B-PAR-V-L}).

Let us now consider the contributions from processes involving the
creation of a single active scattering center away from the
$\alpha\nu =c0,\,s1$ {\it Fermi points} and $\alpha\nu\neq
c0,\,s1$ limiting bare-momentum values. These contributions lead
to the next more divergent singular features, which arise in the
vicinity of {\it branch lines}. In this case the energy
eigenstates which span a reduced J-CPHS subspace are associated
with a $(k,\,\omega)$-plane line that corresponds to a
discontinuity in the values of the functional (\ref{Omega0}). By
changing the bare momentum $q$ of the active $\alpha\nu$
pseudofermion or pseudofermion hole scattering center, one
generates a branch line in the $(k,\,\omega)$-plane. Often one of
the end points (or both end points) of such a line coincides with
a $(lk_0,\,l\omega_0)$ point associated with a point-subspace. For
each CPHS ensemble subspace, at least one of the end points of the
most divergent of such branch lines coincides with a first
point-subspace point $(lk_0,\,l\omega_0)$. The active $\alpha\nu$
pseudofermion (and $\alpha\nu$ pseudofermion hole) scattering
center can belong to any branch (to the $\alpha\nu =c0,\,s1$
branches). A $\alpha\nu =c0,\,s1$ branch line is generated by
creation of the $\alpha\nu$ pseudofermion (or $\alpha\nu$
pseudofermion hole) for all available bare-momentum values in the
domain $\vert q\vert\in [q^0_{F\alpha\nu},\,q^0_{\alpha\nu}]$ (or
$\vert q\vert\in [0,\,q^0_{F\alpha\nu}]$).

The parametric equations that define the $(k,\,\omega)$-plane
points belonging to a $\alpha\nu$ pseudofermion (or $\alpha\nu$
pseudofermion hole for the $\alpha\nu =c0,\,s1$ bands) branch line
is of the general form,

\begin{eqnarray}
k & = & lk_{\alpha\nu} (q) = l[k_0 + c_1\,q] \, ; \hspace{1cm}
\omega = l\omega_{\alpha\nu} (q) = l[\omega_0
+c_1\,\epsilon_{\alpha\nu} (q)] \, , \hspace{1cm}
\alpha\nu = c0,\,s1 \, ; \nonumber \\
k & = & lk_{c\nu} (q) =l[k_0 + k_{c\nu} - q] \, ; \hspace{1cm}
\omega = l\omega_{c\nu} (q) = l[\omega_0 + \epsilon^0_{c\nu} (q)]
\, ,
\hspace{1cm} c\nu\neq c0 \, ; \nonumber \\
k & = & lk_{s\nu} (q) =l[k_0 + q] \, ; \hspace{1cm} \omega =
l\omega_{s\nu} (q) = l[\omega_0 + \epsilon^0_{s\nu} (q)] \, ,
\hspace{1cm} s\nu\neq s1 \, . \label{*line}
\end{eqnarray}
Here the index $l=\pm 1$ is that of the corresponding spectral
function $B_{{\cal{N}}}^{l} (k,\,\omega)$ of Eq.
(\ref{ABONjl-J-CPHS}), the momentum $k_{c\nu}$, energy $\omega_0$,
and momentum $k_0$ are given in Eqs. (\ref{DP-NI}),
(\ref{omega0}), and (\ref{k0-J}), respectively, and the constant
$c_1$ is such that $c_1=+1$ (and $c_1=-1$) for creation of a
$\alpha\nu$ pseudofermion (and a $\alpha\nu$ pseudofermion hole).

Let us consider a $(k,\omega)$-plane point located just above
($l=+1$) or below ($l=-1$) the branch line whose momentum $k$
obeys the relation given in Eq. (\ref{*line}) and the energy
$\omega$ is such that $l(\omega - l\omega_{\alpha\nu}(q))$ is
small and positive. The weight-distribution expression at that
point is controlled by the elementary processes (C), which
generate from the initial excited energy eigenstates corresponding
to the line a set of tower states whose momentum and energy
relative to the ground state are precisely $k$ and $\omega$. By
performing the summations of the general expression
(\ref{B-PAR-J-CPHS-sum-0}) over the initial excited energy
eigenstates generated from the ground state by the elementary
processes (A) which correspond to line points in the vicinity of
such a point, in Appendix B we derive the following
spectral-weight distribution expression,

\begin{eqnarray}
B_{{\cal{N}}}^{l} (k,\,\omega) & \approx & \Theta \Bigl(\Omega
-l[\omega-l\omega_{\alpha\nu}(q)]\Bigr)\,\Theta
\Bigl(l[\omega-l\omega_{\alpha\nu}(q)]\Bigr)\,C_{\alpha\nu}
(q)\Bigl({l[\omega - l\omega_{\alpha\nu}(q)]\over
4\pi\sqrt{v_{c0}\,v_{s1}}}\Bigr)^{-1+\zeta_0 (q)} \, ;
\hspace{0.50cm} \zeta_0 (q) > 0 \, , \label{Ichiun}
\end{eqnarray}
where

\begin{eqnarray}
C_{\alpha\nu} (q) & = & {{\rm sgn} (q)\,c_{\alpha\nu}\over
2\pi\,C_c\,C_s}\Bigl\{\Theta\Bigl(v_{s1}\,[1-{l[\omega -
l\omega_{\alpha\nu}(q)]\over\Omega}]-\vert v_{\alpha\nu}
(q)\vert\Bigr)\int_{-{{\rm sgn} (q)\,c_{\alpha\nu}\over
v_{s1}}}^{{{\rm sgn}
(q)\,c_{\alpha\nu}\over v_{s1}}}dz\nonumber \\
& + & \theta\Bigl(\vert v_{\alpha\nu}
(q)\vert-v_{s1}\,[1-{l[\omega -
l\omega_{\alpha\nu}(q)]\over\Omega}]\Bigr) \int_{-{{\rm sgn}
(q)\,c_{\alpha\nu}\over v_{s1}}}^{{1\over v_{\alpha\nu}
(q)}\Bigl(1-{l[\omega-l\omega_{\alpha\nu}(q)]\over\Omega}\Bigr)}dz\Bigr\}{F_0
(z)\over [1 -z\,v_{\alpha\nu}(q)]^{\zeta_0(q)}} \, . \label{Can}
\end{eqnarray}
This expression does not apply to $k$ and $\omega$ values such
that $\omega\approx \iota\,v_{\alpha\nu}(k-lk_0)+l\omega_0$, where
$\alpha\nu = c0,\,s1$ and $\iota =\pm 1$. The functional
$\zeta_0=\zeta_0 (q)$ appearing in expression (\ref{Can}) is given
in Eqs. (\ref{zeta0}) and $F_0 (z)$ is the function (\ref{C-v}).
Furthermore, $c_{\alpha\nu}=1$ for $\alpha\nu =c0,\,s\nu$,
$c_{\alpha\nu}=-1$ for $\alpha\nu =c\nu\neq c0$.

The unity added to the exponent $-2 + \zeta_0$ of the convolution
function (\ref{B-breve-asym}) and spectral function expression
(\ref{B-PAR-V-L}) to reach the branch-line exponent $-1 + \zeta_0
(q)$ of expression (\ref{Ichiun}), stems from the summations
performed in Appendix B over the suitable initial excited energy
eigenstates generated by the elementary processes (A). For a
singular $\alpha\nu$ branch line the corresponding power-law
exponent $-1 + \zeta_0 (q)$ is negative. Since when $-1 + \zeta_0
(q)<0$ the weight distribution shows singularities at the branch
line, we expect that in this case spectral peaks will be observed
in experiments. This is confirmed for one-electron removal in Ref.
\cite{blinde} for both the TTF and TCNQ spectral dispersions of
the quasi-1D organic compound TTF-TCNQ, following the preliminary
results of Refs. \cite{spectral1,spectral} for the TCNQ stacks of
molecules. When the region above ($l=+1$) or below ($l=-1$) the
branch line is not contained in a $(k,\,\omega)$-plane domain
associated with a two-active-scattering-center reduced J-CPHS
subspace, the expression (\ref{Ichiun}) describes the
spectral-function behavior when $-1 + \zeta_0 (q)<0$, whereas
$0<-1 + \zeta_0 (q)<1$ is a sign of near absence of spectral
weight. In contrast, if that region is contained in such a domain,
the expression (\ref{Ichiun}) describes a spectral-function weight
edge  when $0<-1 + \zeta_0 (q)<1$. Finally, $-1 + \zeta_0 (q)>1$
is always a sign of near absence of spectral weight in the
vicinity of the line.

For momentum values $k$ and energies $\omega$ such that
$\omega\approx \iota\,v_{\alpha\nu}(k-lk_0)+l\omega_0$, where
$\alpha\nu = c0,\,s1$ and $\iota =\pm 1$, the spectral function
corresponds to the vicinity of a $\alpha\nu = c0,\,s1$ branch line
end point. However, in this case the expression of the spectral
function in the vicinity of the branch-line is not that of Eq.
(\ref{Ichiun}). This results from a resonance effect: the branch
line group velocity $v_{\alpha\nu} (q)$ equals the velocity
$\iota\,v_{\alpha\nu}$ associated with the $\alpha\nu,\,\iota$
pseudofermion particle-hole excitation sub-branch generated by the
elementary processes (C). Moreover, the $\alpha\nu = c0,\,s1$
branch line is in this limiting situation generated by these
elementary processes. Indeed, one can consider that in a first
step the branch-line $\alpha\nu = c0,\,s1$ pseudofermion or hole
is created at the $q=\iota\,q^0_{F\alpha\nu}$ {\it Fermi point} by
the elementary processes (B). This first step corresponds to a
given reduced J-CPHS subspace. One then accesses the corresponding
J-CPHS subspace by moving the $\alpha\nu = c0,\,s1$ pseudofermion
or hole along its energy and momentum dispersion. This second
small-momentum and low-energy process does not involve creation of
any quantum object and thus corresponds to an elementary process
(C). When moving the $\alpha\nu = c0,\,s1$ pseudofermion or hole
in the vicinity of the $q=\iota\,q^0_{F\alpha\nu}$ {\it Fermi
point}, its energy dispersion is linear in the momentum. It
follows that for these values of $k$ and $\omega$ the elementary
processes (C) replace the elementary processes (A) in what the
generation of the $\alpha\nu$ branch line is concerned. While the
elementary $\alpha\nu = c0,\,s1$ pseudofermion particle-hole
processes (C) generate the branch line, the elementary
${\bar{\alpha}\bar{\nu}}$ pseudofermion particle-hole processes
(C) play the usual role of these processes in the general
convolution function (\ref{B-l-i-breve}). (As in Sec. IV, here we
have used the index ${\bar{\alpha}\bar{\nu}}$ such that
${\bar{c}\bar{0}}=s1$ and ${\bar{s}\bar{1}}=c0$.) It follows that
for values of $k$ and $\omega$ obeying the relation $\omega\approx
\iota\,v_{\alpha\nu}(k-lk_0)+l\omega_0$, where $\iota =\pm 1$, the
convolution function associated with the $\alpha\nu = c0,\,s1$
spectral function is given by,

\begin{equation}
{\breve{B}}^{l,0}_{\alpha\nu,\,\iota} (\Delta\omega,v) \approx
{\iota 1\over 2\pi}\int_{0}^{\Delta\omega}d\omega'
\int_{\iota\omega'/v_{\alpha\nu}-\iota\Delta
q_{line}}^{\iota\omega'/v_{\alpha\nu}}
dk'\,B^{l,0}_{Q_{\bar{\alpha}\bar{\nu}}} \Bigl(\Delta\omega/v -
k',\Delta\omega-\omega'\Bigr)\,B^{l,0}_{Q_{\alpha\nu}}
\Bigl(k',\omega'\Bigr) \, ; \hspace{0.50cm} l =\pm 1 \, ,
\label{B-iota-bar}
\end{equation}
where $\Delta q_{line}=l[2\pi/L]\,x$ such that $0<x<1$ is the
effective momentum {\it width} of the branch line. Here we
consider the more general case of the functional
$2\Delta_{\alpha\nu}^{-\iota}$ having values in the domain
$0<2\Delta_{\alpha\nu}^{-\iota}<1$ and define $x$ as
$x=[1/N_a]^{(1/2\Delta_{\alpha\nu}^{-\iota}-1)}$. The introduction
of the branch-line momentum effective width $\Delta q_{line}$
provides a good approximation for the value for the power-law
multiplicative constant given below in Eq. (\ref{Can-bar}). In
turn, the power-law $k$ and $\omega$ dependence obtained below is
exact. The function (\ref{B-iota-bar}) differs from the general
convolution function (\ref{B-l-i-breve}) in the momentum
integration domain. Indeed, now the role of one of the elementary
$\alpha\nu = c0,\,s1$ pseudofermion particle-hole processes (C) is
to generate the $\alpha\nu$ branch line.

By use of the same methods as for the general spectral function
expression (\ref{Ichiun}) in the vicinity of a branch line, in
Appendix B it is found that for $k$ and $\omega$ values such that
$\omega\approx \iota\,v_{\alpha\nu}(k-lk_0)+l\omega_0$ the
spectral-weight distribution expression is given by,

\begin{eqnarray}
B_{{\cal{N}}}^{l} (k,\,\omega) & \approx & \Theta \Bigl(\Omega
-l(\omega-l[\omega_0+\iota\,v_{\alpha\nu}(k-lk_0)])\Bigr)\,\Theta
\Bigl(l(\omega-l[\omega_0+\iota\,v_{\alpha\nu}(k-lk_0)])\Bigr)
\nonumber \\
& \times & C_{\alpha\nu,\,\iota} \Bigl({l(\omega -
l[\omega_0+\iota\,v_{\alpha\nu}(k-lk_0)])\over
4\pi\sqrt{v_{c0}\,v_{s1}}}\Bigr)^{-1+\zeta^0_{\alpha\nu,\,\iota}}
\, ; \hspace{0.5cm} \alpha\nu = c0,\, s1 \, ; \hspace{0.15cm}
\iota =\pm 1 \, ; \hspace{0.50cm} l =\pm 1 \, , \label{Ichiun-bar}
\end{eqnarray}
for $\zeta^0_{\alpha\nu,\,\iota}>0$, where the exponent
$\zeta^0_{\alpha\nu,\,\iota}$ is given in Eq. (\ref{zetasc}) of
Appendix B,

\begin{eqnarray}
C_{\alpha\nu,\,\iota} & = & {\iota 1\over 2\pi\,C_c\,C_s}\Bigl\{
\theta\Bigl(v_{\alpha\nu}-v_{\bar{\alpha}\bar{\nu}}\,
[1-{l(\omega-l[\omega_0+\iota\,v_{\alpha\nu}(k-lk_0)])\over\Omega}]
\Bigr)\int_{-{\iota\,1\over
v_{\bar{\alpha}\bar{\nu}}}}^{{\iota\,1\over
v_{\alpha\nu}}\Bigl(1-{l(\omega-l[\omega_0+\iota\,v_{\alpha\nu}(k-lk_0)])
\over\Omega}\Bigr)}dz\nonumber \\
& + & \Theta\Bigl(v_{\bar{\alpha}\bar{\nu}}\,
[1-{l(\omega-l[\omega_0+\iota\,v_{\alpha\nu}(k-lk_0)])\over\Omega}]-v_{\alpha\nu}
\Bigr)\int_{-{\iota\,1\over
v_{\bar{\alpha}\bar{\nu}}}}^{{\iota\,1\over
v_{\bar{\alpha}\bar{\nu}}}}dz\Bigr\}{F^0_{\alpha\nu,\,\iota}
(z)\over (1
-\iota\,z\,v_{\alpha\nu})^{\zeta^0_{\alpha\nu,\,\iota}}} \, ,
\label{Can-bar}
\end{eqnarray}
and the function $F^0_{\alpha\nu,\,\iota} (z)$ is provided in Eq.
(\ref{C-v-iota}) of that Appendix. This weight-distribution
expression refers to the proximity of a $\alpha\nu =c0,\,s1$
pseudofermion or hole branch-line end point and differs from the
corresponding general spectral function expression given in Eq.
(\ref{Ichiun}). The latter expression applies in the vicinity of
the $\alpha\nu =c0,\,s1$ branch lines when the branch-line group
velocity $v_{\alpha\nu}(q)$ is such that $v_{\alpha\nu}(q)\neq
\iota\, v_{\alpha\nu}$ for $\iota =\pm 1$. As
$v_{\alpha\nu}(q)\rightarrow \iota\, v_{\alpha\nu}$, the spectral
function $k$ and $\omega$ values correspond to the vicinity of a
branch-line end point and thus it is given by expression
(\ref{Ichiun-bar}) instead of (\ref{Ichiun}). Moreover, in the
proximity of the $(lk_0,\,l\omega_0)$ point but for values of $k$
and $\omega$ corresponding to the region where the expression
(\ref{B-PAR-V-L}) is valid, the spectral function is given by that
expression. There are very small cross-over regions between the
$(k,\,\omega)$-plane regions where these different expressions
apply.

\subsection{THE $\delta$-FUNCTION SINGULARITIES}

For some spectral-weight distributions there are isolated
$(k,\,\omega)$-plane points, $(lk_0,\,l\omega_0)$, associated with
point-subspaces such that $\Delta N_{\alpha\nu} =0$ for $\alpha\nu
=c0,\,s1$, $N_{\alpha\nu}=0$ for $\alpha\nu \neq c0,\,s1$, and
$L_{c,\,-1/2}>0$ and/or $L_{s,\,-1/2} > 0$. These point-subspaces
correspond to $(lk_0,\,l\omega_0)$ points such that,

\begin{equation}
k_0 = \pi\, L_{c,\,-1/2} \, ; \hspace{1cm} \omega_0 =
\Bigl[\,2\mu\, L_{c,\,-1/2} + 2\mu_0 H\, L_{s,\,-1/2}\Bigr] \, ,
\hspace{1.0cm} L_{c,\,-1/2}\hspace{0.2cm} {\rm
and/or}\hspace{0.2cm}L_{s,\,-1/2} > 0\, , \label{k0-omega0}
\end{equation}
where the energies $2\mu$ and $2\mu_0 H$ are given in Eq.
(\ref{mu}). Interestingly, the four functionals
$2\Delta_{\alpha\nu}^{\iota}$ of Eq. (\ref{Delta0}) and thus the
value of the functional (\ref{Omega0}) vanish at these points for
all values of $U/t$. In this case the expression of the general
operator ${\hat{\Theta}}_{{\cal{N}},\,k}^l$ defined in Eq. (27) of
Ref. \cite{V} does not include any pseudofermion operators, since
no pseudofermions are created nor annihilated under the
ground-state - excited-energy transition. Moreover, the elementary
processes (C) give no contribution, since when the four
functionals $2\Delta_{\alpha\nu}^{\iota}$ of Eq. (\ref{Delta0})
vanish the energy $\Omega$ which defines the energy range of these
processes also vanishes. In this case the operator
${\hat{\Theta}}_{{\cal{N}},\,k}^l$ is such that in the matrix
element $\langle f;\,CPHS.L\vert
{\hat{\Theta}}_{{\cal{N}},\,k}^{l} \vert GS\rangle$ of Eq. (26) of
Ref. \cite{V} the energy eigenstate $\vert f;\,CPHS.L\rangle$ is
the ground state itself. Thus, for such point subspaces the matrix
elements derived in Sec. V of Ref. \cite{V} must be replaced by
$\langle GS\vert {\hat{\Theta}}_{{\cal{N}},\,k}^l\vert GS\rangle$
and the spectral function has at the corresponding
$(k,\,\omega)$-plane points the following $\delta$-function
behavior,

\begin{equation}
B_{{\cal{N}}}^{l} (k,\,\omega) = {\vert\langle GS\vert
{\hat{\Theta}}_{{\cal{N}},\,k}^l\vert GS\rangle\vert^2\over
C_c\,C_s}\, \delta \Bigl(\omega -
l\omega_0\Bigr)\,\delta_{k,\,lk_0} \, ; \hspace{0.50cm} l =\pm 1
\, . \label{Ichiun-U-0}
\end{equation}

The $\delta$-peak singularities of Eq. (\ref{Ichiun-U-0})
correspond to isolated $(lk_0,\,l\omega_0)$ points and occur for
all finite values of $U/t$. Another example of $\delta$-peak
singularities occurs for the one-electron spectral-weight
distributions. Such a behavior occurs for some $c0$ or $s1$ branch
lines (\ref{*line}), in whose vicinity the spectral function has
the form (\ref{Ichiun}) for $U/t>0$, and for $U/t\rightarrow 0$
the value of the functional (\ref{Omega0}) behaves as $\beth_0
(k,\,\omega)=\beth_0 (k,\,\omega_{\alpha\nu} (q))\rightarrow 0$.
Also the overall phase shifts $Q_{\alpha\nu}(\iota
q^0_{F\alpha\nu})/2$ behave as $Q_{\alpha\nu}(\iota
q^0_{F\alpha\nu})/2\rightarrow 0$ for $\alpha\nu = c0$ or
$\alpha\nu =s1$ and $\iota =\pm 1$. In this case the
spectral-function expressions derived in this paper remain valid,
provided that one considers in these expressions the limit
$2\Delta_{\alpha\nu}^{\iota}\rightarrow 0$ for the four
functionals given in Eq. (\ref{Delta0}). The energy $\Omega$ which
defines the energy range of the elementary processes (C) behaves
in this case as $\Omega\rightarrow 0$ for $\beth_0
(k,\,\omega)\rightarrow 0$ and, therefore, the weight distribution
becomes $\delta$-function like. The function (\ref{fan}) is such
that $f_{\alpha\nu,\,\iota} = \sqrt{f(-\pi)}\rightarrow 1$ as
$Q_{\alpha\nu}(\iota q^0_{F\alpha\nu})\rightarrow 0$. Indeed, $f
(Q)=f(-Q)$ is the function on the right-hand side of Eq. (24) of
Ref. \cite{Penc97} such that $f (\pi)=f(-\pi)=1$. Moreover,
$S^0_{c0}\,S^0_{s1}\rightarrow 1$ as $U/t\rightarrow 0$ and thus
the $\delta$-peak weight $2\pi(1/N_a)^{\zeta_0}D_0$ of Eq.
(\ref{B-l-i-breve-int-0}) is such that
$(1/N_a)^{\zeta_0}D_0\rightarrow 2\pi$ as $U/t\rightarrow 0$, once
for the above branch lines also $\zeta_0\rightarrow 0$. It follows
that in this case the convolution function (\ref{B-l-i-breve}) has
not the form given in Eq. (\ref{B-breve-asym}). Instead, as
$U/t\rightarrow 0$ it is given by the first expression of Eq.
(\ref{B-l-i-breve-int-0}) and the asymptotic expression
(\ref{Ichiun}) is not valid for $U/t\rightarrow 0$. By performing
calculations similar to those of Appendix B but with the
convolution function given by the first expression of Eq.
(\ref{B-l-i-breve-int-0}), instead of its asymptotic expression
(\ref{B-breve-asym}), one arrives to the following expression for
the spectral-weight distribution at the branch line,

\begin{eqnarray}
B_{{\cal{N}}}^{l} (k,\,\omega) & = & {D_0\over C_c\,C_s}\,\int
dq'\,\left({1\over N_a}\right)^{\zeta_0 (q')}\,\delta
\Bigl(\omega-l\omega_{\alpha\nu}(q')\Bigr)\,\delta
\Bigl(k-lk_{\alpha\nu}(q')\Bigr)
\nonumber \\
& = & {1\over C_c\,C_s}\,\delta \Bigl(\omega - l\omega_{\alpha\nu}
(q)\Bigr) \, ; \hspace{0.25cm} \zeta_0 (q') \rightarrow 0 \, ;
\hspace{0.50cm} l =\pm 1 \, , \label{Ichiun-0}
\end{eqnarray}
where the functions $k_{\alpha\nu}(q)$ and $\omega_{\alpha\nu}(q)$
are defined in Eq. (\ref{*line}) and the $q'$ integration is over
the branch line points such that $\zeta_0 (q') \rightarrow 0$.

For two-electron spectral-weight distributions a similar behavior
occurs for the weight generated by two active-scattering-center
creation away from the $\alpha\nu =c0,\,s1$ {\it Fermi points} and
$\alpha\nu\neq c0,\,s1$ limiting bare-momentum values in
two-dimensional regions of the $(k,\,\omega)$-plane where $\beth_0
(k,\,\omega)\rightarrow 0$. Again, for such regions the
convolution function (\ref{B-l-i-breve}) has not the form given in
Eq. (\ref{B-breve-asym}) but that corresponding to the first
expression of Eq. (\ref{B-l-i-breve-int-0}). Thus, as $\beth_0
(k,\,\omega)\rightarrow 0$ the expression (\ref{B-2D-inte}) is not
valid and instead the spectral-weight distribution has the
following behavior,

\begin{eqnarray}
B_{{\cal{N}}}^{l} (k,\,\omega) & = & {D_0\over
2\pi\,C_c\,C_s}\,\int dq\,\int dq'\,\left({1\over
N_a}\right)^{\zeta_0 (q,q')}\,\delta
\Bigl(\omega-l\omega_{\alpha\nu,\,\alpha'\nu'}(q,\,q')\Bigr)\,\delta
\Bigl(k-lk_{\alpha\nu,\,\alpha'\nu'}(q,\,q')\Bigr)
\nonumber \\
& = & {1\over 2\pi\,C_c\,C_s}\,\int dq'\,\delta
\Bigl(\omega-l\omega_{\alpha\nu,\,\alpha'\nu'}(q
(k,\,q'),\,q')\Bigr) \, ; \hspace{0.25cm} \zeta_0 (q,\,q')
\rightarrow 0  \, ; \hspace{0.50cm} l =\pm 1 \, . \label{B-zeta-0}
\end{eqnarray}
Here the functions $k_{\alpha\nu,\,\alpha'\nu'}(q,\,q')$ and
$\omega_{\alpha\nu,\,\alpha'\nu'}(q,\,q')$ are defined in Eq.
(\ref{*plane}), the bare-momentum integrations correspond to the
two-dimensional regions of the $(k,\,\omega)$-plane where $\zeta_0
(q,\,q') \rightarrow 0$, and the function $q=q (k,\,q')$ is
uniquely defined by the parametric equation
$k=lk_{\alpha\nu,\,\alpha'\nu'}(q,\,q')$.

An example of the type of singularity (\ref{Ichiun-U-0}) is the
$\delta$-peak which appears in the spin-singlet Cooper pair $l=1$
addition spectral function at the point
$(k_0,\omega_0)=(\pi,2\mu)$, as a result of the $\eta$-pairing
mechanism \cite{Yang89}. One-electron spectral-function
singularities of the type (\ref{Ichiun-0}) are studied in Ref.
\cite{spectral}. It is found in that reference that some of the
$c0$ and $s1$ pseudofermion branch lines behave as in Eq.
(\ref{Ichiun-0}) for $U/t\rightarrow 0$ with $C_c=C_s=1$, where
the correct $U/t=0$ non-interacting one-electron spectrum is
recovered. The two-electron weight distribution described by
expression (\ref{B-zeta-0}) is valid when $U/t\rightarrow 0$ with
$C_c=C_s=1$. Interestingly, for the dynamical structure factor it
is also valid when $U/t\rightarrow\infty$, where such a spectral
function is that of the spin-less fermions which describe the
charge degrees of freedom in that limit. There is a qualitative
difference between the spectral-function behavior described by Eq.
(\ref{Ichiun-U-0}) and that described by Eqs. (\ref{Ichiun-0}) and
(\ref{B-zeta-0}): the former behavior is valid at isolated points
for finite value of $U/t$, whereas the latter behavior is valid
for $U/t\rightarrow 0$ or $U/t\rightarrow\infty$ only. Moreover,
in the case of expressions (\ref{Ichiun-0}) and (\ref{B-zeta-0}),
as $\zeta_0\rightarrow 0$ all the spectral weight becomes located
onto the $(k,\,\omega)$-plane regions where $\beth_0
(k,\,\omega)\rightarrow 0$.

Finally, for the $n=1$ Mott-Hubbard insulator phase there are
other approaches to the study of finite-energy spectral-weight
distributions for finite values of $U/t$. A method for the
description of the weight distribution in the vicinity of specific
spectral features was proposed in Ref. \cite{Sorella}. Also for
$n=1$ but for very small values of $U/t$ the 1D Hubbard model can
be mapped onto the sine-Gordon model. In this limit the form
factor approach was used in the study of the zero-momentum
two-electron charge spectral-weight distribution associated with
the frequency-dependent optical conductivity \cite{Controzzi}. In
general, the form factor approach does not apply to the 1D Hubbard
model.

\section{DISCUSSION AND CONCLUDING REMARKS}
\label{SecVI}

In this paper we derived general closed-form analytical
expressions for the finite-energy one- and two-electron
spectral-weight distributions of a 1D correlated metal with
on-site electronic repulsion. Our results also provide general
expressions for the one- and two-atom spectral functions of a
correlated quantum system of cold fermionic atoms in a 1D optical
lattice with on-site atomic repulsion. In order to solve that
problem, we addressed the issue of the dominant spectral-weight
processes. That allowed the derivation of the basic
weight-distribution pieces corresponding to surface-like,
line-like, and point-like spectral features, which involved the
introduction of the concepts of a border line and a singular and
edge branch line. Our study reveals that the functional
(\ref{Omega0}) plays a key role in the control of the
spectral-weight distribution. Indeed, most singularities are of
power-law type and correspond to discontinuities of the value such
a functional in the $(k,\,\omega)$-plane. Moreover, that
functional also controls the exceptional occurrence of
$\delta$-function singularities, which correspond to the
$(k,\,\omega)$-plane points where its value vanishes.

For the one- and two-electron problems considered in this paper,
the spectral-function expressions generated by processes
associated with excited J-CPHS subspaces with none, one, and two
active scattering centers away from the $\alpha\nu =c0,\,s1$ {\it
Fermi points} and $\alpha\nu\neq c0,\,s1$ limiting bare-momentum
values describe nearly the whole $(k,\,\omega)$-plane weight
distribution. In addition to other spectral features, here we
presented a detailed derivation of the spectral-weight
distribution expressions used in the studies of Refs.
\cite{super,blinde}. These expressions are used in Ref.
\cite{blinde} in the study of both the spectral-weight
distributions observed for the TTF and TCNQ stacks of molecules in
the quasi-1D organic compound TTF-TCNQ, following the preliminary
predictions of Refs. \cite{spectral0,spectral1,spectral} for the
TCNQ spectral dispersions. Interestingly, the studies of Ref.
\cite{blinde} find quantitative agreement with the observed
spectral features for the whole experimental energy band width.
Recently, the dynamical density matrix renormalization group
method was used in the study of the one-electron spectral function
of the 1D Hubbard model \cite{Eric} for electronic densities
$n<1$. These numerical studies found spectral features consistent
with those of our spectral-weight distribution expressions.
Moreover, the results of Ref. \cite{super} are consistent with the
phase diagram observed in the (TMTTF)$_2$X and (TMTSF)$_2$X series
of organic compounds and explain the absence of superconductive
phases in TTF-TCNQ. Other applications of our finite-energy
spectral-weight-distribution expressions to several materials,
correlated quantum systems, and spectral functions are in
progress. This includes the use of our theoretical results in the
two-atom spectral-weight distributions measured in 1D optical
lattices \cite{NEW}. Finally, it is shown elsewhere \cite{CFT}
that the general spectral-weight distribution expressions
introduced in this paper reproduce the low-energy behavior
obtained by conformal-field theory \cite{Frahm}.

\begin{acknowledgments}
We thank Natan Andrei, Yong Chen, Ralph Claessen, Francisco (Paco)
Guinea, Vladimir E. Korepin, Patrick A. Lee, Sung-Sik Lee, Jo\~ao
M. B. Lopes dos Santos, Lu\'{\i}s M. Martelo, Tiago C. Ribeiro,
Pedro D. Sacramento, and Xiao-Gang Wen for stimulating
discussions. J.M.P.C. and D.B. thank the hospitality and support
of MIT and the financial support of the Gulbenkian Foundation and
FCT under the grant POCTI/FIS/58133/2004, J.M.P.C. thanks the
support of the Fulbright Commission, K.P. thanks the support of
the OTKA grants T038162 and T049607, and D.B. thanks the
hospitality of the Research Institute for Solid State Physics and
Optics in Budapest and the financial support of the Portuguese FCT
grant SFRH/BD/6930/2001.
\end{acknowledgments}
\appendix

\section{GROUND-STATE RAPIDITY FUNCTIONS}

In this Appendix we introduce simple expressions for the inverse
of the ground-state rapidity functions $k^{0}(q)$,
$\Lambda^{0}_{c\nu}(q)$, and $\Lambda^{0}_{s\nu}(q)$ that appear
in the pseudofermion energy band expressions
(\ref{epc})-(\ref{epsn}). These expressions are derived by
solution of the integral equations obtained by introducing in Eqs.
(13)-(16) of Ref. \cite{I} the ground-state distribution functions
given in Eqs. (C1)-(C3) of the same reference. The solution of
these equations can be written in terms of the inverse functions
of $k^0 (q)$ and $\Lambda_{\alpha\nu}^0 (q)$ for $\nu >0$, which
we call $q^0_{c} (k)$ and $q^0_{\alpha\nu} (\Lambda)$,
respectively. Here $k$ and $\Lambda$ are the rapidity-momentum
coordinate and rapidity coordinate, respectively. These are odd
functions such that $q^0_{c} (k) = - q^0_{c} (-k)$ and
$q^0_{\alpha\nu} (\Lambda) = -q^0_{\alpha\nu} (-\Lambda)$. The
domains of the rapidity-momentum coordinate $k$ and rapidity
coordinate $\Lambda$ are such that $-\pi\leq k\leq +\pi$ and
$-\infty \leq \Lambda\leq \infty$, respectively. It follows that
$k^0 (\pm\pi )=\pm\pi$, $\Lambda^0_{c0}(\pm\pi ) = 0$, and
$\Lambda^0_{\alpha\nu}(\pm q^0_{\alpha\nu})=\pm\infty$ for
$\alpha\nu\neq c0$. Such relations are equivalent to $q^0_{c}
(\pm\pi )=\pm\pi$, $q^0_{s1} (\pm\infty ) = \pm k_{F\uparrow}$,
$q^0_{c\nu} (\pm\infty ) = \pm [\pi - 2k_F]$ for $c\nu\neq c0$,
and $q^0_{s\nu} (\pm\infty ) = \pm [k_{F\uparrow}
-k_{F\downarrow}]$ for $s\nu\neq s1$. By suitable manipulation of
Eqs. (13)-(16) and (C1)-(C3) of Ref. \cite{I} we find,

\begin{equation}
q^0_c (k) = k + \int_{-Q}^{+Q}dk'\,\widetilde{\Phi }_{c0,\,c0}
\left(k',k\right) \, ; \hspace{0.5cm} q^0_{s\nu} (\Lambda ) =
\int_{-Q}^{+Q}dk'\, \widetilde{\Phi }_{c0,\,s\nu}
\left(k',\Lambda\right)  \, ; \hspace{0.5cm} \nu =1,2, ... \, ,
\label{kcGS}
\end{equation}

\begin{equation}
q^0_{c\nu} (\Lambda ) = 2\,{\rm Re}\,\{\arcsin \Bigl(\Lambda + i
\nu U/4t\Bigr)\} -\int_{-Q}^{+Q}dk'\,\widetilde{\Phi }_{c0,\,c\nu}
\left(k',\Lambda\right)  \, ; \hspace{0.5cm} \nu =1,2, ... \, ,
\label{GcnGS}
\end{equation}
where the two-pseudofermion phase shifts $\tilde{\Phi}$ are such
that,

\begin{equation}
\tilde{\Phi}_{c0,\,c0}(k,k') = \bar{\Phi
}_{c0,\,c0}\left({4t\,\sin k\over U}, {4t\,\sin k'\over U}\right)
\, ; \hspace{1cm} \tilde{\Phi}_{c0,\,\alpha\nu}(k,\Lambda') =
\bar{\Phi }_{c0,\,\alpha\nu}\left({4t\,\sin k\over U},
{4t\,\Lambda'\over U}\right) \, , \label{tilPcc}
\end{equation}

\begin{equation}
\tilde{\Phi}_{\alpha\nu,\,c0}(\Lambda,k') = \bar{\Phi
}_{\alpha\nu,\,c0}\left({4t\,\Lambda\over U}, {4t\,\sin k'\over
U}\right) \, ; \hspace{1cm}
\tilde{\Phi}_{\alpha\nu,\,\alpha\nu'}\left(\Lambda,\Lambda'\right)
= \bar{\Phi }_{\alpha\nu,\,\alpha\nu'}\left({4t\,\Lambda\over U}
,{4t\,\Lambda'\over U}\right) \, , \label{tilPanan}
\end{equation}
and the phase shifts $\bar{\Phi }_{\alpha\nu,\,\alpha'\nu'}
(r,r')$ are defined by the integral equations (A1)-(A13) of Ref.
\cite{IIIb}. Moreover, the parameters $Q \equiv k^{0}(2k_F)$ and
$B \equiv \Lambda^{0}_{s1}(k_{F\downarrow})$ are such that
$q^0_{c} (\pm Q) = \pm 2k_F$ and $q^0_{s1} (\pm B) = \pm
k_{F\downarrow}$ and are self-consistently defined by the solution
of the following equations,

\begin{equation}
2k_F = Q + \int_{-Q}^{+Q}dk\,\widetilde{\Phi }_{c0,\,c0}
\left(k,Q\right) \, ; \hspace{0.5cm} k_{F\downarrow} =
\int_{-Q}^{+Q}dk\, \widetilde{\Phi }_{c0,\,s1} \left(k,B\right) \,
. \label{Beq}
\end{equation}

\section{SPECTRAL-WEIGHT DISTRIBUTIONS AND RELATED QUANTITIES}

In this Appendix we provide complementary information about the
evaluation of the finite-energy spectral-weight distribution
expressions studied in this paper. We start by expressing the
generators appearing in the matrix elements given in Eqs. (59) and
(60) of Ref. \cite{V} in terms of the sets of canonical-momentum
and bare-momentum values of the corresponding distribution
function deviations provided in Eqs.
(\ref{DN-gen})-(\ref{DN-NF-an}). For the generators associated
with $\alpha\nu = c0,\,s1$ pseudofermion occupancy configurations
this leads to,

\begin{equation}
F_{J-NF,\,\alpha\nu}^{\dag} =
\prod_{j_=1}^{N^{phNF}_{\alpha\nu}}\Bigl[
f^{\dag}_{{\bar{q}}_j,\,\alpha\nu} \,f_{{\bar{q}'}_j,\,\alpha\nu}
\Bigr] \,\prod_{i=1}^{\vert \Delta
N^{NF}_{\alpha\nu}\vert}\Bigl[\Theta \Bigl({\rm sgn}(\Delta
N^{NF}_{\alpha\nu})\Bigr)f^{\dag}_{{\bar{q}}_{i},\,\alpha\nu}
+\Theta \Bigl(-{\rm sgn}(\Delta
N^{NF}_{\alpha\nu})\Bigr)f_{{\bar{q}}_{i},\,\alpha\nu}\Bigr] \, ,
\label{J-NF}
\end{equation}

\begin{equation}
F_{-GS,\,\alpha\nu}^{\dag} = \prod_{\iota =\pm
1}F_{-GS,\,\alpha\nu,\,\iota}^{\dag} \, ; \hspace{0.5cm}
F_{-GS,\,\alpha\nu,\,\iota}^{\dag} = \delta_{\iota,\,+1}
[\prod_{q_j=0}^{q_{F\alpha\nu}^0+\Delta
q_{F\alpha\nu,\,+1}}\,f^{\dag }_{q_j,\,\alpha\nu}] +
\delta_{\iota,\,-1}[\prod_{q_j=-q_{F\alpha\nu}^0+\Delta
q_{F\alpha\nu,\,-1}}^{0}\,f^{\dag }_{q_j,\,\alpha\nu}] \, ,
\label{GS-SD-0}
\end{equation}

\begin{equation}
F_{J-GS,\,\alpha\nu}^{\dag}  = \prod_{\iota =\pm 1}
F_{J-GS,\,\alpha\nu,\,\iota}^{\dag} \, ; \hspace{0.5cm}
F_{J-GS,\,\alpha\nu,\,\iota}^{\dag} =
\delta_{\iota,\,+1}[\prod_{{\bar{q}}_j=0}^{q_{F\alpha\nu}^0+\Delta
{\bar{q}}_{F\alpha\nu,\,+1}}\,f^{\dag }_{{\bar{q}_j},\,\alpha\nu}]
+\delta_{\iota,\,-1}[\prod_{{\bar{q}}_j=-q_{F\alpha\nu}^0 + \Delta
{\bar{q}}_{F\alpha\nu,\,-1}}^{0}\,f^{\dag
}_{{\bar{q}_j},\,\alpha\nu}] \, , \label{GS-SD}
\end{equation}

\begin{equation}
F_{p-h,\,\alpha\nu}^{\dag} = \sum_{\iota =\pm
1}F_{p-h,\,\alpha\nu,\,\iota}^{\dag} \, ; \hspace{1cm}
F_{p-h,\,\alpha\nu,\,\iota}^{\dag} =
\prod_{j_{\iota}=1}^{N^{phF}_{\alpha\nu,\,\iota}}\Bigl[
f^{\dag}_{{\bar{q}}_{j_{\iota}},\,\alpha\nu}
\,f_{{\bar{q}'}_{j_{\iota}},\,\alpha\nu} \Bigr] \, ;
\hspace{0.5cm} \iota = \pm 1 \, . \label{Fph}
\end{equation}
For the generator corresponding to $\alpha\nu\neq c0,\,s1$
pseudofermion occupancy configurations we find,

\begin{equation}
F_{NF,\,\alpha\nu}^{\dag} =
\prod_{i=1}^{N_{\alpha\nu}^{NF}}f^{\dag}_{{\bar{q}'}_{i},\,\alpha\nu}
\, ; \hspace{0.5cm} \alpha\nu\neq c0,\,s1 \, . \label{F-non-c0-s1}
\end{equation}
The operators (\ref{J-NF}) and (\ref{F-non-c0-s1}) generate the
elementary processes (A), whereas the operator (\ref{Fph})
generates the elementary processes (C). The operators
$F_{-GS,\,\alpha\nu}^{\dag}$ and $F_{J-GS,\,\alpha\nu}^{\dag}$
create the whole $\alpha\nu = c0,\,s1$ pseudofermion {\it Fermi
sea}, including the {\it Fermi point} shifts generated by the
elementary processes (B), for the discrete canonical-momentum
values of the ground state and reduced-subspace excited state,
respectively. In the second expression of both Eqs.
(\ref{GS-SD-0}) and (\ref{GS-SD}) the products refer to the
$\alpha\nu = c0,\,s1$ pseudofermions belonging to the $\iota ={\it
sgn} (q)\,1$ sub-branch and $\Delta q_{F\alpha\nu ,\,\iota}$ and
$\Delta {\bar{q}}_{F\alpha\nu ,\,\iota}$ are the bare-momentum and
canonical-momentum shifts provided in Eqs. (\ref{qiFan}) and
(\ref{barqanF}), respectively.

The weight of the peak corresponding to the $N^{ph}_{\alpha\nu}$
$\alpha\nu =c0,\,s1$ $\alpha\nu$ pseudofermion particle-hole
excitation mentioned in Sec. IV reads,

\begin{equation}
A^{[N^{ph}_{\alpha\nu}]}_{\alpha\nu}=
A^{(0,\,0)}_{\alpha\nu}\,a^{[N^{ph}_{\alpha\nu}]}_{\alpha\nu} \, ;
\hspace{0.5cm}
a^{[N^{ph}_{\alpha\nu}]}_{\alpha\nu}=a^{[N^{ph}_{\alpha\nu}]}_{\alpha\nu}
(q_{1},\cdots,q_{N^{ph}_{\alpha\nu}},
{q'}_{1},\cdots,{q'}_{N^{ph}_{\alpha\nu}}) \, ; \hspace{0.25cm}
\alpha\nu = c0,\,s1 \, . \label{aNN}
\end{equation}
Here $A^{(0,\,0)}_{\alpha\nu}$ is the lowest-peak weight given in
Eq. (\ref{A00}), $a^{[N^{ph}_{\alpha\nu}]}_{\alpha\nu}$ is the
relative weight, and $q_{1},\cdots,q_{N^{ph}_{\alpha\nu}}$ and
${q'}_{1},\cdots,{q'}_{N^{ph}_{\alpha\nu,\,\iota}}$ are the
corresponding set of $\alpha\nu =c0,\,s1$ pseudofermion "particle"
and "hole" bare-momentum values, respectively. By use of the
method of Ref. \cite{Penc97}, we find that for
$N^{ph}_{\alpha\nu}=1$ the relative weight (\ref{aNN}) is given
by,

\begin{eqnarray}
a^{[1]}_{\alpha\nu} (q_{1},\,{q'}_{1}) = \frac{ g(q_{1}) }{g
({q'}_{1})} \frac{1}{\sin^2(\frac{[q_{1}-{q'}_{1}]a}{2})} \, ;
\hspace{0.5cm} \alpha\nu = c0,\,s1 \, . \label{a10}
\end{eqnarray}
The function $g(q)$ appearing on the right-hand side of Eq.
(\ref{a10}) reads,

\begin{equation}
g(q_j) = \prod_{{\bar{q}'}_j={\bar{q}}_{F\alpha\nu ,\,-1}}^{{\bar{q}}_{F\alpha\nu ,\,+1}}
\sin^{2} \left(\frac{[{\bar{q}}_j -{\bar{q}'}_j]
a}{2}\right)\,\prod_{{q''}_j=q_{F\alpha\nu ,\,-1}}^{q_{F\alpha\nu ,\,+1}}\,
{1\over\sin^{2}\left(\frac{[{\bar{q}}_j-{q''}_j] a}{2}\right)} \, ; \hspace{0.5cm}
\alpha\nu = c0,\,s1 \, , \label{g}
\end{equation}
where the correspondence between  $q_j$ and ${\bar{q}}_j$ is defined by the equation
${\bar{q}}_j = q_j + Q^{\Phi}_{\alpha\nu} (q_j)/L$ with $Q^{\Phi}_{\alpha\nu} (q_j)/2$
given in Eq. (\ref{qcan1j}), $q_{F\alpha\nu ,\,\pm 1}$ and ${\bar{q}}_{F\alpha\nu ,\,\pm
1}$ are the {\it Fermi points} provided in Eqs. (\ref{qiFan}) and (\ref{barqanF}),
respectively, and the corresponding products run over the discrete ground-state
bare-momentum and excited-state canonical-momentum values, except for
${\bar{q}'}_j={\bar{q}}_j$ and ${q''}_j=q_j$. Similarly, for for $N^{ph}_{\alpha\nu}=2$
we find,

\begin{equation}
a^{[2]}_{\alpha\nu} (q_{1},\,q_{2},\,{q'}_{1},\,{q'}_{2}) = \frac{
g(q_{1}) g(q_{2}) } { g({q'}_{1}) g({q'}_{2})}\, \frac{
\sin^2(\frac{[{q'}_{1} \!-\! {q'}_{2}]a}{2}) \sin^2(\frac{[q_{1}
\!-\! q_{2}]a}{2})} { \sin^2(\frac{[q_{1} \!-\! {q'}_{1}]a}{2})
\sin^2(\frac{[q_{1} \!-\! {q'}_{2}]a}{2}) \sin^2(\frac{[q_{2}
\!-\! {q'}_{1}]a}{2}) \sin^2(\frac{[q_{2} \!-\! {q'}_{2}]a}{2})}
\, , \label{a20}
\end{equation}
where $\alpha\nu = c0,\,s1$. The expression of the function
(\ref{aNN}) is similar for increasing number of $\alpha\nu
=c0,\,s1$ pseudofermion particle-hole processes, but since it is
longer, for simplicity we do not give it here.

We proceed by deriving the spectral-function expression
(\ref{B-2D-inte}) associated with creation of two active
scattering centers away from the $\alpha\nu =c0,\,s1$ {\it Fermi
points} and $\alpha\nu\neq c0,\,s1$ limiting bare-momentum values
by the elementary processes (A). For simplicity, we consider that
the two created active pseudofermion and/or hole scattering
centers belong to the $\alpha\nu = c0,\,s1$ branches.
Generalization to active scattering centers associated with
$\alpha\nu\neq c0,\,s1$ branches is straightforward. In order to
arrive to the specific form of the general $i=0$ spectral function
expression (\ref{B-PAR-J-CPHS-sum-0}) for the excitations
associated with the latter processes, we start by considering
excited energy eigenstates generated by the elementary processes
(A) whose energy $\Delta E$ and momentum $\Delta P$ relative to
the initial ground state have values such that $[\omega-l\Delta
E]$ and $[k-l\Delta P]$ are small and $0<[\omega-l\Delta
E]<\Omega$ and $-1/v_{s1} < 1/v <+1/v_{s1}$. Here $k$ and $\omega$
are the momentum and energy values provided in Eq. (\ref{*plane}),
$v$ is the velocity defined in Eq. (\ref{V-ok}), and the
corresponding energy $\Delta E=\Delta E (q+\Delta q,\,q'+\Delta
q')$ and momentum $\Delta P = \Delta P (q+\Delta q,\,q'+\Delta
q')$ are the general functionals given in Eqs. (\ref{DE-T}) and
(\ref{DP-T}), respectively, which for the present case read,

\begin{equation}
\Delta P = [k_0 + c_1\,(q+\Delta q)+ c_1'\,(q'+\Delta q')] \, ;
\hspace{0.5cm}  \Delta E = [\omega_0 +c_1\,\epsilon_{\alpha\nu}
(q+\Delta q)+c_1'\,\epsilon_{\alpha'\nu'} (q'+\Delta q')] \, ,
\label{*plane-d}
\end{equation}
where the bare-momentum deviations $\Delta q$ and $\Delta q'$
relative to the bare momentum values of Eq. (\ref{*plane}) are
small.

The excited energy eigenstates which contribute to the
weight-distribution expression (\ref{B-2D-inte}) have momentum $k$
and energy $\omega$ relative to the initial ground state. Such
states are generated by the elementary processes (C) from the
states whose momentum and energy spectra are given in Eq.
(\ref{*plane-d}). In order to reach the momentum $k$ and energy
$\omega$ given in Eq. (\ref{*plane}), the elementary processes (C)
generate excitations whose momentum and energy are given by
$[k-l\Delta P]$ and $[\omega-l\Delta E]$, respectively. From use
of Eqs. (\ref{*plane}) and (\ref{*plane-d}) we find,

\begin{equation}
[k-l\Delta P] = -l[c_1\,\Delta q+ c_1'\,\Delta q'] \, ;
\hspace{0.5cm} [\omega-l\Delta E] = -l[c_1\,v_{\alpha\nu}
(q)\,\Delta q+c_1'\,v_{\alpha'\nu'} (q')\,\Delta q'] \, ,
\label{*plane+delta}
\end{equation}
where $\alpha\nu = c0,\,s1$ and $\alpha'\nu' = c0,\,s1$. Use of
these expressions leads to,

\begin{equation}
\Delta q = -l\,[\omega-l\Delta E]\,{c_1[1-v_{\alpha'\nu'}
(q')\,z]\over [\,v_{\alpha\nu} (q)-v_{\alpha'\nu'} (q')]} \, ;
\hspace{1cm} \Delta q' = l\,[\omega-l\Delta
E]\,{c_1'[1-v_{\alpha\nu} (q)\,z]\over [\,v_{\alpha\nu}
(q)-v_{\alpha'\nu'} (q')]} \, ; \hspace{0.50cm} z\equiv 1/v \, .
\label{dq-dq'}
\end{equation}
For small but finite values of $[\omega-l\Delta E]$ we can use the
asymptotic expression (\ref{B-breve-asym}) in the general $i=0$
spectral function expression (\ref{B-PAR-J-CPHS-sum-0}), which
corresponds to the dominant contribution to the spectral function
(\ref{ABONjl-J-CPHS}) for $i=0$. We thus find,

\begin{eqnarray}
B_{{\cal{N}}}^{l} (k,\,\omega) & \approx & {1\over
4\pi^2\,C_c\,C_s}\,\int dq''\,\int dq'''\,\Theta
\Bigl(\Omega-l[\omega-l\Delta E]\Bigr)\,\Theta
\Bigl(l[\omega-l\Delta E]\Bigr)\nonumber \\
& \times & {F_0 ([k-l\Delta P]/[\omega-l\Delta E])\over
4\pi\sqrt{v_{c0}\,v_{s1}}}\Bigl({l[\omega-l\Delta E]\over
4\pi\sqrt{v_{c0}\,v_{s1}}} \Bigr)^{-2+\zeta_0 (q,\,q')} \, ;
\hspace{0.5cm} l= \pm 1 \, . \label{B-2D}
\end{eqnarray}
Here the $q''=q+\Delta q$ and $q'''=q'+\Delta q'$ integration
domains correspond to the excitation domains such that
$0<[\omega-l\Delta E]<\Omega$ and $-1/v_{s1} < 1/v <+1/v_{s1}$,
$\zeta_0=\zeta_0 (q,\,q')$ is the functional (\ref{zeta0}), and
$F_0 (z)$ is given in Eq. (\ref{C-v}). By use of Eq.
(\ref{dq-dq'}), we perform a transformation from $q''$ and $q'''$
to the integration variables $\omega'=[\omega-l\Delta E]$ and $v$,
what leads to,

\begin{equation}
B_{{\cal{N}}}^{l} (k,\,\omega) \approx {l\over
4\pi^2\,C_c\,C_s\,\vert\,v_{\alpha\nu} (q)-v_{\alpha'\nu'}
(q')\vert}\,\int_0^{l\Omega}
d\omega'\,\int_{-1/v_{s1}}^{+1/v_{s1}} dz\,F_0
(z)\Bigl({l\omega'\over 4\pi\sqrt{v_{c0}\,v_{s1}}}
\Bigr)^{-1+\zeta_0 (q,\,q')} \, ; \hspace{0.5cm} l= \pm 1 \, .
\label{B-2D-omz}
\end{equation}
Finally, by performing the $\omega'$ integration we reach the
general expression (\ref{B-2D-inte}).

Our next goal is the derivation of the branch-line expression
defined by Eqs. (\ref{Ichiun}) and (\ref{Can}). For simplicity,
let us consider a $\alpha\nu =c0,\,s1$ branch line. The
expressions correspond to $k= l[k_0 +c_1\,q]$ and $\omega\approx
l[\omega_0 +c_1\,\epsilon_{\alpha\nu} (q)]$ with the
$(k,\,\omega)$-plane point being located in the vicinity and just
above $(l=+1)$ or below $(l=-1)$ the branch line. In this case the
initial state integrations of Eq. (\ref{B-PAR-J-CPHS-sum-0})
correspond to points belonging to a branch-line segment associated
with $q'=q+\Delta q$ values such that $\Delta q$ is small and has
both negative and positive values. The elementary processes (C)
generate excited states of momentum $k=(\Delta P+k')$ and energy
$\omega=(\Delta E+\omega')$ relative to the initial ground state.
The general energy spectrum $\Delta E$ and momentum spectrum
$\Delta P$ of the corresponding states of the reduced subspace is
given in Eqs. (\ref{DE-T}) and (\ref{DP-T}), respectively. In the
present case these spectra are functions of the bare momentum
$q'=q+\Delta q$ only and read,

\begin{equation}
\Delta E (q+\Delta q) = \omega_0 +c_1\,\epsilon_{\alpha\nu} (q) +
\Delta q\,c_1\,v_{\alpha\nu} (q) \, ; \hspace{1cm} \Delta P
(q+\Delta q) = k_0 +c_1\,q + c_1\,\Delta q\, . \label{spectra-bl}
\end{equation}
The velocity (\ref{V-ok}) is then given by,

\begin{equation}
v = v_{\alpha\nu} (q) - c_1\,{l(\omega - l[\omega_0
+c_1\,\epsilon_{\alpha\nu} (q)])\over \Delta q} \, ; \hspace{1cm}
\Delta q = -c_1\,{l(\omega - l[\omega_0 +c_1\,\epsilon_{\alpha\nu}
(q)])\over v-v_{\alpha\nu} (q)} \, , \label{v-bl}
\end{equation}
where we also provided the corresponding value of $\Delta q$ as a
function of $v$.

For ${\rm sgn} (v)=-{\rm sgn} (q)$ one has that $v\in
(-\infty,\,-v_{s1})$ if ${\rm sgn} (q)\,1=+1$ and $v\in
(v_{s1},\,\infty)$ when ${\rm sgn} (q)\,1=-1$. By use of Eq.
(\ref{v-bl}) we then find that the limiting values $-{\rm sgn}
(q)\,\infty$ and $-{\rm sgn} (q)\,v_{s1}$ correspond to $\Delta
q=0$ and $\Delta q={\rm sgn} (q)\,c_1\,l(\omega - l[\omega_0
+c_1\,\epsilon_{\alpha\nu} (q)])/[v_{s1}+\vert v_{\alpha\nu}
(q)\vert]$, respectively. In turn, for ${\rm sgn} (v)={\rm sgn}
(q)$ there are two cases. When $\vert v_{\alpha\nu} (q)\vert\leq
v_{s1}\,[1-l(\omega - l[\omega_0 +c_1\,\epsilon_{\alpha\nu}
(q)])/\Omega]$ we find that $v\in (-\infty,\,-v_{s1})$ for ${\rm
sgn} (q)\,1=-1$ and $v\in (v_{s1},\,\infty)$ for ${\rm sgn}
(q)\,1=1$. In contrast, if $\vert v_{\alpha\nu} (q)\vert\geq
v_{s1}\,[1-l(\omega - l[\omega_0 +c_1\,\epsilon_{\alpha\nu}
(q)])/\Omega]$ we find that $v\in (-\infty,\,-\vert v_{\alpha\nu}
(q)\vert/[1-l(\omega - l[\omega_0 +c_1\,\epsilon_{\alpha\nu}
(q)])/\Omega])$ for ${\rm sgn} (q)\,1=-1$ and $v\in (\vert
v_{\alpha\nu} (q)\vert/[1-l(\omega - l[\omega_0
+c_1\,\epsilon_{\alpha\nu} (q)])/\Omega],\,\infty)$ for ${\rm sgn}
(q)\,1=1$. The limiting values ${\rm sgn} (q)\,\infty$, ${\rm sgn}
(q)\,v_{s1}$, and $v_{\alpha\nu} (q)/[1-l(\omega - l[\omega_0
+c_1\,\epsilon_{\alpha\nu} (q)])/\Omega]$ correspond to $\Delta
q=0$, $\Delta q=-{\rm sgn} (q)\,c_1\,l(\omega - l[\omega_0
+c_1\,\epsilon_{\alpha\nu} (q)])/[v_{s1}-\vert v_{\alpha\nu}
(q)\vert]$, and $\Delta q=-c_1\,[\Omega-l(\omega - l[\omega_0
+c_1\,\epsilon_{\alpha\nu} (q)])]/v_{\alpha\nu} (q)$,
respectively. By use of the general spectral function expressions
(\ref{ABONjl-J-CPHS}) and (\ref{B-PAR-J-CPHS-sum-0}) for $i=0$ we
then arrive to,

\begin{eqnarray}
B_{{\cal{N}}}^{l} (k,\,\omega) & \approx & {{\rm sgn}
(q)\,c_1\over
2\pi\,C_c\,C_s}\Bigl\{\Theta\Bigl(v_{s1}\,[1-{l(\omega -
l[\omega_0 +c_1\,\epsilon_{\alpha\nu} (q)])\over\Omega}]-\vert
v_{\alpha\nu} (q)\vert\Bigr)\int_{q-{\rm sgn} (q)\,c_1\,{l(\omega
- l[\omega_0 +c_1\,\epsilon_{\alpha\nu} (q)])\over v_{s1}-\vert
v_{\alpha\nu} (q)\vert)}}^{q+{\rm sgn} (q)\,c_1\,{l(\omega -
l[\omega_0 +c_1\,\epsilon_{\alpha\nu}
(q)])\over v_{s1}+\vert v_{\alpha\nu} (q)\vert)}}dq'\nonumber \\
& + & \theta\Bigl(\vert v_{\alpha\nu}
(q)\vert-v_{s1}\,[1-{l(\omega - l[\omega_0
+c_1\,\epsilon_{\alpha\nu}
(q)])\over\Omega}]\Bigr)\int_{-c_1\,{[\Omega-l(\omega - l[\omega_0
+c_1\,\epsilon_{\alpha\nu} (q)])]\over v_{\alpha\nu} (q)}}^{+{\rm
sgn} (q)\,c_1\,{l(\omega - l[\omega_0 +c_1\,\epsilon_{\alpha\nu}
(q)])\over v_{s1}+\vert v_{\alpha\nu} (q)\vert)}}dq'\Bigr\}
\nonumber \\
& \times & {\breve{B}}^{l,0} \left(\omega -l[\omega_0
+c_1\,\epsilon_{\alpha\nu} (q) + \Delta q\,c_1\,v_{\alpha\nu}
(q)],\,v_{\alpha\nu} (q) - c_1\,{l[\omega - l[\omega_0
+c_1\,\epsilon_{\alpha\nu} (q)]\over \Delta q}\right) \, ;
\hspace{0.25cm} l= \pm 1 \, . \label{B-bl}
\end{eqnarray}
Next, we change the integration variable from $q'$ to $v$ and
reach the following spectral-function expression,

\begin{eqnarray}
B_{{\cal{N}}}^{l} (k,\,\omega) & \approx & {{\rm sgn} (q)\,1\over
2\pi\,C_c\,C_s}\,l(\omega - l[\omega_0 +c_1\,\epsilon_{\alpha\nu}
(q)])\Bigl\{\,\int_{-{\rm sgn} (q)\,\infty}^{-{\rm sgn}
(q)\,v_{s1}}dv \nonumber \\
& + & \Theta\Bigl(v_{s1}\,[1-{l(\omega - l[\omega_0
+c_1\,\epsilon_{\alpha\nu} (q)])\over\Omega}]-\vert v_{\alpha\nu}
(q)\vert\Bigr)\int_{{\rm sgn} (q)\,v_{s1}}^{{\rm sgn}
(q)\,\infty}dv\nonumber \\
& + & \theta\Bigl(\vert v_{\alpha\nu}
(q)\vert-v_{s1}\,[1-{l(\omega - l[\omega_0
+c_1\,\epsilon_{\alpha\nu} (q)])\over\Omega}]\Bigr)
\int_{v_{\alpha\nu} (q)\over [1-l(\omega - l[\omega_0
+c_1\,\epsilon_{\alpha\nu} (q)])/\Omega]}^{{\rm sgn}
(q)\,\infty}dv\Bigr\}{1\over (v-v_{\alpha\nu}
(q))^2}\nonumber \\
& \times & {\breve{B}}^{l,0} \left((\omega -l[\omega_0
+c_1\,\epsilon_{\alpha\nu} (q)])\Bigl[{v\over v-v_{\alpha\nu}
(q)}\Bigr],\, v\right) \, ; \hspace{0.5cm} l= \pm 1 \, .
\label{B-bl-v}
\end{eqnarray}
Changing the integration variable from $v$ to $z=1/v$ and using of
the asymptotic expression (\ref{B-breve-asym}) of Eq.
(\ref{B-bl-v}) for small but finite values of $(\omega -l[\omega_0
+c_1\,\epsilon_{\alpha\nu} (q)])$, leads finally to expressions
(\ref{Ichiun}) and (\ref{Can}). (The theta functions added to the
expression (\ref{Ichiun}) define the energy range where it is
valid.) Similar procedures lead to these expressions for the
$\alpha\nu\neq c0,\,s1$ branches provided one uses the
corresponding parametric equations given in Eq. (\ref{*line}).

The derivation of expressions (\ref{Ichiun-bar}) and
(\ref{Can-bar}), which refer to $k$ and $\omega$ values in the
vicinity of the end points of a $\alpha\nu =c0,\,s1$ branch line
such that $\omega\approx \iota\,v_{\alpha\nu}(k-lk_0)+l\omega_0$,
follows the same steps as for the general branch-line
spectral-function expression. When $\Delta\omega$ is small but
finite the convolution function (\ref{B-iota-bar}) can be written
as follows,

\begin{equation}
{\breve{B}}^{l,0}_{\alpha\nu,\,\iota} (\Delta\omega,v) \approx
{1\over N_a}{F_{\alpha\nu,\,\iota}^0 (1/v)\over
4\pi\sqrt{v_{c0}\,v_{s1}}}\,\Theta
\Bigl(l\Delta\omega\Bigr)\,\Bigl({l\Delta\omega\over
4\pi\sqrt{v_{c0}\,v_{s1}}} \Bigr)^{-2+\zeta^0_{\alpha\nu,\,\iota}}
\, ; \hspace{0.5cm} \alpha\nu = c0,\, s1 \, ; \hspace{0.15cm}
\iota =\pm 1 \, ; \hspace{0.50cm} l =\pm 1 \, ,
\label{B-breve-asym-iota}
\end{equation}
where we used the relation $[l\Delta
q_{line}/2\pi]^{2\Delta_{\alpha\nu}^{-\iota}}=1/N_a$. Here the
functional $\zeta^0_{\alpha\nu,\,\iota}$ is given by,

\begin{equation}
\zeta^0_{\alpha\nu,\iota} = 2\Delta_{\alpha\nu}^{\iota}
+2\Delta_{{\bar{\alpha}\bar{\nu}}}^{+1}+2\Delta_{{\bar{\alpha}\bar{\nu}}}^{-1}
= \zeta^0_{\alpha\nu} - 2\Delta_{\alpha\nu}^{-\iota} \, ;
\hspace{0.5cm} \alpha\nu = c0,\, s1 \, ; \hspace{0.15cm} \iota
=\pm 1 \, , \label{zetasc}
\end{equation}
and the function $F^0_{\alpha\nu,\,\iota} (z)$ reads,

\begin{eqnarray}
F^0_{\alpha\nu,\,\iota} (z) & = & {D_0\over
4^{2\Delta_{\alpha\nu}^{-\iota}}}\,\Bigl[\prod_{\alpha'\nu' =
c0,\,s1}\prod_{\iota' =\pm 1}{1\over \Gamma
(2\Delta_{\alpha'\nu'}^{\iota'})}\Bigr]\int_0^1
dx\,\Bigl\{\prod_{\iota'' =\pm 1}\Theta\Bigl(1
-x+\iota''\,v_{\bar{\alpha}\bar{\nu}}\,z-\iota\iota''\,{v_{\bar{\alpha}\bar{\nu}}\over
v_{\alpha\nu}}\,x\Bigr)\nonumber \\
& \times & \left(\sqrt{{v_{\alpha\nu}\over
v_{\bar{\alpha}\bar{\nu}}}}\,\Bigl[1
-x+\iota''\,v_{\bar{\alpha}\bar{\nu}}\,z-\iota\iota''\,{v_{\bar{\alpha}\bar{\nu}}\over
v_{\alpha\nu}}\,x\Bigr]\right)^{2\Delta_{{\bar{\alpha}\bar{\nu}}}^{\iota''}-1}\Bigr\}
\,\left(\sqrt{{v_{\bar{\alpha}\bar{\nu}}\over
v_{\alpha\nu}}}\,2x\right)^{2\Delta_{\alpha\nu}^{\iota}-1} \, .
\label{C-v-iota}
\end{eqnarray}
Furthermore, use of the same procedures as above but with the use
of the functions (\ref{B-iota-bar}), (\ref{B-breve-asym-iota}),
and (\ref{C-v-iota}) instead of (\ref{B-l-i-breve}),
(\ref{B-breve-asym}), and (\ref{C-v}), respectively, leads
straightforwardly to expressions (\ref{Ichiun-bar}) and
(\ref{Can-bar}). The main difference is that the factor $1/N_a$
which multiplies the $q'$ integration factor $L/2\pi$ to give
$1/2\pi$ on the right-hand side of Eq. (\ref{B-bl}) arises from
the state summation of Eq. (\ref{B-PAR-J-CPHS-sum-0}), whereas in
the case of the constant (\ref{Can-bar}) it comes from expression
(\ref{B-breve-asym-iota}).



\end{document}